\documentclass{lmcs}
\pdfoutput=1
\usepackage[utf8]{inputenc}

\usepackage{lastpage}
\lmcsdoi{21}{4}{2}
\lmcsheading{}{\pageref{LastPage}}{}{}%
{Feb.~14,~2023}{Oct.~03,~2025}{}

\usepackage{soul}
\usepackage[noend]{algpseudocode} \usepackage{algorithm}

\usepackage{subcaption}
\makeatletter         
\def\figurecaption#1#2{\noindent\hangindent 40pt
                       \hbox to 36pt {\small\sl #1 \hfil}
                       \ignorespaces {\small #2}}

\long\def\@makecaption#1#2{
  \vskip 10pt 
  \settowidth{\@tempdima}{#2}
  \ifdim\@tempdima>0pt
       \setbox\@tempboxa\hbox{#1: #2}
     \else
       \setbox\@tempboxa\hbox{#1 #2}
   \fi
   \ifdim \wd\@tempboxa >\hsize               
       \begin{list}{#1:}{
       \settowidth{\labelwidth}{#1:}
       \setlength{\leftmargin}{\labelwidth}
       \addtolength{\leftmargin}{\labelsep}
        }\item #2 \end{list}\par   
     \else                                    
       \hbox to\hsize{\hfil\box\@tempboxa\hfil}  
   \fi}
\makeatother

\usepackage{xspace,amstext,amssymb,url}
\usepackage{amsthm}

\usepackage{graphicx}
\usepackage{tikz}
\usetikzlibrary{shapes,backgrounds}
\usetikzlibrary{automata}
\usetikzlibrary{calc}
\usepackage{mathtools}
\usepackage{xspace}
\usepackage{xfrac}
\usepackage{booktabs}
\usepackage{multirow}
\usepackage{scalerel}
\usepackage{pbox}

\def\mid{\enspace|\enspace}

\DeclareFontFamily{U}{matha}{\hyphenchar\font45}
\DeclareFontShape{U}{matha}{m}{n}{
      <5> <6> <7> <8> <9> <10> gen * matha
      <10.95> matha10 <12> <14.4> <17.28> <20.74> <24.88> matha12
      }{}
\DeclareSymbolFont{matha}{U}{matha}{m}{n}

\DeclareMathSymbol{\oplus}        {2}{matha}{"60}
\DeclareMathSymbol{\odot}         {2}{matha}{"64}
\DeclareMathSymbol{\ovoid}        {2}{matha}{"6C}

\newcommand{\oplust}{\mathrel{\oplus_{\vartriangle}}}
\newcommand{\odott}{\mathrel{\odot_{\vartriangle}}}

\newcommand{\cO}{{\mathcal O}}

\newcommand{\nat}{\Bbb{N}}

\newcommand{\balance}{\mathsf{balance}\xspace}
\newcommand{\height}{\mathsf{height}\xspace}
\newcommand{\lp}{\mathsf{lp}\xspace}
\newcommand{\col}{\mathsf{color}\xspace}

\newcommand{\np}{NP\xspace}

\newcommand{\lab}{\mathsf{lab}}
\newcommand{\states}{\ensuremath{\operatorname{states}}}
\newcommand{\nodes}{\ensuremath{\operatorname{Nodes}\xspace}}
\newcommand{\inceval}{\textsc{IncEval}\xspace}

\newcommand{\enuminc}{\textsc{IncEnum}\xspace}
\newcommand{\incenum}{\textsc{IncEnum}\xspace}

\newcommand{\id}{\operatorname{id}}

\newcommand{\xfunc}[1]{\textsc{#1}\xspace}
\newcommand{\out}{\xfunc{Output}}
\newcommand{\enum}{\xfunc{Enum}}
\newcommand{\complete}{\xfunc{Complete}}

\newcommand{\parent}{\ensuremath{\mathsf{Parent}\xspace}}

\newcommand{\Root}{\ensuremath{\mathsf{root}\xspace}}
\newcommand{\numleaves}[1]{||#1||}

\newcommand{\Dom}{\nodes}

\newcommand{\init}{\textsf{Init}\xspace}
\newcommand{\lc}[1]{\ensuremath{\mathsf{child_L}(#1)\xspace}}
\newcommand{\rc}[1]{\ensuremath{\mathsf{child_R}(#1)\xspace}}

\newcommand{\hole}{{\scaleobj{0.75}{\Box}}}
\newcommand{\context}[1]{#1_\hole}

\newcommand{\ta}{NFTA\xspace}
\newcommand{\sta}{NFSTA\xspace}
\newcommand{\etree}{\varepsilon\xspace}
\newcommand{\emptyA}{()\xspace}
\newcommand{\FA}{\mathcal{F}\xspace}
\newcommand{\tra}{\FA\xspace}

\newcommand{\signature}{{\operatorname{SIG}}}
\newcommand{\extsig}{{\operatorname{SIG}^+}}
\newcommand{\signatures}{{\operatorname{SIG}_Q}}

\newcommand{\pre}{{\mathsf{pre}}}
\newcommand{\self}{{\mathsf{self}}}
\newcommand{\post}{{\mathsf{post}}}

\newcommand{\transition}[3]{#1 \xrightarrow[{\raisebox{1.5mm}[0mm][0mm]{\scriptsize $#2$}}]{} #3}

\newcommand{\call}[1]{\textsc{#1}\xspace}
\newcommand{\doRotation}{\call{DoRotation}}

\newcommand{\tryReduceHeight}{\call{TryReduceHeight}}
\DeclareMathSymbol{\mhyphen}{\mathord}{AMSa}{"39}
\newcommand{\deepchild}{\mathsf{child_D}}
\newcommand{\firstchild}{\text{first-child}}
\newcommand{\nextsibling}{\text{next-sibling}}
\newcommand{\roots}{\mathsf{roots}\xspace}

\newcommand{\deficit}{\mathsf{deficit}\xspace}

\newcommand{\fbalance}{\mathcal{B}\xspace}
\newcommand{\psiup}{\Psi_{\uparrow}}\newcommand{\psidown}{\Psi_{\downarrow}}\newcommand{\psistay}{\Psi_{\rightarrow}}\newcommand{\concat}{\oplus}
\newcommand{\conapp}{\odot}
\newcommand{\op}{\ovoid}

\newcommand{\ignore}[1]{}

\tikzset{
  no edge/.style={
    edge from parent/.append style={draw=none}
  },
  topanchor/.style={
    edge from parent/.append style={child anchor=center}
  }
}

\setul{0.5pt}{0.3mm}

\setstcolor{red}

\begin{document}

\title[MSO Queries on Trees: Enumerating Answers under Updates]{MSO Queries on Trees: Enumerating Answers under Updates Using Forest
  Algebras}
\author[S.~Kleest-Mei{\ss}ner]{Sarah Kleest-Mei{\ss}ner\lmcsorcid{0000-0002-4133-7975}}[a]
\author[J.~Marasus]{Jonas Marasus\lmcsorcid{0000-0001-9298-2360}}[a]
\author[M.~Niewerth]{Matthias Niewerth\lmcsorcid{0000-0003-2032-5374}}[b]
\address{Humboldt-Universit\"at zu Berlin, Germany}
\address{Universit\"at Bayreuth, Germany}


\begin{abstract}
  We describe a framework for maintaining forest algebra representations that
  are of logarithmic height for unranked trees. Such representations can be computed in
  $\cO(n)$ time and updated in $\cO(\log(n))$ time. The framework is of
  potential interest for data structures and algorithms for trees whose
  complexity depend on the depth of the tree (representation). We provide an
  exemplary application of the framework to the problem of efficiently
  enumerating answers to MSO-definable queries over trees which are subject to
  local updates. We exhibit an algorithm that uses an $\cO(n)$ preprocessing
  phase and enumerates answers with $\cO(\log(n))$ delay between them. When the
  tree is updated, the algorithm can avoid repeating expensive preprocessing and
  restart the enumeration phase within $\cO(\log(n))$ time. Our algorithms and
  complexity results in the paper are presented in terms of node-selecting tree
  automata representing the MSO queries.
\end{abstract}

\maketitle

\section{Introduction}
Efficient query evaluation is one of the most central problems in
databases. Given a query $Q$ and a database $D$, we are asked to
compute the set $Q(D)$ of tuples of $Q$ on $D$. In general, the number
of tuples in $Q(D)$ can be extremely large: when $Q$ has arity $k$ and
$D$ has size $n$, then $Q(D)$ can contain up to $n^k$ tuples. Since
databases are  typically very large, it may be unfeasible to compute $Q(D)$
in its entirety.

A straightforward solution to this problem is top-$k$ query answering,
where one is interested in the $k$ most relevant answers according to
some metric. Another way to deal with this problem is to produce the
answers one by one without repetition. This is known as \emph{query
  enumeration} (see, e.g.,
\cite{Bagan-CSL06,Courcelle-DAM09,Durand-CSL11,KazanaS-pods13,KazanaS-tocl13,Segoufin-ICDT13}).
More precisely, query enumeration aims at producing a small number of
answers first and then, on demand, producing further small batches of
answers as long as the user desires or until all answers are
depleted. Most existing algorithms for query enumeration consist of
two phases: the \emph{preprocessing phase}, which lasts until the
first answer is produced, and the \emph{enumeration phase} in which
the following answers are produced without repetition. It is natural to try to
optimize two kinds of time intervals in this procedure: the time of
the preprocessing phase and the \emph{delay} between answers, which is
the time required between two answers in the enumeration phase.  Thus,
when one can answer $Q(D)$ with preprocessing time $p$ and delay $d$,
one can compute $Q(D)$ in time $p + d\cdot|Q(D)|$, where $|Q(D)|$ is
the number of answers.

Much attention has been given to finding algorithms that answer
queries with a linear-time preprocessing phase and constant-time delay
between answers~\cite{Segoufin-ICDT13}. The preprocessing phase is
usually used to build an index that allows for efficient enumeration.
Since databases can be subjected to frequent updates and preprocessing
typically costs linear time, it is usually not an option to recompute
the index after every update.  We want to address this concern and
investigate what can be done if one wants to deal with such updates
more efficiently than simply re-starting the preprocessing phase.

We cannot expect efficient algorithms for query enumeration in the general case,
as even deciding whether a (Boolean) query has an answer is \np-hard. Therefore,
we have to restrict the problem. In the literature, there are two general
directions towards tractable query enumeration under updates: restricting the class of queries
or restricting the structure of the data.

There has been research on enumerating certain classes of conjunctive
queries with constant delay and sublinear update
time~\cite{BerkholzKS-pods17,BerkholzKS-icdt17}.
Similarly, there are algorithms that can enumerate FO+MOD queries
with constant delay and constant update time on bounded-degree
databases~\cite{BerkholzKS-tods18}.

In this article, we follow the second
approach and study the enumeration problem for MSO queries with free node
variables over \emph{trees}. Furthermore, the trees can be subjected
to local updates. We consider updates that relabel a node or
insert/delete a leaf. Our aim is to provide an index structure that
can be efficiently updated, when the underlying tree changes. This
makes the enumeration phase \emph{insensitive to such updates}: when
our algorithm is producing answers with a small delay in the
enumeration phase and the underlying data $D$ is updated, we can
update the index and re-start enumerating on the new data within the
same delay.

The complexity results in this article are presented in terms of the
size of the tree; the arity $k$ of the query; and the number of
states of a non-deterministic node-selecting finite tree automaton for
the query.  The connection between run-based node-selecting automata
and MSO-queries is well-known, see,
e.g.~\cite{NiehrenPTT-dbpl05,ThatcherW-mst68}.

When measuring complexity in terms of query size, we have to keep in mind that
MSO queries can be non-elementarily smaller than their equivalent
non-deterministic node-selecting tree automata. Therefore, our
enumeration algorithm is non-elementary in terms of the MSO formula,
which cannot be avoided unless P $=$ NP~\cite{FrickG-apal04}.
For this reason, MSO is usually not used as a query language in practice;
although it is widely regarded as a good yardstick for
expressiveness. 

Our complexities are exponential in the arity $k$ of the
queries. However for practical scenarios, $k$ is usually very small.
We note that $k=2$ suffices for modeling XPath queries, which are
central in XML querying.

Although we do not obtain constant-delay algorithms as in previous
work on static trees, we can prove that, in the dynamic setting
$\cO(\log(n))$ delay is possible. This means that, after
receiving an update, we do not need to restart the $\cO(n)$
preprocessing phase but only require $\cO(\log(n))$ time to produce the
first answer on the updated tree and continue enumerating from there.
We allow updates to arrive at any time: If an update arrives during
the enumeration phase, we immediately start the enumeration phase for
the new structure.

In~\cite{AmarilliBMN-pods19} it has been shown that the constant delay
enumeration approach for MSO queries over static trees
from~\cite{AmarilliBJM-icalp17,AmarilliBM-icdt18} can be extended to allow
updates, if the trees are represented by balanced forest algebra formulas that
are described in this article and were already presented in a preliminary
version~\cite{Niewerth-lics18}.

\subsection*{Previous Results on MSO Queries on Trees}
\begin{table*}
  \begin{tabular}{cclr}
    Update & Delay & Remarks & Reference \\ \toprule
    $\cO(\log^2(n))$ & --- & only Boolean queries; $\cO(\log(n))$ on strings & \cite{BalminPV-tods04} \\
    $\cO(\log^2(n))$ & --- & Boolean XPath queries & \cite{BjorklundGM-tods10}  \\
    --- & $\cO(1)$ & updates in $\cO(n)$ by recomputation & \cite{Bagan-CSL06}  \\
    --- & $\cO(1)$ & different proof using decomposition forests &  \cite{KazanaS-tocl13} \\
    --- & $\cO(1)$ & different proof using circuits &  \cite{AmarilliBJM-icalp17} \\
    --- & $\cO(1)$ & streaming algorithm & \cite{MunozR24} \\
    $\cO(\log^2(n))$ & $\cO(\log^2(n))$ & complexities drop to $\cO(\log(n))$ on strings & \cite{LosemannM-LICS14} \\
    $\cO(\log(n))$ & $\cO(1)$ & only works on strings; huge constants & \cite{NiewerthS-pods18} \\
    $\cO(\log(n))$ & $\cO(1)$ & only relabel updates; uses circuits & \cite{AmarilliBM-icdt18} \\
    $\cO(\log(n))$ & $\cO(1)$ & uses the techniques developed in this article &\cite{AmarilliBMN-pods19}\\ 
    $\cO(\log(n))$ & $\cO(\log(n))$ & uses forest algebras & this work \\
    \bottomrule
  \end{tabular}
\caption{Data complexity of existing solutions. Preprocessing time is always in $\cO(n)$.}\label{table:related}
\end{table*}

We have collected previous results on evaluation and enumeration of
MSO queries on strings and trees in Table~\ref{table:related}.

For MSO sentences, this problem has been studied by Balmin,
Papakonstantinou, and Vianu~\cite{BalminPV-tods04}. Balmin et al.\
show how one can efficiently maintain satisfaction of a finite tree
automaton (and therefore, an MSO property) on a tree $t$ which is
subjected to updates. More precisely, when an update transforms $t$ to
$t'$, they want to be able to decide very quickly after the update
whether $t'$ is accepted by the automaton.  Taking $n$ as the size of
$t$, they show that, using a one-time preprocessing phase of time
$\cO(n)$ to construct an auxiliary data structure, one can always
decide within time $\cO(\log^2(n))$ after the update whether $t'$ is
accepted. The delay between answers is irrelevant in the setting of
Balmin et al.\ since their queries always have a Boolean
answer. Bj{\"o}rklund et al.~\cite{BjorklundGM-tods10} show a similar result for XPath queries,
which are less expressive than MSO but can be exponentially more
succinct than tree automata, which leads to better constants.
Losemann and Martens~\cite{LosemannM-LICS14} extended the
work of Balmin et al.\ to enumeration of $k$-ary queries under updates
with $\cO(\log^2(n))$ delay and update time. Our goal is to improve the
delay and update time to $\cO(\log(n))$.

The enumeration problem
of static trees was studied by Bagan~\cite{Bagan-CSL06}, who showed
that (fixed) monadic second-order~(MSO) queries can be evaluated with
linear time preprocessing and constant delay over structures of
bounded tree-width. Independently, another constant delay algorithm
(but with $\cO(n \log(n))$ preprocessing time) was obtained
by Courcelle et {al.}~\cite{Courcelle-DAM09}. Kazana and Segoufin~\cite{KazanaS-tocl13} provided
an alternative proof of Bagan's result based on a deterministic
factorization forest theorem by Colcombet~\cite{Colcombet-ICALP07}, which is
itself based on a result of Simon~\cite{Simon-tcs90}.  Such (deterministic)
factorization forests provide a good divide-and-conquer strategy for
words and trees, but it is unclear how they can be maintained under
updates. It seems that they would have to be recomputed entirely after
an update which is too expensive for our purposes.

With exception of
\cite{Bagan-CSL06}, which presents an algorithm that is cubic in terms
of the tree automaton, these papers present complexities in terms
of the size of the trees only, that is, they consider the MSO formula
to be constant. To the best of our knowledge, the data structures in
these approaches cannot be updated efficiently if the underlying tree
is updated. An overview of enumeration algorithms with constant
delay was given in \cite{Segoufin-ICDT13}.

Mu\~{n}oz and Riveros~\cite{MunozR24} present a streaming algorithm for MSO
  queries on trees that enumerates answers with constant delay.

\subsection*{Heavy Path Decomposition vs.\ Forest Algebras}
A main idea in~\cite{BalminPV-tods04} is a decomposition of trees
into heavy paths which allows one to decompose the problem for trees
into $\cO(\log(n))$ similar problems on words, for which a solution
was given by Patnaik and Immerman in~\cite{PatnaikI-jcss97}. This
allows to solve the incremental evaluation problem with $\cO(n)$
preprocessing time and $\cO(\log^2(n))$ update time, where one $\log$
factor stems from the heavy path decomposition and the other from
solving the problem over strings using monoids of finite string automata.

The approach of~\cite{BalminPV-tods04} was later extended to
enumeration of MSO queries by Losemann and
Martens~\cite{LosemannM-LICS14}. They tweaked the monoid to contain
additional information needed to find the symbols that appear in query
results, which allows logarithmic delay and update time. Just as
Balmin et al., Losemann and Martens use heavy path decomposition to
lift the algorithm from words to trees resulting again in an
additional logarithmic factor in the delay and update time.

We adapt the algorithm of Losemann and Martens from monoids to forest algebras
in Section~\ref{sec:enum}. This saves us a logarithmic factor compared to their
results. We believe that the framework we introduce in
Section~\ref{sec:maintaining} to compute and maintain forest algebra formulas with
logarithmic height can be applied in other areas to lift results from strings to
trees.

\subsection*{Tree Decomposition vs.\ Forest Algebras}
In~\cite{AmarilliBM-icdt18}, Amarilli et {al.} use tree
decompositions~\cite{BodlaenderH-siam98} to convert arbitrary trees to trees
of logarithmic height. Having a tree of logarithmic height is central in their
algorithm to allow enumeration of MSO queries over trees under relabeling
updates. According to Amarilli et {al.}, the biggest obstacle in generalizing
their work to allow structural updates (insertion/deletion of nodes) of the tree
was the inability to update the tree decomposition when the input tree changes.

Following up on the preliminary version of this article~\cite{Niewerth-lics18},
it has been shown in~\cite{AmarilliBMN-pods19} that the enumeration algorithm
from~\cite{AmarilliBM-icdt18} can be generalized to allow structural updates on
the tree using the methods we develop in sections~\ref{sec:maintaining}
and~\ref{sec:automata}.

Korhonen et al.~\cite{KorhonenMNPS23} present an algorithm that maintains a tree
decomposition of a given graph under updates. If the graph is guaranteed to never
exceed treewidth $k$, they maintain a tree decomposition of treewidth at most
$6k+5$ with an amortized update time of $\mathcal{O}_k(2^{\sqrt{\log(n)}
  \log(\log(n))})$, where $\mathcal{O}_k$ hides factors that depend on $k$. They
can also maintain satisfaction of CMSO$_2$-sentences in the same time bounds.

\subsection*{Further Related Work}
There are implementations for and experimental results on incremental
evaluation of XML documents \mbox{wrt.} DTDs~\cite{BarbosaMLMA-icde04}
and regular expressions with counters on
strings~\cite{BjorklundMT-cikm15}. 
The query evaluation problem has also been studied from a descriptive
complexity point of view, e.g., for conjunctive
queries~\cite{ZeumeS-jcss17} and the reachability query on
graphs~\cite{DattaKMSZ-jacm18}.

\subsection*{Contributions}
We describe a mechanism for representing trees by the means of forest algebra
formulas in Section~\ref{sec:maintaining}. Furthermore, we show how these
representations can be maintained after updates that can relabel, insert, or
delete individual nodes of the tree. We believe that these techniques are
interesting on their own. Building on top of a preliminary version of this
article,~\cite{AmarilliBMN-pods19} already uses the techniques provided here to
achieve further improvements in the enumeration of MSO queries on trees.

Section~\ref{sec:automata} shows how forest algebras are related to tree
automata through their transition algebra. This relationship is exactly the same
as the relationship between monoids and string automata.

We provide a first application of our framework in
Section~\ref{sec:inceval}, where we show how to solve the incremental evaluation problem for Boolean MSO formulas
on trees by means of forest algebra formulas. This approach allows for
logarithmic update time after linear time preprocessing.

Finally, in Section~\ref{sec:enum}, we show how forest algebras can be used to
enumerate MSO queries over trees with logarithmic delay and logarithmic update
time after a linear preprocessing step. Even though there is an improved algorithm
available in~\cite{AmarilliBMN-pods19}, we believe that the presented algorithm
is useful. First, it gives a quite simple demonstration how the general
techniques presented in Section~\ref{sec:maintaining} can be applied, and second
it is a much simpler algorithm that might actually be more efficient for medium
sized trees than the more complex algorithm presented
in~\cite{AmarilliBMN-pods19}.

\subsection*{Earlier Version}
An earlier version of this article was presented at the 2018 Symposium on Logic
in Computer Science (LICS 2018)~\cite{Niewerth-lics18}.
The earlier version does not contain formal proofs. Unfortunately, the
unpublished proofs turned out to have a flaw. Therefore we modified the
algorithms slightly in order to show the claimed complexity bounds. We believe
that the algorithms as presented in~\cite{Niewerth-lics18} also meet the same
complexity bounds, but we were not able to prove this.

\section{Preliminaries}\label{sec:definitions}

\subsection*{Trees, Forests, Contexts}
In this article, trees and forests are labeled by some finite alphabet $\Sigma$,
rooted, and ordered. Contexts are forests which contain a single hole, denoted
by $\hole$, which may not be part of any forest. In detail:
A \emph{forest} is a tuple $F = (V, E, \leq, \lab)$, where $V$ is a
finite set of nodes with $\hole \notin V$, $E \subseteq V \times V$ is the set of edges, ${\leq}
\subseteq V \times V$ is the sibling order, and $\lab \colon V \to
\Sigma$ is the labeling function.

If $(v,w) \in E$, we say $w$ is a \emph{child} of $v$ and $v$ is the
\emph{parent} of $w$. If $(v,w)$ is within the transitive closure of $E$, we say
$w$ is a \emph{descendant} of $v$ and $v$ is an \emph{ancestor} of $w$. A node
without parents is called a \emph{root} and we denote the set of all roots of
$F$ by $\roots(F)$. A node without children is called a \emph{leaf}. The
relation $E$ must be acyclic and no node may have more than one
parent. If $v_1 \leq v_2$, we say, $v_1$ lies \emph{left} of $v_2$. For every
node $v$, $\leq$ must be a linear order on the set of all children of $v$.
Likewise, $\leq$ needs to be a linear order on $\roots(F)$. Other than that, $\leq$
must have no additional elements. A forest with a single root is called a
\emph{tree}.

A \emph{context} $C = (V, E, \leq, \lab)$ is a forest with the exception
that $V$ contains the hole $\hole$. The hole must be a leaf and does not carry
a label, i.e., $\lab$ is a function on $V \setminus \{\hole\}$.

We overload notation and use a symbol $a\in \Sigma$ to denote the forest that
contains a single node with label $a$. Similarly we use $\context a$ with $a \in
\Sigma$ to denote the context with an $a$-labeled root that has the hole as its
only child. We use $\varepsilon$ and $\hole$ to denote the empty forest and the
context that only consists of the hole, respectively.

We usually will denote forests by $F$ and contexts by $C$ possibly with
subscripts. We use $D$ (with subscripts) to denote that the object can be either
a forest or a context.

We will usually assume \mbox{w.l.o.g.} that for different forests
and contexts the sets of nodes  are disjoint (except $\hole$). This can be achieved by renaming
the nodes of one of the operands.

We define two operations on contexts and forests: The (horizontal)
\emph{concatenation} $\oplust$ places two forests or a forest and a
context next to each other and the (vertical) \emph{context application} $\odott$ which
inserts a forest or context into a context by replacing the hole with the roots
of the inserted forest or context. Not all combinations of operands are allowed
in order for the result to be well defined:
\begin{itemize}
\item It is not possible to concatenate two contexts; and
\item the left operand of a context application has to be a context.
\end{itemize}
We depict examples of the two operations in Figure~\ref{fig:operations}. In
detail the operations are defined as follows:

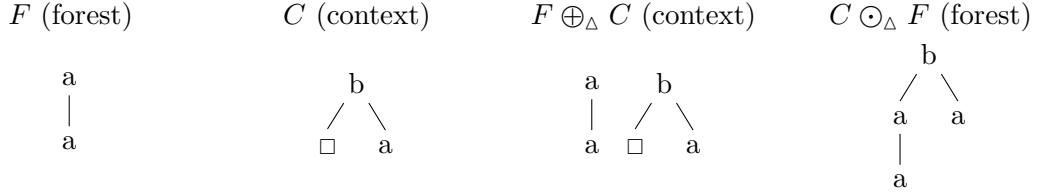
\begin{figure}
  \hfill
	\begin{minipage}[t]{0.2\textwidth}
		\centering
$F$ (forest)\\[1em]
		\begin{tikzpicture}[child anchor=north, level distance=2.2em, sibling distance=1em, node distance=1.9em]
      \node(a) {a}
        child { node {a}};
		\end{tikzpicture}
	\end{minipage}
  \hfill
	\begin{minipage}[t]{0.2\textwidth}
		\centering
$C$ (context)\\[1em]
		\begin{tikzpicture}[child anchor=north, level distance=2.2em, sibling distance=2em, node distance=1.9em]
		\node[] {b}
		child { node {$\hole$}}
		child { node {a}}
		;
		\end{tikzpicture}
	\end{minipage}
  \hfill
	\begin{minipage}[t]{0.2\textwidth}
		\centering
$F \oplust C$ (context)\\[1em]
		\begin{tikzpicture}[child anchor=north, level distance=2.2em, sibling distance=2em, node distance=2.5em]
		\node(a) {a}
		child { node {a}}
		;
		\node[right of=a] {b}
		child { node {$\hole$}}
		child { node {a}}
		;
		\end{tikzpicture}
	\end{minipage}
  \hfill
	\begin{minipage}[t]{0.2\textwidth}
		\centering
		$C\odott F$ (forest)\\
		\begin{tikzpicture}[child anchor=north, level distance=2.2em, sibling distance=2em]
		\node[] {b}
		child { node {a}
			child {node {a}}}
		child { node {a}}
		;
		\end{tikzpicture}
	\end{minipage}
  \hfill
	\caption{Examples for concatenation and context application}
	\label{fig:operations}
\end{figure}

Let $D_1=(V_1,E_1,\leq_1,\lab_1)$ and $D_2=(V_2,E_2,\leq_2,\lab_2)$ be two
forests or one forest and a context in either order. We compute $D = D_1 \oplust
D_2$ by taking the disjoint componentwise union and specifying that all roots of
$D_2$ have to be right of all roots of $D_1$. Formally we define
$D_1 \oplust D_2 = (V,E,\leq,\lab)$, where
\begin{align*}
  V \;\;&=\;\; V_1 \cup V_2 \\
  E \;\;&=\;\; E_1 \cup E_2 \\
  \leq \;\;&=\;\; \leq_1 \;\cup\; \leq_2 \;\cup\;\; \{(v_1,v_2)\mid v_1 \in \roots(D_1), v_2 \in \roots(D_2)\} \\
  \lab \;\;&=\;\; \lab_1 \;\cup\; \lab_2 
\end{align*}
We note that $D_1 \oplust D_2$ is a forest if $D_1$ and $D_2$ are
both forests and it is a context if one of $D_1$, $D_2$ is a context. If $D_1$
and $D_2$ are both contexts, then $D_1 \oplust D_2$ is undefined.

Let $C_1=(V_1,E_1,\leq_1,\lab_1)$ be a context and $D_2=(V_2,E_2,\leq_2,\lab_2)$ be a context or a forest. We compute $D = C_1 \odott D_2$
by first removing the hole from $C_1$, then taking the componentwise disjoint
union and afterwards tweaking the set of edges and sibling order of $D$ such that the ordered
list of roots of $D_2$ appears where the hole in $C_1$ was before. Let
$C_1^\boxtimes = (V_1^{\boxtimes},E_1^\boxtimes,\leq_1^\boxtimes,\lab_1)$ be the forest derived from $C_1$ by removing the hole.
Formally we define $C_1 \odott D_2 = (V,E,\leq,\lab)$, where 
\begin{align*}
V \;\;&=\;\; V_1^\boxtimes \cup V_2 \\
E \;\;&=\;\; E_1^\boxtimes \cup E_2 \cup \{(v_1,v_2) \mid (v_1,\hole) \in E_1, v_2 \in \roots(C_2)\} \\
  \leq \;\;&=\;\; {\leq_1^\boxtimes} \cup {\leq_2} \cup 
                \{(v_1,v_2) \mid (v_1,\hole) \in {\leq_1}, v_2 \in \roots(C_2) \text{ or } v_1 \in \roots(C_2), (\hole,v_2) \in {\leq_1}\}\\
\lab \;\;&=\;\; \lab_1 \cup \lab_2
\end{align*}
We note that $C_1 \odott D_2$ is a context if $D_2$ is a context and it is a
forest if $D_2$ is a forest.

\subsection*{Forest Algebras}
Here, we introduce forest algebras that were first described by Boja{\'n}czyk
and Walukiewicz~\cite{BojanczykW-2007}.  We prefer the syntax used in
the Handbook of Automata Theory~\cite{bojanczyk2021algebra} that also
provides a nice introduction.

A \emph{forest algebra} \[\FA\;\;=\;\;(H, V, \concat_{HH}, \concat_{HV}, \concat_{VH},
  \conapp_{VV}, \conapp_{VH}, \varepsilon, \hole )\] consists of two
monoids, $(H,\concat_{HH},\etree)$ and $(V,\conapp_{VV},\hole)$
along with three monoidal actions:
\begin{align*}
\concat_{HV} &\colon H \times V \to V & &\text{(left monoidal action of $H$ on
  $V$),}\\
\concat_{VH} &\colon V \times H \to V & &\text{(right monoidal action of $H$ on
  $V$), and}\\
\conapp_{VH} &\colon V \times H \to H & &\text{(right monoidal action of $V$ on
  $H$).}
\end{align*}
  
Intuitively, each element of $H$ represents a forest and each element in $V$
represents a context. The monoid operations correspond to concatenation of
forests and context application (on a context), respectively. The neutral
elements of $H$ and $V$ correspond to the empty forest and the context consisting
only of the hole. The monoidal actions correspond to concatenation of a forest
and a context (or the other way round) and context application of a context on a
forest.

\begin{table}
  \begin{tabular}{c@{\hspace{.5cm}}r@{\;\;}c@{\;\;}l@{\hspace{.5cm}}c}
    \toprule
     & & \clap{Axiom} & & Name \\ \midrule
    (A1) & $\etree \concat_{HH} F_1 \;\; = $ & $F_1$ & $ = \;\; F_1 \concat_{HH} \etree$      & \multirow{4}{*}{neutral element} \\
    (A2) & $ \etree \concat_{HV} C_1 \;\; = $ & $C_1$ & $ = \;\; C_1 \concat_{VH} \etree $ \\ 
    (A3) & $\hole \conapp_{VV} C_1 \;\; = $ & $C_1$ & $ = \;\; C_1 \conapp_{VV} \hole$   & \\ 
    (A4) & $\hole \conapp_{VH} F_1 \;\; = $ & $F_1$ &    & \\ \midrule
    (A5) & $(F_1 \concat_{HH} F_2) \concat_{HH} F_3 $ & $=$ & $ F_1 \concat_{HH} (F_2 \concat_{HH} F_3)$ & \multirow{6}{*}{associativity}\\
    (A6) & $(F_1 \concat_{HH} F_2) \concat_{HV} C_1 $ & $=$ & $ F_1 \concat_{HV} (F_2 \concat_{HV} C_1)$ \\
    (A7) & $(C_1 \concat_{VH} F_1) \concat_{VH} F_2 $ & $=$ & $ C_1 \concat_{VH} (F_1 \concat_{HH} F_2)$ \\
    (A8) & $(F_1 \concat_{HV} C_1) \concat_{VH} F_2 $ & $=$ & $ F_1 \concat_{HV} (C_1 \concat_{VH} F_2)$ \\
    (A9) & $(C_1 \conapp_{VV} C_2) \conapp_{VV} C_3 $ & $=$ & $ C_1 \conapp_{VV} (C_2 \conapp_{VV} C_3)$ \\
    (A10) & $(C_1 \conapp_{VV} C_2) \conapp_{VH} F_1 $ & $=$ & $ C_1 \conapp_{VH} (C_2 \conapp_{VH} F_1)$ \\\midrule
    (A11) & $(F_1 \concat_{HV} C_1) \conapp_{VH} F_2 $ & $=$ & $ F_1 \concat_{HH} (C_1 \conapp_{VH} F_2)$ & \multirow{4}{*}{interaction} \\
    (A12) & $(F_1 \concat_{HV} C_1) \conapp_{VV} C_2 $ & $=$ & $ F_1 \concat_{HV} (C_1 \conapp_{VV} C_2)$ \\
    (A13) & $(C_1 \concat_{VH} F_1) \conapp_{VH} F_2 $ & $=$ & $ (C_1 \conapp_{VH} F_2) \concat_{HH} F_1$ \\
    (A14) & $(C_1 \concat_{VH} F_1) \conapp_{VV} C_2 $ & $=$ & $ (C_1 \conapp_{VV} C_2) \concat_{VH} F_1$ \\
\bottomrule
  \end{tabular}
  \caption{Axioms of Forest Algebras}\label{tab:axioms}
\end{table}

In Table~\ref{tab:axioms}, we depict the axioms that forest algebras need to
satisfy. The two upper groups of axioms are the usual axioms for monoids and
monoidal actions stating the effect of the neutral elements and the laws of
associativity. The axioms in the bottom group describe the interactions of both
monoids and are special for forest algebras. The intuitive explanation of these
axioms is that if one concatenates a forest $F_1$ and a context $C_1$ (in either
order), it is solely the context that is responsible for
what happens in a subsequent context application. Therefore a subsequent context
application should have the same effect as first doing the context application
on $C_1$ and afterwards doing the concatenation with $F_1$.

We note that the Axioms A11--A14 are not presented
in~\cite{bojanczyk2021algebra}. This seems to be an oversight. Without these
axioms it is possible to construct a forest algebra $\FA$, such that there does not
exist a morphism from the free forest algebra (see below) to $\FA$.
However,~\cite{bojanczyk2021algebra} states that such a morphism exists for
every forest algebra.

We will often drop the indices of the monoid operations and monoidal
actions, i.e., we will just use $\concat$ and $\conapp$. Which
operation is needed is clear from the operands. In some cases, we do
not even specify, whether we refer to concatenation or context
application. In this case, we use $\op$. Given a formula $\Psi$ and an
inner node $v$ of the parse tree, we denote by $\op_v$ the operation
at node $v$.

\subsection*{Free Forest Algebra}
The \emph{free forest algebra} over an alphabet $\Sigma$ is defined as
\[\tra_\Sigma=(H,V, \oplust, \oplust, \oplust, \odott, \odott, \varepsilon, \hole)\;,\] where $(H, \oplust,
\varepsilon)$ is the monoid of all forests over $\Sigma$ with the concatenation
operation and $(V, \odott, \hole)$ is the monoid of all contexts over $\Sigma$
with the context application operation. We note that all five operations of the
algebra are provided by the two operations $\oplust$ and $\odott$ by
restricting them to the respective domains.
\begin{lem}
  The free forest algebra over any alphabet $\sigma$ satisfies all axioms of
  forest algebras.
\end{lem}
\begin{proof}[Proof sketch]
  All axioms can be shown by applying the definition of $\oplust$ and $\odott$.
  The axioms for the neutral elements $\varepsilon$ and $\hole$ (axioms A1--A4)
  hold, as the operations $\oplust$ and $\odott$ are defined by the means of
  unions where one side of each union is empty.

  For the associativity axioms of the horizontal concatenation (axioms
  A5--A8) it suffices to observe that $\leq$ is constructed in a way that
  is compatible with associativity. All other components of the resulting forest
  or context are constructed by taking the union, which is associative.

  The associativity of the context application (Axioms A9 and A10) can be
  shown by distinguishing the holes in $C_1$ and $C_2$. The context application
  involving $C_3$ or $F_1$ is always applied to the hole that originates from
  $C_2$. Analogously, in the axioms A11--A14, the context application
  involving $C_2$ or $F_2$ is always applied to the hole originating from $C_1$.
\end{proof}

\subsection*{Parse Trees (of Formulas)}
The \emph{parse tree} of a forest algebra formula (from some algebra $\FA=(V,H)$) is
a binary tree with inner nodes labeled by
$\{\concat_{HH},\concat_{HV},\concat_{VH},\conapp_{VV},\conapp_{VH}\}$ and leaf
nodes marked by some element of $V \cup H$. In most cases we will omit the
indices from $\concat$ and $\conapp$, as the concrete operation is implied by
the types of the operands. However, we will always assume that the concrete
operation is stored in the node. This way it is always possible to tell which
operand of a concatenation contains the hole (if any).

In our application, leaf
nodes of parse trees will always correspond to exactly one node of the represented tree.
That is, we do not use the empty forest $\varepsilon$ or the context
consisting solely of the hole $\hole$. Nor do we have leaves in a parse tree that
correspond to more than one node of the represented tree.
In the case of formulas from the free forest algebra,
leaves will be labeled by elements from $\Sigma \cup \{\context a \mid a \in
\Sigma\}$, where elements from $\Sigma$ indicate individual nodes with the
corresponding label and elements of the form $\context a$ indicate a node with
the given label that has as its only child the hole. As a consequence, the 
contexts from Figure~\ref{fig:operations} will not occur in our application.

Given a formula $\Psi$, we denote by $\numleaves{\Psi}$ the number of leaves in
(the parse tree of) $\Psi$, which---in our application---is always
equal to the number of nodes in the represented forest or context. As parse
trees are always binary trees, they will have one inner node less than leaves.
Therefore, a parse tree is always roughly twice the size of the represented
forest or context.

The \emph{balance} $\balance(v)$ of an inner node $v$ of the parse tree is the
height difference of the two trees, rooted at the children of $v$. Positive
values for $\balance(v)$ denote that the right subtree is higher than the left
subtree. By $\fbalance(\Psi)$ we denote the sum of all absolute balances, i.e.,
\[\fbalance(\Psi) \;\;=\;\; \sum_{v \in \Psi} |\balance(v)|\;.\]

To avoid ambiguities in the language---the balancedness of a formula increases
when the (absolute) balances decrease---we say
that the balance of a formula or node \emph{improves} (by some amount) to denote that the
absolute value of balance decreases (by some amount). Similarly we use the verb
\emph{worsen} in the opposite case.

For each node $v$ in a parse tree, we denote by $\lc v$ and $\rc v$, the left
and right child of $v$, respectively. If $\balance(v) \neq 0$, we denote with
$\deepchild(v)$ the child of $v$ that belongs to the deeper subtree.
For every node $v$ of a formula $\Psi$, the \emph{long path} of $v$ denoted by
$\lp(v)$ is defined recursively as follows:
\[
  \lp(v) \;\;=\;\; \begin{cases}
    v & \text{if $\balance(v)=0$} \\
    v \cdot \lp(\lc{v}) & \text{if $\balance(v) < 0$} \\
    v \cdot \lp(\rc{v}) & \text{if $\balance(v) > 0$}
  \end{cases}
\]
We use $\lp^{-1}(v)$ to denote the upwards path starting at $v$ that contains
all nodes $u$ with $v$ being on $\lp(u)$.

\begin{figure}
	\centering
	\begin{minipage}{.40\textwidth}
		\begin{tikzpicture}[level distance=2em, sibling distance=2.5em, 
		level 1/.style={sibling distance=8em},
		level 2/.style={sibling distance=4em},
		level 3/.style={sibling distance=2em}]
		\node{$\odot$}
		child { node{$\odot$}
			child {node {$a_\hole$}}
			child {node {$\oplus$}
				child {node {$b_\hole$}}
				child {node {$b$}}
			}
		}
		child {node{$\odot$}
			child {node{$\odot$}
				child {node{$a_\hole$}}
				child {node{$b_\hole$}}
			}
			child {node{$\oplus$}
				child {node{$a$}}
				child {node{$a$}}
			}
		}
		;	
		\end{tikzpicture}
	\end{minipage}
  \hspace{1cm}
	\begin{minipage}{.15\textwidth}
		\begin{tikzpicture}[sibling distance=3em, level distance=2em]
		\node [] {$a$}
		child { node [] {$b$}
			child { node [] {$a$}
				child { node [] {$b$}
					child { node [] {$a$}}
					child { node [] {$a$}}
				}
			}
		}
		child { node [] {$b$}}
		;
		\end{tikzpicture}
	\end{minipage}
	\caption{Parse tree of a forest-algebra formula and its generated forest.}
	\label{fig:example_formula}
\end{figure}
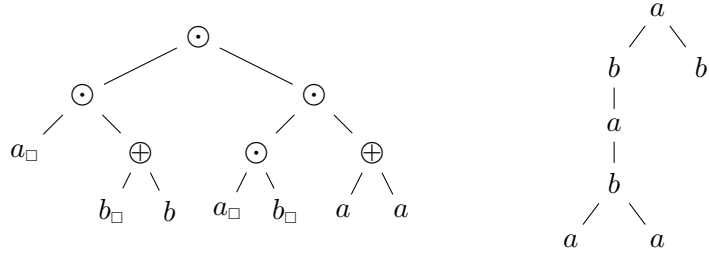

We use forest-algebra formulas synonymously with their parse trees. In this
sense the height of a forest-algebra formula $\height(\Psi)$ is defined as the height
of its parse tree, i.e., the maximal number of nodes on path from the root to a
leaf.

We use $v$ and $w$ to denote nodes of either a given tree $T$, or a parse tree
of a forest algebra formula $\Psi$, respectively. To keep the notation clean, we identify
leaves of $\Psi$ with nodes of $T$, whenever we have a formula $\Psi$ describing
a tree $T$. Given a node $v$ of the parse tree of $\Psi$, we use $\Psi_v$ to
denote the subformula of $\Psi$ rooted at $v$ and $T_v$ to denote the forest or
context described by $\Psi_v$.

\subsection*{Morphisms of Forest Algebras}

A homomorphism $h=(h_H,h_V)$ from a forest algebra $\tra_1=(H_1,V_1)$ to a
forest algebra $\tra_2=(H_2,V_2)$ is given by two monoid morphisms $h_H \colon
H_1 \to H_2$, $h_V \colon V_1 \to V_2$ that additionally satisfy
\[h_H(C \conapp F)\;\;=\;\;h_V(C) \conapp h_H(F)\] for all $C \in V_1$ and
$F \in H_1$. To simplify notation, we will omit the indices $H$ and
$V$ and use $h$ for both morphisms.

For every forest algebra $\FA=(V,H)$ one can define a morphism from the free
forest algebra $\FA_\Sigma$ by providing a function $h \colon \{\context a \mid a \in
\Sigma\} \rightarrow V$, i.e., by providing the mappings for all contexts that
have a single root with the hole as its only child. The function $h$ can be
extended to a morphism from the free forest algebra by applying the algebra
operations. For example, the mapping for a single $a$-labeled node can be computed
as $h(\context a) \conapp_{VH} \varepsilon$.

\subsection*{Updates of Forests and Formulas}

We consider the following \emph{updates} on trees:
\begin{enumerate}[(i)]
\item $\mathsf{relab}(v,a)$: Replace the current label of node $v$ by $a$.
\item $\mathsf{subdiv}(v,a)$: Insert a new $a$-node as the only child of $v$ making all existing children of $v$ children of the
  new node.
\item $\mathsf{insert_L}(v,a)$, $\mathsf{insert_R}(v,a)$: Insert a new $a$-node
  as left or right sibling of node $v$.
\item $\mathsf{delete}(v)$: Delete node $v$, where either $v$ must be a
  leaf or $v$ must have no siblings. In the latter case, all children of $v$
  will be placed at the position of $v$ in the list of children of $\parent(v)$.\end{enumerate}
We refer to the updates as \emph{relabeling}, \emph{subdivision}, \emph{leaf
  insertion}, and \emph{deletion}. Usually we subsume leaf insertions and
subdivisions under the broader term \emph{insertions}. We note that the insertion of a node below a
leaf is a special case of a subdivision. The update of deleting a node 
which has no siblings is also called edge contraction in graph
theory and can be seen as the inverse of a subdivision update.

\begin{figure}
  \begin{subfigure}[b]{0.45\textwidth}
		\begin{center}
			\begin{tikzpicture}[level distance=2.2em, sibling distance=2.5em]
				\node {$a$}
					child { node {$\vphantom{b}c$}}
					child { node {$d$}};

        \node at (1.5,-1.1em) {$\longrightarrow$};

        \node at (3,0) {$a$}
					child { node {$b$}
						child { node {$\vphantom{b}c$}}
						child { node {$d$}}
					};

			  \node at (0,-7em) {$\odot$}
					child { node {$a_\Box$}}
					child {  node {$\oplus$}
						child { node {\vphantom{b}$c$}}
						child { node {$d$}}
					};

        \node at (1.5,-8.9em) {$\longrightarrow$};

				\node at (3,-7em) {$\odot$}
					child[sibling distance=3.5em] { node {$\odot$}
						child[sibling distance=2em] { node {$a_\Box$}}
						child[sibling distance=2em] { node {$b_\Box$}}
					}
					child[sibling distance=3.5em] {  node {$\oplus$}
						child[sibling distance=2em] { node {\vphantom{b}$c$}}
						child[sibling distance=2em] { node {$d$}}
					};
			\end{tikzpicture}
		\end{center}
		\caption{Subdividing the $a$-node with a new $b$-node}
		\label{fig:update-subdiv}
  \end{subfigure}
  \hspace{3em}
  \begin{subfigure}[b]{0.44\textwidth}
		\begin{center}
			\begin{tikzpicture}[level distance=2.2em, sibling distance=2.5em]
				\node {$a$}
				child { node {$b$}}
				child { node {\vphantom{b}$c$}};

        \node at (1.3,-1.1em) {$\longrightarrow$};

        \node at (3,0) {$a$}
					child { node {$b$}}
					child { node {\vphantom{b}$c$}}
					child { node {$d$}};

				\node at (0,-5em)  {$\odot$}
					child { node {$a_\Box$}}
					child {  node {$\oplus$}
						child { node {$b$}}
						child { node {\vphantom{b}$c$}}
					};

        \node at (1.5,-8.3em) {$\longrightarrow$};

				\node at (3,-6em) {$\odot$}
					child { node {$a_\Box$}}
					child {  node {$\oplus$}
						child { node {$b$}}
						child {  node {$\oplus$}
							child { node {\vphantom{b}$c$}}	
							child { node {$d$}}}
				};
			\end{tikzpicture}
		\end{center}
		\caption{Inserting a $d$-node right of the $c$-node}
		\label{fig:update-insertLeaf}
  \end{subfigure}

  \vspace{3ex}

  \begin{subfigure}[b]{0.44\textwidth}
		\begin{center}
			\begin{tikzpicture}[level distance=2.2em, sibling distance=2.5em]
				\node {$a$}
					child { node {$b$}
						child { node {$c$}}
				};
        \node at (2,-1.1em) {$\longrightarrow$};
        \node at (4,0) {$a$}
					child { node{$c$}};
        \node at (0,-6.6em) {$\odot$}
					child {  node {$\odot$}
					child { node {$a_\Box$}}
						child { node {$b_\Box$}}}
						child { node {\vphantom{b}$c$}};
        \node at (2,-7.7em) {$\longrightarrow$};
        \node at (4,-6.6em) {$\odot$}
					child { node {\vphantom{b}$a_\Box$}}
					child { node {$c$}};
      \end{tikzpicture}
    \end{center} 
		\caption{Deleting the $b$-node (inner node without siblings)}
		\label{fig:update-delete-1}
  \end{subfigure}  
  \hspace{4em}
  \begin{subfigure}[b]{0.44\textwidth}
		\begin{center}
			\begin{tikzpicture}[level distance=2.2em, sibling distance=2.5em]
				\node {$a$}
					child { node {$b$}
						child { node {$c$}}
				};

        \node at (2,-1.1em) {$\longrightarrow$};

        \node at (4,0) {$a$}
					child { node{$b$}};

        \node at (0,-6.6em) {$\odot$}
					child {  node {$\odot$}
					child { node {$a_\Box$}}
						child { node {$b_\Box$}}}
						child { node {\vphantom{b}$c$}};

        \node at (2,-7.7em) {$\longrightarrow$};

        \node at (4,-6.6em) {$\odot$}
					child { node {\vphantom{b}$a_\Box$}}
					child { node {$b$}};
      \end{tikzpicture}
    \end{center} 
		\caption{Deleting the $c$-node (leaf without siblings)}
		\label{fig:update-delete-2}
  \end{subfigure}
  
  \caption{Examples depicting the tree updates considered in this paper and their corresponding rewrites in the forest algebra formula}
  \label{fig:updates}
\end{figure}
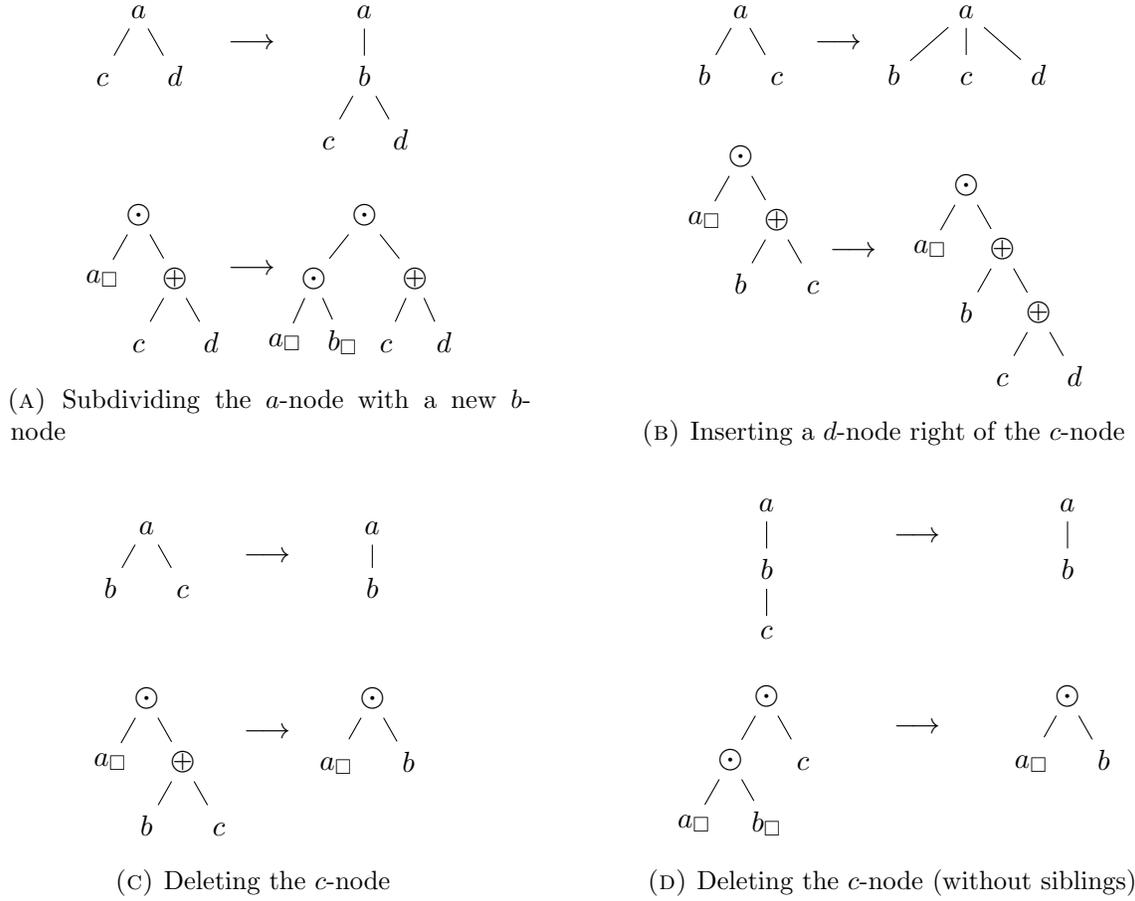

We depict examples of every update (except relabeling) in
Figure~\ref{fig:updates}. The figure always depicts an update in a tree and the
corresponding update in the formula (see below for a description). We use
unique node labels in each tree, such that corresponding nodes in the tree and
the formula can be easily identified.

We now describe how each update can be applied to a formula $\Psi$ that
represents the tree $T$. We assume that given a node $v$ in $T$ we can locate
the corresponding node $v'$ in the formula in constant time, e.g., by adding
pointers to each node in $T$. In the following, $v'$ is the node in $\Psi$ that
corresponds to $v$.
\begin{enumerate}[(i)]
\item $\mathsf{relab}(v,a)$: It suffices to locate $v'$ in the formula and change
  its label to $a$ or $\context a$, depending on whether $v$ is a leaf or an
  inner node of $T$.
\item $\mathsf{subdiv}(v,a)$: We locate $v'$ in the formula and add a new node
  $w$ that is labeled $\conapp_{VH}$ (if $v$ is a leaf) or $\conapp_{VV}$ (if
  $v$ is an inner node) at the place of $v'$. We make $v'$ the left child of
  $w$. The right child of $w$ is a new node that has label $a$ (if $v$ was a
  leaf) or $\context a$ (if $v$ was an inner node). If $v$ was a leaf, the label
  of $v'$ needs to be changed from some label $b$ to the label $\context b$. An
  example is depicted in
  Figure~\ref{fig:update-subdiv}. 
\item $\mathsf{insert_L}(v,a)$, $\mathsf{insert_R}(v,a)$: We replace $v'$ with
  $(a \oplus v')$ or $(v' \oplus a)$, respectively. An example is depicted in
  Figure~\ref{fig:update-insertLeaf}. 
\item $\mathsf{delete}(v)$: We distinguish three cases. 
	\begin{itemize}	
  \item Case 1: The node $v$ is a leaf that has a left sibling or a
    right sibling. That is, the smallest subformula that strictly contains $v$
    is of the form $(\Psi' \oplus v)$ or of the form $(v \oplus \Psi')$,
      respectively, where $\Psi'$ is some subformula.

    We replace
    the whole subformula with $\Psi'$, i.e., the $\oplus$-node and $v$ are
    removed and $\Psi'$ is placed at the previous position of the $\concat$-node.
  \item Case 2: The node $v$ is an inner node with no siblings. In this case the
    smallest subformula that strictly contains $v$ has the form $(v_\Box \odot
    \Psi')$ or $(\Psi' \conapp v_\Box)$ and we remove $v_\Box$ and the $\odot$
    like in case 1.
    An example is depicted in Figure~\ref{fig:update-delete-1}.
  \item Case 3: The node $v$ has no sibling and is a leaf. That is,
    the smallest subformula that strictly contains $v$ is of the
    form $(\Psi' \odot v)$. We
    note, that $\Psi'$ is necessarily a context, while $(\Psi' \odot v)$ is a
    forest. Thus, we cannot simply replace $(\Psi' \odot v)$ by $\Psi'$. Instead
    we first have to locate the node $w'$ corresponding to the parent $w$ of $v$
    and change its label from some label $\context a$ to $a$ (i.e., remove the
    hole below the node). Furthermore, we have to adapt the labels of ancestors
    of $w'$ in $\Psi'$, as all of these ancestors change their type from context
    to forest. Therefore the labels $\concat_{VH}$ and $\concat_{HV}$ have to be
    changed to $\concat_{HH}$ and the label $\conapp_{VV}$ has to be changed to
    $\conapp_{VH}$. At the end, $v$ has to be removed and the parent of $v$ has to be replaced by its
    left subformula. An example is depicted in Figure~\ref{fig:update-delete-2}.
\end{itemize}
\end{enumerate}

We note that the above cases for $\mathsf{delete}(v)$ are indeed complete.
If $v$ has a sibling, it has to be under a $\concat$-node in the formula
and if $v$ has no sibling, its parent in the formula cannot be a
$\concat$-node. All of the cases are also handled correctly. Replacing $(\Psi'
\oplus v)$ or $(v \oplus \Psi')$ with $\Psi'$ simply removes $v$ from the list
of children under its parent in the tree. Similarly, replacing $(v \conapp
\Psi')$ with $\Psi'$ puts $\Psi'$ in the place of $v$. Finally, in the formula
$(\Psi' \conapp v)$ with $v$ being a context, there has to be a $\conapp$-node
$u$ above whose right subformula describes the forest below $v$. Thus
replacing the subformula with $\Psi'$, $u$ will put this forest at the place
where $v$ was before, as this is exactly where the hole of $\Psi'$ is.

The updates relabeling, subdivision and leaf insertion are local to the node
$v$. This does not hold for deletion, as the parent node $w$ of the deleted node
will be affected if its last child is a leaf and is deleted. In this case we have to change
all nodes on the path between $v$ and $w'$ in $\Psi$. We note that it is easy
to find the node $w'$ within the parse tree in this case. We are in the case 3
described above, i.e., there is a subformula $(\Psi' \odot v)$ in $\Psi$. We
can perform a top-down search in $\Psi'$ always going to the left subformula at each
$\concat_{VH}$-node and the right subformula at each $\concat_{HV}$- and
$\conapp_{VV}$-node. There can be no $\concat_{HH}$- and $\conapp_{VH}$-nodes on
this path.

\section{Maintaining Forest Algebra Formulas under Updates}\label{sec:maintaining}
This section shows how (parse trees for) forest algebra formulas can be
maintained when the represented tree is updated. That is, the main result of
this section is:
\begin{thm}\label{thm:update}
  Given a tree $T$, it is possible to compute in time $\cO(|T|)$ a forest algebra
  formula $\Psi$ in the free forest algebra, such that
  \begin{itemize}
  \item $\Psi$ represents $T$;
  \item the parse tree of $\Psi$ is of height at most $10\log(|T|)$; and
  \item each update to $T$ can be translated to an update of $\Psi$,
    such that the new formula can be computed in time $\cO(\log(|T|))$
    and has height at most $10\log(|T|)$+1.
  \end{itemize}
\end{thm}

\subsection{High Level Description of the Proof}\label{sec:fa-highlevel}
To keep the parse tree of $\Psi$ shallow, we use rotations
similar to those used in
AVL trees~\cite{AdelsonVL62}. Unfortunately, the well-known rotations used to
balance AVL trees only work if the underlying algebra is fully associative,
which does not hold for forest algebras, as, e.g., $c \conapp (f_1 \concat f_2)
\neq (c \conapp f_1) \concat f_2$. Therefore, we provide additional rotations
that can be used where the traditional rotations fail.

Towards our invariant we assign colors to nodes of a formula $\Psi$ as follows:
\[
  \col(v) \;=\; \left\{{\arraycolsep=0pt
      \begin{array}{llll}
        \text{green }\;\; & \text{ if }\;\; & \height(\Psi_v) &{} \;\leq\; 10 \log(\numleaves{\Psi_v}) \\
        \text{yellow} & \text{ if }\;\; & \height(\Psi_v)-1 &{} \;\leq\; 10
                                                           \log(\numleaves{\Psi_v})
                                                           \;<\; \height(\Psi_v)\\
        \text{red} & \text{otherwise}
    \end{array}} \right.                                                                
\]

The equality can be rearranged to give bounds on the number of
leaves of a subformula depending on the height. A green node $v$ satisfies
$\numleaves{\Psi_v} \geq 2^{\frac{\height(\Psi_v)}{10}}$ and a yellow node $v$
satisfies $2^{\frac{\height(\Psi_v)}{10}} > \numleaves{\Psi_v} \geq 2^{\frac{\height(\Psi_v)-1}{10}}$.

\medskip\noindent \textbf{Strong Invariant} Every node is green.

\medskip\noindent \textbf{Weak Invariant} There are no red nodes and there is a
node $v$, such that all yellow nodes are on $\lp^{-1}(v)$.

\begin{obs}
  A formula $\Psi$ that satisfies either invariant has logarithmic height.
\end{obs}

The strong invariant will hold up between runs of the insertion algorithm, but
not necessarily during the runtime of the insertion algorithm. The weak
invariant is always satisfied.

We note that if the height of a subformula is increased by one (e.g., due to an
insertion), then green nodes can become yellow and yellow nodes can become red.
Similarly, if we reduce the height of a subformula by a rotation (introduced in
the next subsection), then yellow nodes will turn green.

In Section~\ref{sec:rotations} we introduce the rotations used by our algorithm
and prove some technical results. The main intuition behind our proof is given
by the lemmas~\ref{lem:noinvariant-unbalanced-path}
and~\ref{lem:noinvariant-rotation}. Together they imply that whenever there is a
yellow node $v$, then we can apply some rotation. We will see that this
rotation can lead to another node (strictly below $v$) becoming yellow. Thus it
might be necessary to apply several rotations. In the
sections~\ref{sec:preprocessing},~\ref{sec:insertions}, and~\ref{sec:deletions},
we show how we can construct a formula for a given tree in linear time, how we
can handle insertions, and how we can handle deletions, respectively.

\subsection{Rotations}\label{sec:rotations}
Rotations of formulas follow a similar spirit as rotations in AVL trees. They
rewrite the formula (preserving equivalence) in such a way that one subformula
(i.e., a subtree of the parse tree) is moved one level up and another subformula
is moved one level down. Formally, rotations are defined on the algebraic level
as follows:

\begin{table}
  \[
    \begin{array}{r@{\;\;}c@{\;\;}ll} \toprule
  \multicolumn{3}{c}{\text{Rotations 1a (left-to-right) and 1b (right-to-left)}} \\ \midrule
     (x_1 \concat_{HH} x_2) \concat_{HH} x_3 &=& x_1 \concat_{HH} (x_2 \concat_{HH} x_3)\\
      (x_1 \concat_{VH} x_2) \concat_{VH} x_3 &=& x_1 \concat_{VH} (x_2 \concat_{HH} x_3)\\
      (x_1 \concat_{HV} x_2) \concat_{VH} x_3 &=& x_1 \concat_{HV} (x_2 \concat_{VH} x_3)\\
      (x_1 \concat_{HH} x_2) \concat_{HV} x_3 &=& x_1 \concat_{HV} (x_2 \concat_{HV} x_3)\\
      (x_1 \conapp_{VV} x_2) \conapp_{VV} x_3&=& x_1 \conapp_{VV} (x_2 \conapp_{VV} x_3)\\
      (x_1 \conapp_{VV} x_2) \conapp_{VH} x_3 &=& x_1 \conapp_{VH} (x_2
                                 \conapp_{VH}
                                                         x_3)\\[1.5ex]
\begin{tikzpicture}[>=latex,level distance=8mm, level 1/.style={sibling distance=14mm}, level 2/.style={sibling distance=8mm}]
\node (r1a) at (2.3,0) {$\op$}
    child [sibling distance=10mm] { node (down1) {$x_1$} edge from parent[child anchor=north]}
    child { node (1a) {$\op$}
      child { node (x2a) {$x_2$} edge from parent[child anchor=north]}
      child { node (up1) {$x_3$} edge from parent[child anchor=north]}
    };

    \node (r1b) at (0,0) {$\op$}
    child { node (1b) {$\op$}
      child { node (up2) {$x_1$} edge from parent[child anchor=north]}
      child { node (x2b) {$x_2$} edge from parent[child anchor=north]}
    }
    child [sibling distance=10mm] { node (down2) {$x_3$} edge from parent[child anchor=north] };
\node at ($(up2) +(0.3,0)$) {$\uparrow$};
    \node at ($(down2) +(0.3,0)$) {$\downarrow$};
    \draw[->] ($(x2b) +(1,0)$) -- ($(x2a) +(-1,0)$) node [above,midway] {\scriptsize 1a}; 
\node at ($(1a) +(+0.4,0.1)$) {$w'$};
    \node at ($(1b) +(-0.35,0.1)$) {$w$};
    \node at ($(r1a) +(+0.35,0.1)$) {$v'$};
    \node at ($(r1b) +(-0.3,0.1)$) {$v$};
  \end{tikzpicture} & &
  \begin{tikzpicture}[>=latex,level distance=8mm, level 1/.style={sibling distance=14mm}, level 2/.style={sibling distance=8mm}]
\node (r1a) at (2.3,0) {$\op$}
    child [sibling distance=10mm] { node (down1) {$x_1$} edge from parent[child anchor=north]}
    child { node (1a) {$\op$}
      child { node (x2a) {$x_2$} edge from parent[child anchor=north]}
      child { node (up1) {$x_3$} edge from parent[child anchor=north]}
    };

    \node (r1b) at (0,0) {$\op$}
    child { node (1b) {$\op$}
      child { node (up2) {$x_1$} edge from parent[child anchor=north]}
      child { node (x2b) {$x_2$} edge from parent[child anchor=north]}
    }
    child [sibling distance=10mm] { node (down2) {$x_3$} edge from parent[child anchor=north] };
    \node at ($(up1) +(0.3,0)$) {$\uparrow$};
    \node at ($(down1) +(0.3,0)$) {$\downarrow$};
\draw[->] ($(x2a) +(-1,-0)$) -- ($(x2b) +(1,-0)$) node [above,midway] {\scriptsize 1b}; 
    \node at ($(1a) +(+0.35,0.1)$) {$w$};
    \node at ($(1b) +(-0.4,0.1)$) {$w'$};
    \node at ($(r1a) +(+0.3,0.1)$) {$v$};
    \node at ($(r1b) +(-0.35,0.1)$) {$v'$};
  \end{tikzpicture} \\
      \bottomrule\\[4ex]
      \toprule
  \multicolumn{3}{c}{\text{Rotations 2a (left-to-right) and 2b (right-to-left)}} \\ \midrule
      (x_1 \concat_{HV} x_2) \conapp_{VH} x_3 &=& x_1 \concat_{HH} (x_2 \conapp_{VH} x_3) \\
       (x_1 \concat_{HV} x_2) \conapp_{VV} x_3 &=& x_1 \concat_{HV} (x_2
                                      \conapp_{VV} x_3) \\[1.5ex]
\begin{tikzpicture}[>=latex,level distance=8mm, level 1/.style={sibling distance=14mm}, level 2/.style={sibling distance=8mm}]
\node (r1a) at (2.3,0) {$\concat$}
    child [sibling distance=10mm] { node (down1) {$x_1$} edge from parent[child anchor=north]}
    child { node (1a) {$\conapp$}
      child { node (x2a) {$x_2$} edge from parent[child anchor=north]}
      child { node (up1) {$x_3$} edge from parent[child anchor=north]}
    };
    \node (r1b) at (0,0) {$\conapp$}
    child { node (1b) {$\concat_{\mathrlap{HV}}$}
      child { node (up2) {$x_1$} edge from parent[child anchor=north]}
      child { node (x2b) {$x_2$} edge from parent[child anchor=north]}
    }
    child [sibling distance=10mm] { node (down2) {$x_3$} edge from parent[child anchor=north] };
\node at ($(up2) +(0.3,0)$) {$\uparrow$};
    \node at ($(down2) +(0.3,0)$) {$\downarrow$};
    \draw[->] ($(x2b) +(1,0)$) -- ($(x2a) +(-1,0)$) node [above,midway] {\scriptsize 2a}; 
\node at ($(1a) +(+0.4,0.1)$) {$w'$};
    \node at ($(1b) +(-0.35,0.1)$) {$w$};
    \node at ($(r1a) +(+0.35,0.1)$) {$v'$};
    \node at ($(r1b) +(-0.3,0.1)$) {$v$};
  \end{tikzpicture} & &
  \begin{tikzpicture}[>=latex,level distance=8mm, level 1/.style={sibling distance=14mm}, level 2/.style={sibling distance=8mm}]
\node (r1a) at (2.3,0) {$\concat$}
    child [sibling distance=10mm] { node (down1) {$x_1$} edge from parent[child anchor=north]}
    child { node (1a) {$\conapp$}
      child { node (x2a) {$x_2$} edge from parent[child anchor=north]}
      child { node (up1) {$x_3$} edge from parent[child anchor=north]}
    };
    \node (r1b) at (0,0) {$\conapp$}
    child { node (1b) {$\concat_{\mathrlap{HV}}$}
      child { node (up2) {$x_1$} edge from parent[child anchor=north]}
      child { node (x2b) {$x_2$} edge from parent[child anchor=north]}
    }
    child [sibling distance=10mm] { node (down2) {$x_3$} edge from parent[child anchor=north] };
    \node at ($(up1) +(0.3,0)$) {$\uparrow$};
    \node at ($(down1) +(0.3,0)$) {$\downarrow$};
\draw[->] ($(x2a) +(-1,0)$) -- ($(x2b) +(1,0)$) node [above,midway] {\scriptsize 2b}; 
    \node at ($(1a) +(+0.35,0.1)$) {$w$};
    \node at ($(1b) +(-0.4,0.1)$) {$w'$};
    \node at ($(r1a) +(+0.3,0.1)$) {$v$};
    \node at ($(r1b) +(-0.35,0.1)$) {$v'$};
  \end{tikzpicture} \\
      \bottomrule\\[4ex]
      \toprule
  \multicolumn{3}{c}{\text{Rotations 3a (left-to-right) and 3b (right-to-left)}} \\ \midrule
      (x_1 \concat_{VH} x_2) \conapp_{VH} x_3 &=& (x_1 \conapp_{VH} x_3)
                                        \concat_{HH} x_2 \\
      (x_1 \concat_{VH} x_2) \conapp_{VV} x_3 &=& (x_1 \conapp_{VV} x_3)
                                                         \concat_{VH} x_2\\[1.5ex]
\begin{tikzpicture}[>=latex,level distance=8mm, level 1/.style={sibling distance=14mm}, level 2/.style={sibling distance=8mm}]
\node (r1a) at (2.9,0) {$\concat$}
    child { node (1a) {$\conapp$}
      child { node (x2a) {$x_1$} edge from parent[child anchor=north]}
      child { node (up1) {$x_3$} edge from parent[child anchor=north]}
    }
    child [sibling distance=10mm] { node (down1) {$x_2$} edge from parent[child anchor=north]};
    \node (r1b) at (0,0) {$\conapp$}
    child { node (1b) {$\concat_{\mathrlap{VH}}$}
      child { node {$x_1$} edge from parent[child anchor=north]}
      child { node (x2b) {$x_2$} edge from parent[child anchor=north]}
    }
    child [sibling distance=10mm] { node (down2) {$x_3$} edge from parent[child anchor=north] };
\node at ($(x2b) +(0.3,0)$) {$\uparrow$};
    \node at ($(down2) +(0.3,0)$) {$\downarrow$};
    \draw[->] ($(x2b) +(0.6,0)$) -- ($(x2a) +(-0.5,0)$) node [above,midway] {\scriptsize 3a}; 
\node at ($(1a) +(+0.4,0.1)$) {$w'$};
    \node at ($(1b) +(-0.35,0.1)$) {$w$};
    \node at ($(r1a) +(+0.35,0.1)$) {$v'$};
    \node at ($(r1b) +(-0.3,0.1)$) {$v$};
  \end{tikzpicture} & &
  \begin{tikzpicture}[>=latex,level distance=8mm, level 1/.style={sibling distance=14mm}, level 2/.style={sibling distance=8mm}]
\node (r1a) at (2.9,0) {$\concat$}
    child { node (1a) {$\conapp$}
      child { node (x2a) {$x_1$} edge from parent[child anchor=north]}
      child { node (up1) {$x_3$} edge from parent[child anchor=north]}
    }
    child [sibling distance=10mm] { node (down1) {$x_2$} edge from parent[child anchor=north]};
    \node (r1b) at (0,0) {$\conapp$}
    child { node (1b) {$\concat_{\mathrlap{VH}}$}
      child { node {$x_1$} edge from parent[child anchor=north]}
      child { node (x2b) {$x_2$} edge from parent[child anchor=north]}
    }
    child [sibling distance=10mm] { node (down2) {$x_3$} edge from parent[child anchor=north] };
    \node at ($(up1) +(0.3,0)$) {$\uparrow$};
    \node at ($(down1) +(0.3,0)$) {$\downarrow$};
\draw[->] ($(x2a) +(-0.5,0)$) -- ($(x2b) +(0.6,0)$) node [below,midway] {\scriptsize 3b}; 
    \node at ($(1a) +(+0.35,0.1)$) {$w$};
    \node at ($(1b) +(-0.4,0.1)$) {$w'$};
    \node at ($(r1a) +(+0.3,0.1)$) {$v$};
    \node at ($(r1b) +(-0.35,0.1)$) {$v'$};
  \end{tikzpicture} \\
      \bottomrule
    \end{array}
  \]
  \caption{Rotations as equations and parse trees. Vertical arrows at nodes in
    parse trees indicate that the respective subformula is moved one level
    up or down in the rotation. We will refer to individual rotations using the
    numbers on the horizontal arrows.}\label{tab:rotations}
\end{table}
 \begin{defi}\label{def:rotation}
  A \emph{rotation} $\alpha$ is a rewriting of a forest algebra formula
  $\Psi$ into an equivalent forest algebra formula
  $\alpha(\Psi)$ using one of the equations in Table~\ref{tab:rotations},
  where $x_1$ to $x_3$ are variables that can be replaced by arbitrary formulas
  of the correct type.
\end{defi}

A rotation $\alpha$ is applicable at a subformula $\Psi_v$ if one side
of the defining equation can be matched to $\Psi_v$ by replacing $x_1$, $x_2$,
and $x_3$ with subformulas $\Psi_1$, $\Psi_2$, and $\Psi_3$ from $\Psi$.
Applying the rotation $\alpha$ at $\Psi_v$ is the process of replacing
$\Psi_v$ with $\alpha(\Psi_v)$, which results from the other side of the
equation by also replacing $x_1$, $x_2$, and $x_3$. We note that all rotations
are applicable in both directions.

In Table~\ref{tab:rotations}, next to the equations
defining the rotations, we depict graphical representations of the rotations. We
have annotated the nodes labeled $x_1$, $x_2$, and $x_3$ with arrows $\uparrow$,
$\downarrow$ to denote that the subformula is moved upwards or downwards in the
rotation. For the rotations 1a and 1b, all operations depicted by $\op$-nodes
need either to be concatenations or context applications, but no mixture. E.g.,
the rotations 1a in the figure depict 6 different rotations, as can be seen in
Table~\ref{tab:rotations}. All $\op$-nodes could be $\concat_{HH}$ in case that
$x_1$, $x_2$, and $x_3$ are forests, and there are three other possibilities for
concatenations, where exactly one of $x_1$, $x_2$, and $x_3$ depicts a context.
Additionally there are two possibilities involving context applications.
The rotations 2a, 2b, 3a, and 3b change the order in which a concatenation and a
context application are performed. This also changes the relative order of the
operands $x_2$ and $x_3$ in the rotations 3a and 3b.

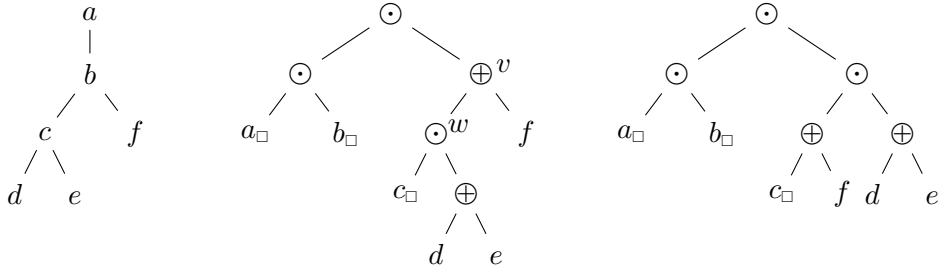
\begin{figure}
  \begin{tikzpicture}[level distance=8mm,level 1/.style={sibling distance=24mm}, level 2/.style={sibling distance=12mm}, level 3/.style={sibling distance=8mm}]
    \node {$a$}
    child { node {$b$}
      child { node {$c$}
        child { node {$d$} }
        child { node {$\vphantom{d}e$} }
      }
      child { node {$f$} }
    };

    \node at (4,0) {$\conapp$}
    child { node {$\conapp$}
      child { node {$\context a$} }
      child { node {$\context b$} }
    }
    child { node (v) {$\concat$}
      child { node (w) {$\conapp$}
        child { node {$\context c$} }
        child { node {$\concat$}
          child { node {$d$} }
          child { node {$\vphantom{d}e$} }
        }
      }
      child { node {$f$} }
    };

    \node at (9,0) {$\conapp$}
    child { node {$\conapp$}
      child { node {$\context a$} }
      child { node {$\context b$} }
    }
    child { node {$\conapp$}
      child { node {$\concat$}
        child { node {$\context c$} }
        child { node {$f$} }
      }
      child { node {$\concat$}
        child { node {$d$} }
        child { node {$\vphantom{d}e$} }
      }
    };

    \node at ($(v) +(+0.3,0.1)$) {$v$};
    \node at ($(w) +(+0.3,0.1)$) {$w$};
      
  \end{tikzpicture}
  \caption{Example of rotation 3b applied at node $v$}\label{fig:rot-example}
\end{figure}

\begin{exa}
  Figure~\ref{fig:rot-example} depicts a tree $T$, a formula $\Psi$ representing
  $T$ and the formula that results from applying rotation 3b at
  the node marked with $v$.
\end{exa}

\begin{lem}\label{lem:rotations-sound}
  The rotations depicted in Table~\ref{tab:rotations} are valid operations, i.e.,
  the equations hold for all possible substitutions of $x_1$ to $x_3$ with
  formulas of the correct type.
\end{lem}
\begin{proof}
  The equations in Table~\ref{tab:rotations} follow immediately from the axioms
  of associativity and interaction in Table~\ref{tab:axioms}.
\end{proof}

Technically, a rotation is applicable at a node $v$ of a formula $\Psi$ whenever
the structure of $\Psi_v$ corresponds to the structure required by the rotation.
E.g., rotation 1a is applicable whenever the node $v$ and its left child are
both labeled with $\concat$ or both are labeled with $\conapp$. Obviously,
applying a rotation does not always improve the ``balancedness'' of the
formula. In order to improve the balancedness of a formula, we only apply a
rotation, if the balances of $v$ and $w$ are as indicated in
Table~\ref{tab:balances}. As depicted in the table, depending on the balances, we
classify the rotations into height reducing rotations and height preserving
rotations. We do not specify height preserving conditions for rotations 2b and 3b, as we
will only use the rotations 2b and 3b if they strictly reduce the height.

\begin{table}
  \renewcommand{\arraystretch}{1.1}
  \begin{tabular}{ccccccc}
    \toprule
    rotation & \hspace{2ex} & \multicolumn{2}{c}{height reducing} & \hspace{2ex} & \multicolumn{2}{c}{height preserving} \\ \cmidrule{3-4} \cmidrule{6-7}
    &  & $\balance(v)$ & $\balance(w)$ & & $\balance(v)$ & $\balance(w)$ \\ \midrule
1a & & $\leq -2$ & $< 0$ & & $\leq -2$ & $> 0$\\
    1b & & $\geq +2$ & $> 0$ & & $\mathllap{(} \geq +2$ & $< 0 \mathrlap{\;)}$\\
2a & & $\leq -2$ & $< 0$ & & $\leq -2$& $> 0$ \\
    2b & & $\geq +2$ & $> 0$ & & & \\ 
    3a & & $\leq -2$ & $> 0$ & & $\leq -2$& $< 0$ \\
    3b & & $\leq -2$ & $> 0$ & & & \\ \bottomrule 
  \end{tabular}
  \caption{Balance conditions for height reducing and height preserving
    rotations. Height preserving rotation 1b is only used as part of a
      double rotation (see Figure~\ref{fig:dr}).}\label{tab:balances}
\end{table}

We now have two lemmas further characterizing the effects of height reducing and
height preserving rotations. To ease the presentation of the lemmas and their
proofs we introduce some more notation. We depict the subformulas of the
rotations by their movement, i.e., we depict by $\psiup$ the subformula (among
$x_1$, $x_2$, and $x_3$, see Table~\ref{tab:rotations}) that is moved one level
up. Similarly we depict by $\psidown$ the subformula that is moved one level
down and by $\psistay$ the subformula that stays on the same level. E.g.\, for
rotation 1a, $\psiup=x_1$, $\psidown=x_3$, and $\psistay=x_2$ and for rotation
1b, the roles of $\psiup$ and $\psidown$ are reversed, i.e., $\psiup=x_3$ and
$\psidown=x_1$. As indicated in Table 3, we use $v'$ and $w'$ to denote the nodes $v$ and
$w$ after the rotation.

Using this notation, we have that for all rotations it holds that prior to the
rotation, $\psiup$ and $\psistay$ were the two children of $w$ and $\psidown$ was
the other child of $v$ (i.e., the sibling of $w$). Similarly, after the rotation,
$\psiup$ becomes a child of $v'$ and $\psidown$ and $\psistay$
become the children of $w'$.

\begin{lem}\label{lem:hr-rotate}
  Let $\alpha$ be a height reducing rotation that is applicable at $v$, then
  \begin{enumerate}[(a)]
  \item $\alpha$ reduces the height of $\Psi_v$ by one, i.e., $\height(\alpha(\Psi_v))=\height(\Psi_v) -1$;
  \item $\alpha$ improves the overall balance of $\Psi_v$ by at least
    three, i.e., $\fbalance(\alpha(\Psi_v)) \leq \fbalance(\Psi_v) - 3$;
  \item all yellow nodes on $\lp^{-1}(v)$ change their color to green; and 
  \item all nodes on $\lp^{-1}(v)$ except $v$ improve their balance by
    one.
\end{enumerate}
\end{lem}
We remind that $\fbalance(\Psi)$ denotes the sum of the absolute balances of all
nodes in $\Psi$.
\begin{proof}
   We first observe that by the condition on the balances, we
  have that
  \begin{equation}
    \height(\psiup) \quad>\quad \max\big(\height(\psidown),\height(\psistay)\big)\;. \tag{$\dagger$}\label{eq:hr-heights}
  \end{equation}
  \begin{enumerate}[(a)]
  \item The statement follows from~\eqref{eq:hr-heights} and the fact that
    $\psiup$ is moved one level up.
  \item We now compute the change of $\fbalance(\Psi_v)$. As the balances of the
  subformulas $\psiup$, $\psidown$, and $\psistay$ do not change, it is enough
  to look at the balances of $v$ and $w$. As we are only interested in changes of the
  absolute value of the balances, it does not matter which child is the
  left/right child of $v$ and $w$.
  \begin{eqnarray*}
    |\balance(v)| &=& \height(\psiup)+1-\height(\psidown) \\
    |\balance(w)| &=& \height(\psiup) - \height(\psistay) \\
    |\balance(v')| &=& \height(\psiup)-\max\big(\height(\psidown),\height(\psistay)\big)-1 \\
    |\balance(w')| &=& |\height(\psistay)-\height(\psidown)| 
  \end{eqnarray*}
  We note that the right sides of the first three equations cannot be negative
  because of~\eqref{eq:hr-heights}, which is why we can avoid taking the
  absolute value.
Using the fact that $a + b + |a-b| = 2 \max(a,b)$ for arbitrary numbers $a$
  and $b$, we can now compute the change of overall balance as
  \begin{align*}
    \fbalance(\Psi_v)-\fbalance(\Psi_{v'}) \;\; &=\;\; |\balance(v)| + |\balance(w)| - |\balance(v')| - |\balance(w')| \\
                                                &=\;\; \height(\psiup) - \max\big(\height(\psidown),\height(\psistay)\big) + 2\;.
  \end{align*}
  By~\eqref{eq:hr-heights}, we get that $\fbalance(\Psi_v)$ is improved by at least 3.
  \item As by~(a), the height of $\Psi_v$ is reduced by one, the height of each $\Psi_u$ with
    $u \in \lp^{-1}(v)$ is decreased by one. As the number of leaves does not
    change, the definition of colors implies that yellow nodes on $\lp^{-1}(v)$
    turn green.
  \item It is a direct consequence of~(a), that all nodes on $\lp^{-1}(v)$
    except $v$ improve their balance by one. \qedhere
\end{enumerate}
\end{proof}

The overall idea of our algorithm is to use height reducing rotations whenever
the height of some (sub)formula was increased by the insertion of a new node or
some (sub)formula needs to be reduced in height after some deletions. However,
there are some situations in which no height reducing rotation can be applied,
even if the formula is severely imbalanced. In these cases we use height
preserving rotations in order to restructure the formula in such a way that
afterwards height reducing rotations can be applied.

\begin{lem}\label{lem:hp-rotate}
  Let $\alpha$ be a height preserving rotation that is applicable at $v$
  (including the case where it is applied as part of a double rotation), then
  \begin{enumerate}[(a)]
  \item $\height(\alpha(\Psi_v))=\height(\Psi_v)$;
  \item $|\balance(v')| = |\balance(w)|+1$;
  \item $|\balance(w')| = |\balance(v)|-1$; 
  \item $\fbalance(\alpha(\Psi_v)) = \fbalance(\Psi_v)$; and
  \item the color of $w'$ is green if there was a node $u \in \lp(v)$ with
    $|\balance(u)| \leq 1$ one to seven levels below $v$ and all nodes in
    $\Psi_u$ were green or yellow.
  \end{enumerate}
\end{lem}
\begin{proof}
  Again, we denote by $\psiup$, $\psidown$, and $\psistay$ the subformulas that
  moves one level up, moves one level down, and stays on the same level,
  respectively. The condition on balances for height preserving rotations state
  that $\height(\psidown) < \max(\height(\psiup),\height(\psistay))$ and
  $\height(\psistay) > \height(\psiup)$. We can conclude
  \begin{equation}
    \height(\psistay) \quad > \quad \max\big(\height(\psiup),\height(\psidown)\big)\;. \tag{$\ddagger$}\label{eq:hp-heights}
  \end{equation}
  Now we show the individual statements:
  \begin{enumerate}[(a)]
  \item The statement follows from~\eqref{eq:hp-heights} and the fact that
    $\psistay$ stays on the lower level.
  \item
    $\begin{aligned}[t] \phantom{|\balance(w')|}\mathllap{|\balance(v')|}
      \;\;&=\;\; |\height(\psiup) - \max\big(1+\height(\psistay), 1+\height(\psidown)\big)| \\
          &=\;\; \height(\psistay) - \height(\psiup) + 1 \\
          &=\;\; |\balance(w)| + 1
    \end{aligned}$
  \item $\begin{aligned}[t]|\balance(w')| \;\;&=\;\; \height(\psistay) -
      \height(\psidown) \\
      &=\;\;  |\max\big(\height(\psistay),\height(\psiup)\big) + 1 -
      \height(\psidown)| -1\\
      &=\;\;|\balance(v)| -1\end{aligned}$
  \item This follows from~(b) and~(c), as no other nodes in $\Psi_v$ change their balance.
  \item We observe that $w' \in \lp(v')$ and thus $w' \in \lp^{-1}(u)$,
      as the lowest level of $\Psi_v$ was in $\psistay$ for all height preserving
      rotations by the balances given in Table~\ref{tab:balances}. We can
      conclude that $\Psi_u$ is a (not necessarily strict) subformula of $\Psi_{w'}$.
      We furthermore use the fact that any node $u'$ that is yellow or green has at
      least $2^{\frac{\height(\Psi_{u'})-1}{10}}$ many leaves to bound the number of leaves in
      $\Psi_{w'}$ as follows: 
\begin{align*}
      \numleaves{\Psi_{w'}} \;\;&\geq\;\; \numleaves{\Psi_{\lc{u}}}+\numleaves{\Psi_{\rc{u}}} & \text{$\Psi_u$ is a subformula of $\Psi_{w'}$}\\
      &\geq\;\; 2^{\frac{\height(\Psi_{\lc{u}})-1}{10}} + 2^{\frac{\height(\Psi_{\rc{u}})-1}{10}} & \text{definition of colors}\\
                         &\geq\;\; 2^{\frac{\height(\Psi_{w'})-8}{10}} + 2^{\frac{\height(\Psi_{w'})-9}{10}} & \text{relative height of nodes\footnotemark} \\
                         &>\;\; 2 \cdot 2^{\frac{\height(\Psi_{w'})-9}{10}} \;\;=\;\; 2^{\frac{\height(\Psi_{w'})+1}{10}} 
    \end{align*}\footnotetext{The node $u$ is at most $6$ levels below $w'$ and both
      children of $u$ have a height difference of at most 1.}
    We note that $u$ cannot be a leaf, as the height of $u$ is larger than the height
      of its sibling. The statement follows from the definition of colors that says, that a node $w'$ is
    green if and only if $\numleaves{\Psi_{w'}} \geq
    2^{\frac{\height(\Psi_{w'})}{10}}$.\qedhere
  \end{enumerate}
\end{proof}

\begin{figure*}
  \begin{tikzpicture}[>=latex,level distance=8mm, level 1/.style={sibling distance=8mm}, level 2/.style={sibling distance=8mm}]
\node (v1) at (0,0) {$\concat$}
    child { node (wl) {$\concat$}
      child { node {$x_1$} }
      child { node (ul) {$\concat$}
        child { node {$x_2$} }
        child { node {$x_3$} }
      }
    }
    child { node {$x_4$} };

    \node (v2) at (3.25,0) {$\concat$}
    child { node {$\concat$}
      child { node {$\concat$}
        child { node {$x_1$} }
        child { node {$x_2$} }
      }
      child { node {$x_3$} }
    }
    child { node {$x_4$} };
    
    \node (v3) at (6.5,0) {$\concat$}
    child[sibling distance=14mm] { node (w1) {$\concat$}
      child { node {$x_1$} }
      child { node {$x_2$} }
    }
    child[sibling distance=14mm] { node (w2) {$\concat$}
      child { node {$x_3$} }
      child { node {$x_4$} }
    };

    \node (v4) at (9.75,0) {$\concat$}
    child { node {$x_1$} }
    child { node {$\concat$}
      child { node {$x_2$} }
      child { node {$\concat$}
        child { node {$x_3$} }
        child { node {$x_4$} }
      }
    };

    \node (v5) at (13,0) {$\concat$}
    child { node {$x_1$} }
    child { node (wr) {$\concat$}
      child { node (ur) {$\concat$}
        child { node {$x_2$} }
        child { node {$x_3$} }
      }
      child { node {$x_4$} }
    };
    
    \draw[->] ($(v1) +(1,-.3)$) -- ($(v2) +(-1,-.3)$) node [above,midway] {\scriptsize 1b at $v_2$}; 
    \draw[->] ($(v2) +(1,-.3)$) -- ($(v3) +(-1,-.3)$) node [above,midway] {\scriptsize 1a at $v_1$}; 
    \draw[<-] ($(v3) +(1,-.3)$) -- ($(v4) +(-1,-.3)$) node [above,midway] {\scriptsize 1b at $v_1$}; 
    \draw[<-] ($(v4) +(1,-.3)$) -- ($(v5) +(-1,-.3)$) node [above,midway] {\scriptsize 1a at $v_2$};

    \draw[->] (v1) edge [bend left=20] node [above] {\scriptsize 4a} (v3);
    \draw[->] (v5) edge [bend right=20] node [above] {\scriptsize 4b} (v3);

    \node at ($(v1) +(-0.4,0.1)$) {$v_1$};
    \node at ($(v2) +(-0.4,0.1)$) {$v_1$};
    \node at ($(v4) +(+0.4,0.1)$) {$v_1$};
    \node at ($(v5) +(+0.4,0.1)$) {$v_1$};
    \node at ($(wl) +(-0.4,0.1)$) {$v_2$};
    \node at ($(wr) +(+0.4,0.1)$) {$v_2$};

    \node at ($(ul) +(+0.4,0.1)$) {$v_3$};
    \node at ($(ur) +(-0.4,0.1)$) {$v_3$};

    \node at ($(w1) +(-0.4,0.1)$) {$w_1$};
    \node at ($(w2) +(+0.4,0.1)$) {$w_2$};
  \end{tikzpicture}

\caption{Double Rotations. The first rotation is always a rotation 1a or 1b at $v_2$
  followed by a rotation in the opposite direction at $v_1$. We denote double
  rotations as rotations 4a and 4b, respectively.}\label{fig:dr}
\end{figure*}
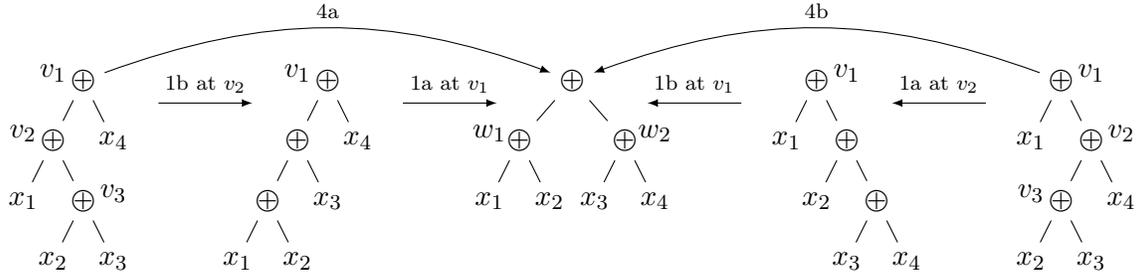

In some cases, we use double rotations, which are exactly the double rotations
used in AVL trees. We only apply double rotations if there are three consecutive
$\concat$-nodes $v_1$, $v_2$, $v_3$ on $\lp(v_1)$ where the direction of the
balances alternates. A double rotation then consists of a height preserving
rotation at $v_2$ followed by a height reducing rotation at $v_1$. We depict the
two possibilities for double rotations in Figure~\ref{fig:dr} and will denote
them as rotations 4a and 4b respectively. We may use the term
  \emph{simple rotation} to emphasize that a rotation is not a double rotation.
From Lemma~\ref{lem:hp-rotate}(e) we
get the following corollary.

\begin{cor}\label{cor:dr-onenodegreen}
  If all nodes were green or yellow and a double rotation was applied at some node $v_1$ that had a node $u$ with
  $|\balance(u)| \leq 1$ at most 7 levels below on $\lp(v_1)$, then the node $w_1$
  as indicated in Figure~\ref{fig:dr} (in case of rotation 4a) or the node $w_2$
  (in case of rotation 4b) is green after the rotation.
\end{cor}

We note that the node $w_1$ or $w_2$ is the node $w'$ of the first of the two
rotations in the double rotation. Therefore, we can directly apply
Lemma~\ref{lem:hp-rotate}(e).

\begin{obs}\label{obs:rot-unique}
  Given a formula $\Psi$ and a node $v$ in $\Psi$, there is at most one rotation
  $\alpha$ that can be applied at $v$ while respecting the balance conditions in
  Table~\ref{tab:balances}.
\end{obs}

In our algorithms, we will use a function $\doRotation(v)$ that applies
the unique possible rotation that respects the balance conditions in
Table~\ref{tab:balances} at node $v$ and returns the node that is labeled
with $w'$ in Table~\ref{tab:rotations}. This function is convenient, as
it avoids case distinctions on whether we have to
continue with the left or the right child of $v$ in the algorithms.
  
The following two lemmas are heavily used in all our algorithms. Together, they
establish that whenever the formula is heavily imbalanced, then the balance can
be improved as some rotation is possible.
Intuitively, Lemma~\ref{lem:noinvariant-unbalanced-path} says that
whenever there is a yellow node $v$, then the long path starting at $v$ is
imbalanced, and Lemma~\ref{lem:noinvariant-rotation} says that whenever there is
an imbalanced path, then we can apply some rotation. Usually both lemmas are
applied together, but sometimes Lemma~\ref{lem:noinvariant-rotation} is used on
its own and we establish the precondition by other means.

\begin{lem}\label{lem:noinvariant-unbalanced-path}
Let $\Psi$ be a formula that only has green and yellow nodes and $v_0$ be a
  yellow node. Let $\lp(v_0)=v_0\cdots v_k$ be the long path starting at $v_0$. Then
  $k \geq 7$ and for $0 \leq i \leq 7$ it holds that
  $|\balance(v_i)| \geq 2$. The conclusion also holds if $v_0$ is red and
  all other nodes are green or yellow.
\end{lem}
\begin{proof}
  Let $h=\height(\Psi_{v_0})$ be the height of $\Psi_{v_0}$ and $j \in \{1,\dots,k\}$ be the smallest
  number such that $|\balance(v_j)|<2$. 
  As every long path ends with a node of
  balance 0, such a $j$ exists. By definition of balances and paths in trees, we
  have that the height of $\Psi_{v_j}$ is $h-j$ and the heights of the two
  subformulas of $\Psi_{v_j}$ are $h-j-1$ and $h-j-1-|\balance(v_j)|$. As both
  subformulas of $\Psi_{v_j}$ only have green and yellow nodes, we can conclude
  that $\Psi_{v_j}$ has at least
  \[
    2^{\frac{h-j-1-1}{10}} + 2^{\frac{h-j-1-|\balance(v_j)|-1}{10}} \;\; \geq \;\; 2
    \cdot 2^{\frac{h-j-3}{10}} \;\; = \;\; 2^{\frac{h-j+7}{10}}
  \]
  many leaves, where $h-j-1$ and $h-j-1-|\balance(v_j)|$ are the heights of the
  children of $v_j$.  The inequality follows by $|\balance(v_j)| \leq 1$. As $v_0$
  is yellow or red, i.e., $\numleaves{\Psi_{v_0}} < 2^{\frac{h}{10}}$, and
  $\numleaves{\Psi_{v_j}} < \numleaves{\Psi_{v_0}}$, we can conclude that $j$ has to
  be at least 8. This concludes the proof.
\end{proof}

\begin{lem}\label{lem:noinvariant-rotation}
  Let $\Psi$ be a formula and $v$ be a node in $\Psi$ such that $v$ and
  the following 7 nodes on $\lp(v)$
  have an absolute balance of at least two. Then there is a node $u \in \lp(v)$, such
  that either
  \begin{itemize}
  \item a height reducing rotation is applicable at $u$;
  \item a height reducing double rotation is applicable at $u$; or
  \item $u$ is a $\conapp$-node, $\balance(u) \leq -2$ and a height
    preserving rotation is applicable at $u$.
  \end{itemize}
  Furthermore, the distance between $v$ and $u$ is at most 6.
\end{lem}  
\begin{proof}
  We denote $v$ by $v_0$. Let $\lp(v_0)=v_0 \cdots v_k$. By precondition, we have
  $|\balance(v_i)| \geq 2$ for $0 \leq i \leq 7$. We show the statement by a
  series of case distinctions on the balances and labels of the nodes
  $v_0,\dots,v_7$.

  The first case is that there is a $\conapp$-node $v_i$ with $i<7$ and
  $\balance(v_i) \leq -2$. In this case we can apply one of the rotations $1a$ (if
  $v_{i+1}$ is a $\conapp$-node), $2a$ (if $v_{i+1}$ is a $\concat_{HV}$-node),
  or $3a$ (if $v_{i+1}$ is a $\concat_{VH}$-node). We note that
  $\balance(v_{i+1}) \neq 0$ and that $v_{i+1}$ cannot
  be a $\concat_{HH}$-node as $\Psi_{v_{i+1}}$ must represent a context. This
  rotation is either height reducing or height preserving, depending on the sign
  of $\balance(v_{i+1})$.

  From now on, we assume that all $\conapp$-nodes from $v_0,\dots,v_6$ have a
  positive balance, as otherwise the first case would apply. We note that by
  assumption none of these nodes can have a balance of $-1$, $0$, or $+1$.

  The second case is that there are two consecutive $\conapp$-nodes $v_i$ and
  $v_{i+1}$ with $i \leq 5$. In this case, we can apply the height reducing
  rotation 1b at $v_i$.

  The third case is that there are three consecutive $\concat$-nodes $v_{i}$,
  $v_{i+1}$, and $v_{i+2}$ with $i \leq 4$. In this case, we can apply one of
  the rotations 1a, 1b, 4a, or 4b, depending on the signs of $\balance(v_i)$ and
  $\balance(v_{i+1})$.
  
  Case 4: If none of the first three cases apply, then there are neither two consecutive
  $\conapp$-nodes nor three consecutive $\concat$-nodes on $v_0,\dots,v_6$. But
  then there have to be two $\concat$-nodes $v_i$ and $v_j$ both immediately
  followed by $\conapp$-nodes $v_{i+1}$ and $v_{j+1}$ with $0 \leq i < j \leq
  5$. 

  Case 4a: Either $v_i$ is a $\concat_{HH}$-node, $v_i$ is a $\concat_{VH}$-node with
  $\balance(v_i)<0$, or $v_i$ is a $\concat_{HV}$-node with $\balance(v_i)>0$.
  In this case we can either apply height reducing rotation 2b or height
  reducing rotation 3b. We remind that the $\conapp$-node $v_{i+1}$ has a
  positive balance, as otherwise case 1 would apply.

  Case 4b: The remaining cases are that $v_i$ is a $\concat_{VH}$-node with
  $\balance(v_i)>0$ or $v_i$ is a $\concat_{HV}$-node with $\balance(v_i)<0$. In
  both cases, $\Psi_{v_{i+1}}$ represents a forest. If $\Psi_{v_{i+1}}$
  represents a forest, then all nodes $v_\ell$ with $i < \ell \leq 6$ are either
  $\concat_{HH}$ or $\conapp_{VH}$ nodes. A formula that represents a forest
  must have either a $\concat_{HH}$-node or a $\conapp_{VH}$-node at the top.
  For a $\concat_{HH}$-node, both subformulas have to represent forests, and for
  a $\conapp_{VH}$-node, the right subformula has to represent a forest. And on
  $\lp(v)$ we are always going to the right subformula at each $\conapp$-node,
  as we assume that these nodes have positive balance. Therefore, $v_j$ has to
  be a $\concat_{HH}$-node and we can either apply 2b or 3b. This concludes the
  proof.
\end{proof}

We want to apply Lemma~\ref{lem:noinvariant-rotation} to reduce the height
of some subformulas. Therefore, in our algorithms, we consider a rotation to
be possible, exactly when one of the three cases of
Lemma~\ref{lem:noinvariant-rotation} applies. The following lemma establishes
that---after a few height preserving rotations---finally a height reducing
operation is possible, if the preconditions of
Lemma~\ref{lem:noinvariant-rotation} are satisfied.

  \begin{lem}\label{lem:noinvariant-hr}
    After applying a height preserving rotation at a $\conapp$-node $v$ which had
    $\balance(v) \leq -2$, the resulting node $v'$ has an absolute balance of at
    least 2 and either $v'$ is a $\concat$-node or $\balance(v')$ is positive.
  \end{lem}
  \begin{proof}
    Applying an applicable height preserving rotation at a node $v$ always
    yields a node $v'$ with absolute balance at least 2 by
    Lemma~\ref{lem:hp-rotate}(b). The only height preserving rotations that can
    be applicable at $v$ are 1a, 2a and 3a. In case of rotation 1a the balance of $v'$
    is positive, as $w'$ is the right child of $v'$ and in case of rotation 2a
    or 3a, the node $v'$ is a $\concat$-node.
  \end{proof}

By Lemma~\ref{lem:noinvariant-hr} we get that after applying height
  preserving rotations at all $\conapp$-nodes $v$ with $\balance(v) \leq -2$,
  there has to be a node where we can apply a height reducing rotation, as the
  third case can no longer apply.

As we cannot prove our desired runtime bounds if we always apply the rotations
on all nodes that satisfy one of the three cases, we sometimes restrict which
rotations we apply. Therefore we use the following tests in our algorithms:
\begin{itemize}
\item $\call{rotationPossible}(v)$: Any of the three cases of
  Lemma~\ref{lem:noinvariant-rotation} applies at node $v$.
\item $\call{hr-rotationPossible}(v)$: A height reducing simple or double
  rotation is applicable\linebreak at $v$.
\item $\call{hp-rotationPossible}(v)$: $v$ is a $\conapp$-node with negative
  balance and a height preserving rotation is applicable at $v$.
\item $\call{simpleRotationPossible}(v)$: A simple height reducing rotation is
  possible at $v$ or $v$ is a $\conapp$-node with negative balance and a height
  preserving rotation is applicable at $v$
\end{itemize}

We would like to stress that we never apply rotations at nodes that satisfy
none of the three cases from Lemma~\ref{lem:noinvariant-rotation}.

\subsection{Preprocessing in Linear Time}\label{sec:preprocessing}
This subsection is devoted to prove the following theorem:

\begin{thm}\label{thm:construct}
  Given a tree $T$, one can construct in time $\mathcal{O}(|T|)$ a formula $\Psi$ of height $\mathcal{O}(\log(|T|))$
  that represents $T$.
\end{thm}

We present the algorithm as Algorithm~\ref{alg:construct}. The main function 
\call{construct} takes $T$ as argument, constructs a formula $\Psi$ corresponding to
the root of the tree, and calls \call{constructRecursive} to insert all
descendants of the root into $\Psi$. At the end it performs some optimizations
(by rotations) before returning the formula $\Psi$. The function
\call{constructRecursive} does a
preorder traversal of $T$ and inserts each node into $\Psi$, using either
subdivision (to insert the first child of some node) or leaf insertion (to
insert all other children). As the insertion
increases the height of the subformula $\Psi_v$, the algorithm calls the
function \call{optimizeUpwards} that goes upwards and searches for a place to
apply a single height reducing rotation. A height reducing rotation at some node $u$
neutralizes the height increase due to the insertions for the ancestors of $u$.
If \call{optimizeUpwards} does not find a possibility for a height reducing
rotation, it does not modify the formula. It stops once it reaches a node whose
height did not change, because the insertion was in the shallower subtree. For a
more detailed description of how \call{optimizeUpwards} works, we refer to the
following proofs, especially the proof of Lemma~\ref{lem:ou-conapp}.

\

After a complete formula is constructed, the algorithm optimizes the complete
formula by doing a postorder traversal of the formula and applying all possible
rotations. This optimizing step is necessary, as the construction does not apply
all possible rotations and might create subformulas that are of height linear in
the number of nodes in the subformula. Especially if we allow deletions later,
this might lead to formulas that are heavily unbalanced. 

\begin{algorithm}[t]
 \caption{Construct formula $\Psi$ from tree $T$\label{alg:construct}}
 \begin{algorithmic}[1]
\Function{\call{construct}}{non-empty tree $T$}
     \State $\Psi \gets \lab(\Root(T))$
     \State $\call{constructRecursive}(\Root(T), \Root(\Psi))$
     \State $\call{optimizeAll}(\Root(\Psi))$
     \State \Return $\Psi$
   \EndFunction
   \medskip
   \Function{\call{constructRecursive}}{$v_t$: a node of $T$, $v$: leaf of $\Psi$
     corresponding to $v_t$}
     \If{$v_t$ has a child}
     \State $\mathsf{subdiv}(v,\lab(\firstchild(v_t)))$ \Comment{see Section~\ref{sec:definitions}}
     \State $\call{optimizeUpwards}(v)$
     \State $\call{constructRecursive}(\firstchild(v_t), \rc{v})$
     \EndIf
     \If{$v_t$ has a right sibling}
     \State $\mathsf{insert_R}(v,\lab(\nextsibling(v_t)))$ \Comment{see Section~\ref{sec:definitions}}
     \State $\call{optimizeUpwards}(v)$
     \State $\call{constructRecursive}(\nextsibling(v_t), \rc{v})$
     \EndIf
   \EndFunction
\medskip
   \Function{\call{optimizeUpwards}}{$v$: node in some formula $\Psi$}
\While{$v \neq \bot$ and no height reducing rotation was done}
        \If{$\call{hr-rotationPossible}(v)$}
          \State $\doRotation(v)$\label{alg:construct:normalize1}
        \ElsIf{$v$ is a $\conapp$-node with $\balance(v) \leq -9$}
          \While{$\call{hp-rotationPossible}(v)$}\label{alg:ou:downwards-while}
            \State $w' \gets \doRotation(v)$\label{alg:construct:rotate-hp}
            \State $v \gets w'$ \Comment{Go to the node $w'$ as
              depicted in Table~\ref{tab:rotations}}\label{alg:construct:assignv1}
\EndWhile
        \Else 
          \State $v \gets \begin{cases} \parent(v) & \text{if $v =
              \deepchild(\parent(v))$} \\ \bot & \text{otherwise} \end{cases}$
          \Comment{\pbox[t]{10cm}{stop if height of $\Psi_v$ is not larger\\than
              height of its sibling}}\label{alg:construct:assignv2}
           \label{alg:ou:goup}
        \EndIf
      \EndWhile 
    \EndFunction
\medskip
   \Function{optimizeAll}{$v$: node in some formula $\Psi$}
     \If{$v$ is an inner node}
       \State $\call{optimizeAll}(\lc{v})$
       \State $\call{optimizeAll}(\rc{v})$
       \State $\call{doAllRotations}(v)$
     \EndIf
   \EndFunction
   \medskip
   \Function{doAllRotations}{$v$: node in some formula $\Psi$}
     \While{$\call{rotationPossible}(v)$}
       \State $\doRotation(v)$
       \State $\call{doAllRotations}(\lc{v})$
       \State $\call{doAllRotations}(\rc{v})$
     \EndWhile
   \EndFunction
 \end{algorithmic}
\end{algorithm}

\begin{lem}\label{lem:construct-sound}
  Algorithm~\ref{alg:construct} is correct, i.e., after the algorithm the
  formula $\Psi$ represents the tree $T$.
\end{lem}
\begin{proof}
  It is straightforward to verify that the function \call{construct} does indeed
  construct a correct formula if one ignores all calls to \call{optimizeUpwards}
  and the call to \call{optimizeAll}. The function does a preorder traversal of
  $T$ and inserts each node it finds into the formula.

  The only changes applied to the formula by the functions
  \call{optimizeUpwards} and \call{optimizeAll} are rotations. By
  Lemma~\ref{lem:rotations-sound}, which says that rotations do not change the
  represented tree, we get that Algorithm~\ref{alg:construct} does indeed
  construct a formula that represents $T$.
\end{proof}

Furthermore, it is easy to see that after the algorithm is finished, no rotation
can be applied. Therefore, lemmas~\ref{lem:noinvariant-unbalanced-path}
and~\ref{lem:noinvariant-rotation} give us the following:

\begin{cor}\label{cor:construct-height}
  Algorithm~\ref{alg:construct} produces a formula that satisfies the strong
  invariant.
\end{cor}

It remains to show that the algorithm runs in linear time in the size of $T$.
Even if phrased differently, the main idea of the proof follows the line of an
amortized analysis that Mehlhorn and Tsakilidis~\cite{MehlhornT-siam1986} did
for AVL trees. We first analyze the behavior of $\call{optimizeUpwards}$ at
$\conapp$-nodes with a big negative balance.

\begin{lem}\label{lem:ou-conapp}
  If the function \call{optimizeUpwards} arrives at a $\conapp$-node $v$ with
  $\balance(v) \leq -9$, then it applies a (possibly empty) series of consecutive height preserving
  rotations starting at $v$, followed by a height reducing rotation on $\lp(v)$.
\end{lem}
\begin{proof}
We use $v_0$ to denote the node $v$ of the initial call to
\call{optimizeUpwards}. We establish the following invariant: As long as the
while-condition is true, the node $v_0$ is always on $\lp(v)$. Indeed this
trivially holds true if $v$ is reassigned in Line~\ref{alg:construct:assignv1} and it also holds
true if $v$ is reassigned in Line~\ref{alg:construct:assignv2}, as the node $w'$ after a height
preserving rotation is always on $\lp(v)$.
 
  Let now $v$ be a $\conapp$-node with $\balance(v) \leq -9$ that is encountered
  by the algorithm. We first observe that at a $\conapp$-node with a balance
  less than minus one, the only scenario where no rotation is
  applicable is that the left child has balance 0. Otherwise we could apply one
  of the rotations~1a, 1b, 2a, or 3a. We now argue that the left child
  of $v$ cannot have a balance equal to $0$. Indeed, $v_0$ that is always on $\lp(v)$
  is the only node on $\lp(v)$ that can have balance 0. However
  $\height(v_0)=2$, as $v_0$ is the parent of the newly inserted node. This
  would imply $|\balance(v)| \leq 1$.

  So we have established that a rotation at $v$ is possible. If the possible
  rotation is height reducing, we are done. Otherwise, the rotation is height
  preserving. The node $w'$ returned by $\doRotation(v)$ in Line~\ref{alg:construct:rotate-hp} is the deep child of $v'$ (i.e., of the new $v$).
  The balance of $w'$ is one more (i.e., it is less negative) than
  the balance of $v$ was before the rotation (Lemma~\ref{lem:hp-rotate}(c)), and
  this node $w'$ has to be a $\conapp$-node. The algorithm continues until
  either a height reducing rotation is possible at the current node or it has
  done at least eight height preserving rotations in a row, as the balance of the
  next node it looks at is always exactly one more than the previous one
  (Lemma~\ref{lem:hp-rotate}(c)). In the former case, we apply the height
  reducing rotation and we are done. In the latter case, the nodes where the
  algorithm applied height preserving rotations are consecutive on
  $\lp^{-1}(u)$, where $u$ is the lowest node where a height preserving
  rotation was performed. By Lemma~\ref{lem:hp-rotate}(b) all those nodes have
  an absolute balance of at least 2. Therefore, by
  Lemma~\ref{lem:noinvariant-rotation}, a rotation is possible
  at one of these nodes. As the algorithm already applied height
    preserving rotations at all of these nodes, this rotation has to be height
    reducing by Lemma~\ref{lem:noinvariant-hr}.
  The algorithm will go upwards again once no height
  preserving rotation is possible any more, apply this height reducing rotation
  and stop. It will therefore not go above the highest node on which a height
  reducing rotation was applied. This concludes the proof.
\end{proof}

We introduce a bit of notation that helps with the following proof. For the
amortized analysis, we use the following potential function:
\[
  \Phi(\Psi) \quad=\quad |\{v \mid \balance(v)=0\}|
  \;\;+\;\; \quad \sum_{\mathclap{\pbox{10cm}{\scriptsize $\begin{aligned}v \in \{w \in \Psi \;|\;
    \lab(w)=\conapp \text{ and }&\\ -8 \leq \balance(w) \leq
    0\}&\end{aligned}$}}} \quad 9 + \balance(v)
\]
The first term counts the number of nodes in $\Psi$ with balance 0. These nodes
can get an absolute balance of 1 by the insertion. They have to be skipped when
going upwards as no rotation is possible. The second term, i.e., the sum,
accounts for $\conapp$-nodes with a balance between $-8$ and $0$. These nodes also
might have to be skipped when going upwards. How often the same node can be
skipped during consecutive insertions depends on its balance. Once the imbalance
is big enough, it will not be skipped because we know that we can perform a series of height preserving
rotations followed by one height reducing rotation. By the condition below the sum symbol, we can conclude that very term below the sum produces numbers
between 1 and 8 inclusive. The potential $9 + \balance(v)$ associated to $\conapp$-nodes with a balance between -8 and 0 corresponds to how often $v$ might need to be
skipped. Obviously, $\Phi(\Psi)$ cannot be negative. 

\begin{lem}\label{lem:ou-runtime}
  The summarized runtime of all calls to \call{optimizeUpwards} in
  Algorithm~\ref{alg:construct} is in $\mathcal{O}(|T|)$.
\end{lem}
\begin{proof}
  We analyze the change of the potential $\Phi(\Psi)$ during the insertion
  of a new node $v$. We need to show two things. First, that the potential is increased by at
  most a constant value by the insertion itself, i.e., not counting the individual increments and decrements of potential in \call{optimizeUpwards}. And second, that the runtime of \call{optimizeUpwards} is
  linear in the overall decrease of potential. The proof does not calculate the net change
  of potential directly. Instead we account for increments
  and decrements separately.

  First of all, we calculate the increments. The insertion adds a new inner node $\parent(v)$
  that has a balance of 0 and could be a $\conapp$-node. This increases the
  potential by at most 10. The only other node $u$ that can increase its
  potential is the parent of the topmost node on $\lp^{-1}(\parent(v))$. This
  node $u$ can improve its balance by 1 and thus possibly increase its contribution to the
  potential by 2.
  All other changes of balance at some node $u$ that
  directly result from the insertion (and not from a rotation) can only worsen
  the balance of $u$, as $\lp(u)$ can only get longer. Therefore, they cannot increase the potential.

  We already know that the algorithm performs at most one height reducing
  rotation (as it will directly stop afterwards) and at most one series of
  height preserving rotations, i.e., it enters the while-loop in Line~\ref{alg:ou:downwards-while} at
  most once. After one series of height preserving operations, either a height
  reducing rotation is possible at the current node, or at most eight nodes above
  (see discussion of the algorithm above). All nodes where we did a height
  preserving rotation will have an absolute balance of at least two and if they
  are a $\conapp$-node they have a positive balance. Therefore, none of the
  rotations did increase the potential contributed by any of these nodes. We only need to account for
  an increased potential due to the height reducing rotation and for the
  $w$ node of the last height preserving rotation. This can be at most constant.
  We note that the balances of the ancestors of the highest node reached in the
  whole insertion operation cannot change if we do perform a height reducing
  rotation, as this rotation cancels out the height increase due to the
  insertion.
  
  We still need to look at the nodes on the path from $u$ to $\parent(v)$, i.e.,
  from the topmost node reached by the algorithm to the parent of the inserted node.
  The balances of these nodes worsen by one. This will
  decrease the potential $\Phi(\Psi)$ by one for every node that changes the
  balance from zero to some non-zero value, i.e., for every node on the path that
  prior to the insertion had a balance of zero. Furthermore it decreases the
  potential by one for every $\conapp$-node that has a balance between $-8$ and $0$
  (inclusive). For all other nodes from $u$ to $\parent(v)$, the potential is unchanged.
  We now show that the decrease in potential is at least linear in the
  runtime of $\call{optimizeUpwards}$.

  In fact, we can show by Lemma~\ref{lem:noinvariant-rotation} that at
  least every eighth node on the path from $u$ to $\parent(v)$ had a
  balance of 0 before the insertion, or was a $\conapp$-node with a balance in the range
  $-8$ to 0. We remind that all nodes on this path that had a non-zero
  balance before the insertion will have an absolute balance of at least 2 after
  the insertion. Also at a $\conapp$-node with a negative balance of at
  least 9, the algorithm would have stopped. Assume towards a contradiction
  that the algorithm arrives at a node $u'$ strictly below $u$, such that the
  first eight nodes on $\lp(u')$ do have an absolute balance of at least two
  (after the insertion) and are not $\conapp$-nodes with a negative balance.
  Then the algorithm can apply a height reducing rotation at $u'$ by
  Lemma~\ref{lem:noinvariant-rotation}. We note that the third case of
  Lemma~3.11 cannot apply, as by assumption there are no $\conapp$-nodes with
  a negative balance at the start of $\lp(u')$. This is a contradiction to
  the assumption that $u$ is the highest node reached by the function.

  As the potential cannot become negative, is increased by at most a constant
  for each insertion, and is decreased by at least a number linear in the
  runtime of \call{optimizeUpwards}, the summarized runtime of all calls to
  $\call{optimizeUpwards}$ is in $ \mathcal{O}(|T|)$.
\end{proof}

It remains to show the runtime of the \call{optimizeAll} function. Towards this
goal, we want to understand how applying a rotation at a node $v$ affects
$\fbalance(\Psi)$.

\begin{lem}\label{lem:rotate}
  Applying a height preserving rotation at some node $v$ in $\Psi$ does not change
  $\fbalance(\Psi)$. Applying a height reducing rotation at some node $v$ in
  $\Psi$  strictly improves $\fbalance(\Psi)$.
\end{lem}
\begin{proof}
  By Lemma~\ref{lem:hp-rotate}(d), a height preserving rotation at $v$  does not change $\fbalance(\Psi_v)$
  and it is easy to see that a height preserving rotation cannot change the
  balance of any node outside of $\Psi_v$.
By Lemma~\ref{lem:hr-rotate}(b), a height reducing rotation at $v$ improves
  $\fbalance(\Psi_v)$ by at least three. There is at most one node $u$ in
  $\Psi$ (the parent of the topmost node on $\lp^{-1}(v)$), such that
  $|\balance(u)|$ worsens. Furthermore, $|\balance(u)|$ can only worsen by one.
  Altogether we can conclude that $\fbalance(\Psi)$ strictly improves.
\end{proof}
  
\begin{lem}\label{lem:oa-runtime}
  The runtime of \call{optimizeAll} is linear in $\mathcal{O}(|T|)$ if called from the
  construction algorithm.
\end{lem}
\begin{proof}
  We first observe that the total imbalance $\fbalance(\Psi)$ of the formula
  $\Psi$ is linear in $\mathcal{O}(|T|)$ just before the call to \call{optimizeAll}. In
  fact, for each insertion, $\fbalance(\Psi)$ is only worsened by one for each
  node on the path from $u$ to $\parent(v)$, where $v$ is the newly inserted
  node and $u$ is the topmost node
  reached by the $\call{optimizeUpwards}$ function. As the runtime of
  $\call{optimizeUpwards}$ is in $\Theta(\height(\Psi_u))$, and the summarized
  runtime of all those calls is linear in $|T|$ by Lemma~\ref{lem:ou-runtime},
  $\fbalance(\Psi)$ is linear in $|T|$. From Lemma~\ref{lem:rotate} we know that rotations performed by
  \call{optimizeUpwards} can only improve $\fbalance(\Psi)$.

  Towards the lemma statement, we show that the runtime of $\call{optimizeAll}$
  is linear in $|T|+\fbalance(\Psi)$. It is obvious that $\call{optimizeAll}$ is
  called exactly once for each inner node in the formula, as this is a classic
  post-order traversal. And the number of calls to $\call{doAllRotations}$ is
  one for each inner node of the formula and additionally two calls for every
  performed rotation. It thus suffices to show that the number of performed
  rotations is linear in $\fbalance(\Psi)$.

  As height reducing rotations strictly improve $\fbalance(\Psi)$ and height
  preserving rotations leave $\fbalance(\Psi)$ unchanged
  (Lemma~\ref{lem:rotate}), it remains to bound the number of height preserving
  rotations. Especially we need to show that $\call{doAllRotations}$
  cannot do more than a constant number of height preserving rotations without
  also doing a height reducing rotation. The only reason why there are two
  recursive calls are double rotations, where rotations might be possible on
  both children afterwards. For simple rotations, the only node where another
  rotation might become possible is $w'$ (see Table~\ref{tab:rotations}), as
  the other child is an unchanged subformula where we already called
  \call{optimizeAll}. So when we have subsequent height preserving rotations,
  these are all on the same long path. By Lemma~\ref{lem:hp-rotate}(b), these
  nodes will all end up with an absolute balance of at least 2. By
  Lemma~\ref{lem:noinvariant-rotation}, a rotation is possible on the lowest 8
  such nodes and by Lemma~\ref{lem:noinvariant-hr}, this rotation cannot be
  again a height preserving rotation. The while-condition will make sure that
  we apply this possible rotation after the recursive calls return. Thus if
  there are more than eight consecutive height preserving rotations, the
  algorithm does a height reducing rotation that yields an improvement of
  $\fbalance(\Psi)$ linear in the number of performed height reducing rotations
  (Lemma~\ref{lem:hr-rotate}(d)). Altogether the runtime is bounded by the
  size of $T$ and the overall imbalance of $\Psi$.
\end{proof}

Theorem~\ref{thm:construct} now follows from Lemma~\ref{lem:construct-sound}
(soundness), Corollary~\ref{cor:construct-height} (height of formula) and
lemmas~\ref{lem:ou-runtime} and~\ref{lem:oa-runtime} (runtime).

\subsection{Maintaining Parse Trees under Insertions}\label{sec:insertions}
Our proof that we can maintain the strong invariant under insertions takes
several steps. In Lemma~\ref{lem:insert}, we show that after any insertion, the
formula still satisfies the weak invariant. We already established in
Lemma~\ref{lem:noinvariant-unbalanced-path} that on the long path of any yellow
node, there are several consecutive nodes that are imbalanced. This is exploited
in Lemma~\ref{lem:noinvariant-rotation} to show that we can apply some rotation
to reduce the height of $\Psi_v$ and change the color of the yellow node back to
green. Unfortunately, the rotation might create (at most) one other yellow node. However, as
this new yellow node is strictly deeper in the parse tree as the previous one,
at most logarithmically many repetitions suffice to reestablish the strong
invariant. This leads to Algorithm~\ref{alg:trh} that we use to recursively
reestablish the strong invariant and Algorithm~\ref{alg:insert} that performs
the insertion and makes use of Algorithm~\ref{alg:trh}.

\begin{lem}\label{lem:insert}
  Let $\Psi$ be a formula that satisfies the strong invariant. Then after
  applying an insertion update $\Delta$ on $\Psi$, the resulting formula
  $\Delta(\Psi)$ satisfies the weak invariant and all yellow nodes of
  $\Delta(\Psi)$ are on $\lp^{-1}(v)$, where $v$ is the parent of the newly
  inserted node.
\end{lem}
\begin{proof}
  In all possibilities of a node insertion in the tree (insertion of a leaf as a
  left/right sibling of an existing node and subdivision), a leaf of the formula
  $\Psi$ is replaced by a subformula with one inner node $v$ and two leaves. One
  of these leaves is the new node. Thus for all nodes on $\lp^{-1}(v)$ in
  $\Delta(\Psi)$ it holds that the height and the number of leaves in the
  subtree has increased by one. For all nodes not on $\lp^{-1}(v)$ the height
  did not change and the number of leaves in the subtree might have increased by
  one. As observed in Section~\ref{sec:fa-highlevel},  
  for nodes on $\lp^{-1}(v)$, the color can change from
  yellow to red, or from green to yellow. As all nodes were green prior to
  the insertion, there are no red nodes after the insertion and all yellow nodes
  after the insertion have to be on $\lp^{-1}(v)$. Thus the weak invariant is satisfied.
\end{proof}

We now state Algorithm~\ref{alg:trh},
which tries to optimize the height of a given (sub-)formula. Algorithm~\ref{alg:trh} does not check the color of any
node. Instead it just starts at a given node $v$, goes downwards on the long path
and performs rotations if possible. The colors are only used in the proofs.
The function $\doRotation(v)$ in lines~\ref{alg:trh:rotate}
and~\ref{alg:trh:finalrotate} applies the unique (see
Observation~\ref{obs:rot-unique}) possible rotation at node $v$ and returns the
node that is labeled with $w'$ in Table~\ref{tab:rotations}. In case of
a double rotation 4a, it returns the node $w_2$ and in case of a double
rotation 4b, it returns the node $w_1$ (see Figure~\ref{fig:dr}). We will
argue later that the preconditions of Corollary~\ref{cor:dr-onenodegreen} are
satisfied whenever we do a double rotation in Algorithm~\ref{alg:trh}. 

There is a little detail we would like to stress: In Line~\ref{alg:trh:recurse2}, the recursive call is only on the subformula $\Psi_{w'}$. This is to ensure that the node $u$ in Line~\ref{alg:trh:testfinal} is not above $w'$ for subsequent recursive calls.

We now have one lemma that characterizes the effects of Algorithm~\ref{alg:trh}
and another lemma talking about the runtime.

\begin{algorithm}[t]
  \caption{Try to optimize the height of a subformula $\Psi_v$}\label{alg:trh}
  \begin{algorithmic}[1]
    \Function{\tryReduceHeight}{$\Psi,v$}
      \If{$|\balance(v)| \geq 2$}
        \If{$\call{simpleRotationPossible}(v)$}\label{alg:trh:testsimple}
          \State $w' \gets \doRotation(v)$\label{alg:trh:rotate}
          \State $\tryReduceHeight(\Psi,w')$ \label{alg:trh:recurse1}
        \Else
          \State $\tryReduceHeight(\Psi,\deepchild(v))$\label{alg:trh:godown}
        \EndIf
      \ElsIf{$\call{rotationPossible}(u)$ for some node $u \in \lp^{-1}(v)$}\label{alg:trh:testfinal}$^\dagger$
        \State $w' \gets \doRotation(u)$ \label{alg:trh:finalrotate}
        \State $\tryReduceHeight(\Psi_{w'},w')$ \label{alg:trh:recurse2}
      \EndIf
    \EndFunction
  \end{algorithmic}
  \raggedright
  \footnotesize{$^\dagger$If there are several nodes $u$ on $\lp^{-1}(v)$ that satisfy the condition, we use
    the lowest such node.}
\end{algorithm}

\begin{lem}\label{lem:trh-reduceheight}
  Let $v$ be a yellow node in a formula $\Psi$ that consists only of green and
  yellow nodes. Let $\Psi_v'$ denote the subformula $\Psi_v$ after the call
  $\tryReduceHeight(\Psi,v)$. Then $\height(\Psi'_v) = \height(\Psi_v) - 1$. If $v$
  was the only yellow node in $\Psi_v$, then all nodes in $\Psi'_v$ are green.
\end{lem}
\begin{proof}
  We use $v_i$ to denote the value of $v$ in the $i$-th recursion step of the
  algorithm, where $v_0$ is the value of $v$ in the topmost call.

  All recursive calls are made on $\lp(v_0)$ unless we perform a height
  reducing rotation at some $v_i$, reducing the height of each $\Psi_{v_j}$ with
  $j \leq i$ by one. Indeed, if at some node no rotation is performed, we stay
  on the long path, as we are going to the deeper child. And if a height
  preserving rotation is performed, the node $w'$ returned by the call
  $\call{doRotation}$ is also on the deeper subformula. 

  By Lemma~\ref{lem:noinvariant-unbalanced-path}, we know that the first 8 nodes
  on $\lp(v_0)$ have an absolute balance of at least 2. This does not change as
  long as we only do height preserving rotations on the path, as we can reapply
  the lemma on the modified path after the rotation. By
  lemmas~\ref{lem:noinvariant-unbalanced-path} and~\ref{lem:noinvariant-rotation}, we know
  that as long as $v$ stays yellow, i.e, as long as the algorithm does not apply
  a height reducing rotation, there will always be a rotation that is applicable
  on $\lp(v_0)$. Thus at some point, the algorithm performs a height reducing
  rotation on $\lp(v_0)$, either in Line~\ref{alg:trh:rotate} or in
  Line~\ref{alg:trh:finalrotate}. This proves the first part of the lemma
  statement. We note, that no height preserving rotation is possible on
  $\lp^{-1}(v_i)$, where $v_i$ is the current node, as a height preserving
  rotation cannot enable another height preserving rotation.

  We now show the second part of the lemma. If $v$ was the only yellow node in
  $\Psi_v$, then there are only green nodes in $\Psi'_v$. From the first part of
  the lemma, it is clear, that $v$ will be green after the call to
  \tryReduceHeight, as the height is reduced by one.

  If after a simple rotation we have a new yellow node, then this node can only be
  the node $w'$ as denoted in Table~\ref{tab:rotations}: The node $v$ cannot turn
  yellow, as the height cannot increase and the subformulas $\psiup$,
  $\psidown$, and $\psistay$ are not touched by the rotation. However, we do a
  recursive call on every node $w'$ of every performed rotation. A simple
  inductive argument yields that a yellow node that was created by a rotation
  turns green by the recursive call on that node. This shows the second part of the lemma 
  for all nodes that result from simple rotations.

  It remains to show the second part of the lemma for nodes resulting from double rotations. 
  	Double rotations can only be performed in Line~\ref{alg:trh:finalrotate}.
  We first argue that the preconditions for
  Corollary~\ref{cor:dr-onenodegreen} are satisfied. The node $v$ has
  $|\balance(v)|<2$ by the if-condition. As $u$ is the lowest node on
  $\lp^{-1}(v)$ where a rotation is possible, we know by
  Lemma~\ref{lem:noinvariant-rotation} that $u$ is at most 7 nodes above $v$. 
  Now we can conclude by Corollary~\ref{cor:dr-onenodegreen} that one of the
  nodes $w_1$ and $w_2$ from Figure~\ref{fig:dr} has to be green. The other
  node from $w_1$ and $w_2$---which is returned by the call
  $\call{DoRotation}$ and called $w'$ in the algorithm---might be yellow. As
  we do a recursive call on $w'$, the argument of the lemma follows, just as
  for rotations in Line~\ref{alg:trh:rotate}. 
\end{proof}

The final property that we need for the \tryReduceHeight function is its
runtime. Again, the lemma statement is more detailed than needed for the
insertion algorithm. The bound of the runtime by the improvement of the balance
is needed later for the deletion algorithm. We note that the runtime bounds are
not subject to any preconditions.

\begin{lem}\label{lem:trh-runtime}
  The runtime of $\tryReduceHeight(\Psi_v, v)$ is in
  $\mathcal{O}(\height(\Psi_v))$. Furthermore,
  the runtime of $\tryReduceHeight(\Psi_v, v)$ is also in $\mathcal{O}(\Delta_\fbalance)$,
  where $\Delta_\fbalance$ is the improvement of the balance of $\Psi$, i.e.,
  $\Delta_\fbalance = \fbalance(\Psi) - \fbalance(\Psi')$ with $\Psi'$
  denoting the formula $\Psi$ after the call to $\tryReduceHeight$.
\end{lem}
\begin{proof}
  We first show that the runtime of $\tryReduceHeight(\Psi_v, v)$ is
  in $\mathcal{O}(\height(\Psi_v))$. Recursive calls in lines~\ref{alg:trh:recurse1} and~\ref{alg:trh:godown} are unproblematic, as they are always on nodes strictly below the current $v$.
  By Lemma~\ref{lem:noinvariant-rotation}, recursive calls in Line~\ref{alg:trh:recurse2} can be up to seven nodes above the current $v$. The number of recursive calls in Line~\ref{alg:trh:recurse2} is still in $\mathcal{O}(\height(\Psi_v))$, as the recursive call will not ascent out of $\Psi_{w'}$, which is a strict subformula of $\Psi_v$.
  We can conclude that the overall runtime is restricted to $\mathcal{O}(\height(\Psi_v))$.

  Towards the second claim,
  we show that $\Delta_\fbalance \geq \max(\frac{x}{7} - 7,0)$, where $x$ is
  the recursion depth counting the initial call as 0. Again, we
  denote by $v_i$ the node $v$ in the $i$-th recursion step. The proof is by induction.
  If the algorithm never performs a height reducing rotation, then the recursion
  depth is bounded by 7 by Lemma~\ref{lem:noinvariant-rotation}: In
  this case the algorithm only performs height preserving rotations. If there
  would be more than 7 height preserving rotations, the algorithm could
  afterwards apply a height reducing rotation, as all these nodes have an
  absolute balance of at least 2 afterwards and no further height preserving
  rotation is possible by Lemma~\ref{lem:noinvariant-hr}.

  If the
  algorithm at some point performs a (first) height reducing rotation in
  Line~\ref{alg:trh:rotate}, then $\fbalance(\Psi)$ is
  improved by at least $y$, where $y$ is the (current) recursion depth: The
  balance of $\Psi_{v_y}$ is improved by at least three
  (Lemma~\ref{lem:hr-rotate}(b)), the balance on $\lp^{-1}(v_y)$ is improved by
  at least $y$ (Lemma~\ref{lem:hr-rotate}(d)), and there is at most one node
  (the parent of the topmost node on $\lp^{-1}(v_y)$) that worsens its 
  balance by one.

  The argument follows by induction on the recursive call in
  Line~\ref{alg:trh:recurse1}. Let $x$ be the total recursion depth, then the
  recursive call has a remaining recursion depth of $x-y$, yielding an
    improvement of balance of at least $\max(\frac{x-y}{7}-7,0)$. The total
  improvement of balance is thus at least $\max(\frac{x-y}{7} - 7,0)+y \geq
  \max(\frac{x}{7} - 7, 0)$.

  Similarly, if the algorithm performs a double rotation in
  Line~\ref{alg:trh:finalrotate} at recursion level $y$, then $\fbalance(\Psi)$
  is improved by at least $\max(1, y - 6) \geq \frac{y}{7}$: We already
  argued in the proof of Lemma~\ref{lem:trh-reduceheight} that the node $u$ is
  at most 7 levels higher than $v_y$. The rotation itself improves the balance by at
  least three (Lemma~\ref{lem:hr-rotate}(b)), the improvement on $\lp^{-1}(u)$
  is at least $y-8$ (Lemma~\ref{lem:hr-rotate}(d)), and there can be at most
  one node that worsens it balance by one. We remind that a double rotation is
  a height preserving rotation (that does not change the overall balance)
  followed by a height reducing rotation where we can apply
  Lemma~\ref{lem:hr-rotate}. This implies that the improvement is at least
  $\frac{y}{7}$. Again, the argument follows by induction, concluding the proof.
\end{proof}

A direct consequence of Lemma~\ref{lem:trh-runtime} is that the runtime is bounded by a
constant if no height reducing rotation is performed. 

The final algorithm for insertions is presented as Algorithm~\ref{alg:insert}. It
inserts the new node and then walks upwards until it either finds a possibility
to reduce the height, or reaches the end of the long path. Nodes above did not
increase height and therefore cannot change color to the worse. The test
\call{rotationPossible} in Line~\ref{alg:insert:if} returns true if any single
or double rotation is applicable at node $v$. The rotation can be height
reducing or height preserving.

\begin{algorithm}[t]
  \caption{Insertion Update on a Formula $\Psi$\label{alg:insert}}
  \begin{algorithmic}[1]
\Function{\call{insert}}{update $\Delta \in \{\mathsf{insert_L}(v,a),
      \mathsf{insert_R}(v,a), \mathsf{subdiv}(v,a)\}$, formula $\Psi$}
      \State apply the update $\Delta$  \label{alg:insert:insert}
      \State $v \gets \parent(v)$ \Comment{$v$ is now the parent of the newly
        inserted $a$-node}
      \While{$v \neq \bot$ and no height reducing rotation was done}
          \If{$\call{rotationPossible}(v)$} \label{alg:insert:if}
          \State $w' \gets \doRotation(v)$
          \Comment{$w'$ is the node from
            Table~\ref{tab:rotations} after the rotation}
          \State $v \gets w'$
        \Else
          \State $v \gets \begin{cases} \parent(v) & \text{if $v =
              \deepchild(\parent(v))$} \\ \bot & \text{otherwise} \end{cases}$
        \EndIf
      \EndWhile\label{line:upwards-endwhile}  
      \If{$v \neq \bot$} $\tryReduceHeight(\Psi_v,v)$ \EndIf \label{alg:insert:trh}
    \EndFunction
  \end{algorithmic}
\end{algorithm}

\begin{lem}\label{lem:insert-runtime}
  Algorithm~\ref{alg:insert} performs the insertion operation correctly,
  maintains the strong invariant, and runs in time $\mathcal{O}(\height(\Psi_v))$, where
  $v$ is the topmost node reached in the while-loop.
\end{lem}
\begin{proof}
  The correctness follows from the facts that we apply the insertion operation
  as described in Section~\ref{sec:definitions} in Line~\ref{alg:insert:insert}
  and afterwards the only changes to
  the formula are rotations that are equivalence-preserving according to
  Lemma~\ref{lem:rotations-sound}. The runtime follows from
  Lemma~\ref{lem:trh-runtime} and the fact that the runtime of the while-loop is
  bounded by the height of the formula:  At any given node $u$, we can have at most
  one height preserving rotation, as afterwards the balance of $u$ is
  positive (rotation 1a or 2a), or the node is a $\concat$-node (rotation 3a).

  It remains to show that the algorithm maintains the strong invariant. We
  assume that all nodes were green prior to the insertion. If there is no yellow
  node immediately after the insertion of a new node $v$, the strong
  invariant is maintained and will not be invalidated by subsequent
  height preserving rotations (Lemma~\ref{lem:hp-rotate}(e)). We note that
  in every iteration of the while-loop there is a node $v'$ with
  $|\balance(v')| \leq 1$ at most 8 levels below the node
  currently stored in $v$: If this would not hold, by
  Lemma~\ref{lem:noinvariant-rotation}, a rotation at one of these nodes would
  be possible. This is a contradiction, as the algorithm applies all possible
  rotations at these nodes.

  If the insertion creates yellow nodes, all of them have to be on
  $\lp^{-1}(\parent(v))$ by Lemma~\ref{lem:insert}. Let $u$ be the lowest yellow
  node on $\lp^{-1}(\parent(v))$.
  By Lemma~\ref{lem:noinvariant-unbalanced-path}, there have to be at least 8
  nodes $v'$ with $|\balance(v')| \geq 2$ at the start of $\lp(u)$. As the
  algorithm applies height preserving rotations to every $\conapp$-node with
  balance less than minus one on its way up, it finally has to find and apply a
  height reducing rotation by Lemma~\ref{lem:noinvariant-rotation}. This reduces
  the height of $\Psi_u$ and changes the color of all yellow nodes back to
  green, as they have the same height as before the insertion operation.
  However, it is possible that the node $w'$ returned by $\call{doRotation}$
  when doing a height reducing rotation
  is a yellow node. We note that this node $w'$ is the only node that
  can possibly be yellow after the while-loop, i.e., after there was a height
  reducing rotation. All other nodes are green. This is why we call
  \tryReduceHeight in Line~\ref{alg:insert:trh}. By
  Lemma~\ref{lem:trh-reduceheight} we can conclude that after the call to
  \tryReduceHeight on $w'$ there are only green nodes in $\Psi_{w'}$ (and thus in $\Psi$) and the strong invariant is satisfied.
\end{proof}

\subsection{Deletions}\label{sec:deletions}
In this section we provide an algorithm for deletions and a slightly modified
invariant that is maintained by this algorithm. We also show that the algorithm
for insertion also maintains the deletion invariant for the case that insertions
and deletions are interleaved.
For a non-green node $v$, we define \[\deficit(v) \quad=\quad 2^{\frac{\height(\Psi_v)}{10}}
  - \numleaves{\Psi_v}\] to be the number of leaves that need to be inserted into
$\Psi_v$ without increasing the height of $\Psi_v$ in order to make $v$ green.
Or the other way around $\deficit(v)$ denotes the number of leaves that have
been removed from $\Psi_v$ without decreasing the height since the node $v$ 
was green.

\medskip
\noindent\textbf{Deletion Invariant:} For every non-green node $v$ it holds that
\[
  \deficit(v) \quad\leq\quad \frac{\height(\Psi_v) \cdot \numleaves{\Psi_v} -
  \fbalance(\Psi_v)}{20\cdot\height(\Psi_v)}\;.
\]

The intention behind this invariant is that the more leaves are missing for a
green node, the smaller $\fbalance(\Psi_v)$ has to become. As $\fbalance(\Psi_v)$
cannot become less than zero, we get an upper bound on $\deficit(v)$ that
implies that every non green node has to be yellow.

\begin{lem}\label{lem:delete-invariant}
  Let $\Psi$ be a formula satisfying the deletion invariant. Then all nodes in
  $\Psi$ are green or yellow. 
\end{lem}
\begin{proof}
  Assume that $v$ is a node that is neither green nor yellow. Then
  $\numleaves{\Psi_v} < 2^{\frac{\height(\Psi_v)-1}{10}}$. By the definition of
  $\deficit(v)$, we get
  \[
    \deficit(v) \quad > \quad  2^{\frac{\height(\Psi_v)}{10}} -
    2^{\frac{\height(\Psi_v)-1}{10}} \quad \geq \quad 0.05 \cdot
    2^{\frac{\height(\Psi_v)}{10}} \quad \geq \quad 0.05 \cdot
    \numleaves{\Psi_v} \;.
  \]
  Now we apply the deletion invariant, multiplying both sides by 20:
  \[
    \numleaves{\Psi_v} \quad< \quad \frac{\height(\Psi_v) \cdot
      \numleaves{\Psi_v} - \fbalance(\Psi_v)}{\height(\Psi_v)} \quad = \quad
    \numleaves{\Psi_v} - \frac{\fbalance(\Psi_v)}{\height(\Psi_v)}
  \]
  We clearly see that $\fbalance(\Psi_v)$ is less than zero, which is a
  contradiction, as $\fbalance(\Psi_v)$ is defined to be a sum of positive values.
\end{proof}

\begin{algorithm}[t]
 \caption{Deletions\label{alg:remove}}
 \begin{algorithmic}[1]
\Function{\call{remove}}{formula: $\Psi$, leaf of $\Psi$: $w$}
     \State $\fbalance \gets \fbalance(\Psi)$
     \State $v \gets \mathsf{sibling}(w)$ \Comment{$v$ will be 
       at the position of $\parent(w)$ after the deletion}
     \State $\mathsf{delete}(w)$
     \While{$v \neq \Root(\Psi)$}
     \State $v \gets \parent(v)$
     \If{height of $\Psi_v$ did not decrease and $\fbalance -
       \fbalance(\Psi) < 21 \cdot \height(\Psi)$}
     \State \tryReduceHeight$(\Psi_v,v)$ \Comment{see Algorithm~\ref{alg:trh}}\label{alg:remove:fix}
     \EndIf
     \EndWhile
   \EndFunction
 \end{algorithmic}
\end{algorithm}

We use Algorithm~\ref{alg:remove} to remove a node from $\Psi$. The algorithm
first performs the actual deletion operation as described in
Section~\ref{sec:definitions}. Afterwards we rebalance the formula as needed.
The algorithm goes upwards starting at the parent of the removed node and checks
at each ancestor $v$ of $w$ whether the height of $\Psi_v$ did decrease. If the
height did decrease, we are fine, as then the color of $v$ has to be green, even
if it was yellow before. Otherwise $v$ might actually be yellow (the height did
not decrease but we removed a leaf). Therefore, we try to reduce the height of
$\Psi_v$ by calling $\tryReduceHeight(\Psi_v,v)$.\footnote{We explicitly limit the call to $\Psi_v$ to prevent $\tryReduceHeight$ of inspecting nodes above $v$ in Line~\ref{alg:trh:testfinal} of Algorithm~\ref{alg:trh}, as these nodes are explicitly tested  in Algorithm~\ref{alg:remove}.} This function will decrease the
height of $\Psi_v$ if $v$ is yellow (Lemma~\ref{lem:trh-reduceheight}) and thus make sure that $v$ is green
afterwards. However, we abort the rebalancing process if the overall balance did
improve by at least 21 times the height of $\Psi$. This abort is to ensure a
logarithmic runtime, even if we have many calls to $\tryReduceHeight$ that
individually have a logarithmic worst-case runtime. The if-condition can
be checked by maintaining a global counter for the change of $\fbalance(\Psi)$.
The counter can be updated whenever a local node balance value is updated.
Nodes that change their balance during a rotation can be easily located via updating the rotation subroutines to account for these changes. Height preserving rotations
only affect the balances of $v'$ and $w'$ (see Table~\ref{tab:rotations}).
Height reducing rotations additionally affect the balances on $\lp^{-1}(v')$
and of the parent of the topmost node of this path. The number of balance
changes is linear in the overall improvement of $\fbalance(\Psi)$: Every
height preserving rotation strictly improves balance and only worsens the
balance of at most one node and we already have established that the number of
height preserving rotations is linear in the overall improvement of $\fbalance(\Psi)$ in the proof of Lemma~\ref{lem:trh-runtime}.

The following is then immediate from Lemma~\ref{lem:trh-runtime} that states that the runtime
of $\tryReduceHeight$ is linear in the improvement of balancedness:
\begin{cor}\label{cor:delete-runtime}
  The runtime of Algorithm~\ref{alg:remove} is
  $\mathcal{O}(\height(\Psi))$.
\end{cor}

In order to show that Algorithm~\ref{alg:remove} maintains the deletion
invariant, we first show that the call to \tryReduceHeight maintains this invariant.

\begin{lem}\label{lem:trh-delete}
  Let $v$ be a yellow node in a formula $\Psi$ such that no node except maybe $v$
  violates the deletion invariant in $\Psi_v$. Then the deletion invariant is
  satisfied in $\Psi_v$ after the call $\tryReduceHeight(\Psi_v,v)$.
\end{lem}
\begin{proof}
  We first show that $v$ satisfies the deletion invariant after the call to
  \tryReduceHeight. Afterwards we apply an inductive argument to show that no
  other node can violate the invariant. Let $\Psi'_v$ be the formula $\Psi_v$
  after the call to \tryReduceHeight. We write $v'$ to denote the root of
  $\Psi'_v$.

  If $v$ satisfied the deletion invariant, then so does $v'$, as the
  balancedness can only improve and the height can only decrease due to
  \tryReduceHeight.

  If $v$ did not satisfy the invariant, then $v$ was yellow and thus $\balance(v)
  \neq 0$ by Lemma~\ref{lem:noinvariant-unbalanced-path}. Let $w$ be
  the deeper child of $v$ and $u$ be the sibling of $w$. As $v$ was yellow, we
  conclude from Lemma~\ref{lem:trh-reduceheight} that $\height(\Psi'_v) = \height(\Psi_v)-1 =
  \height(\Psi_w)$. We bound $\deficit(v')$ as follows, where we use the fact
  that $\height(\Psi'_v)=\height(\Psi_w)$ several times.
  \allowdisplaybreaks
  \begin{align*}
    \deficit(v') \quad&=\quad 2^{\frac{\height(\Psi'_v)}{10}} - \numleaves{\Psi_{v'}} & \text{(definition of $\deficit$)} \\
                 &=\quad \underbrace{2^{\frac{\height(\Psi'_v)}{10}} - \numleaves{\Psi_w}}_{\deficit(w)} - \numleaves{\Psi_u} & \quad (\numleaves{\Psi_{v'}} = \numleaves{\Psi_v} = \numleaves{\Psi_w} + \numleaves{\Psi_u})\\
                 &\leq\quad \frac{\height(\Psi'_v) \cdot \numleaves{\Psi_w} - \fbalance(\Psi_w)}{20\cdot\height(\Psi_w)} - \numleaves{\Psi_u} & \text{($w$ satisfies invariant)} \\
                 &\leq\quad \mathrlap{\frac{\height(\Psi'_v) \cdot \numleaves{\Psi_w} - \fbalance(\Psi_v) + \fbalance(\Psi_u) + |\balance(v)|}{20\cdot\height(\Psi'_v)} - \numleaves{\Psi_u}} &  \\
                 &\leq\quad \frac{\height(\Psi'_v) \cdot \numleaves{\Psi_w} - \fbalance(\Psi_v)}{20 \cdot \height(\Psi'_v)} 
  \end{align*}
  In the last but one inequality, we exploit that $\fbalance(\Psi_v) = \fbalance(\Psi_w)
  + \fbalance(\Psi_u) + |\balance(v)|$. In the last inequality, we exploit that
  $\Psi_u$ has exactly $\numleaves{\Psi_u} - 1$ inner nodes, whose
  absolute balances are bounded by the height. Thus $20 \cdot
  \height(\Psi'_v) \cdot \numleaves{\Psi_u}$  is larger than $\fbalance(\Psi_u) + \balance(v)$.
  We can conclude that $v'$
  satisfies the invariant because $\numleaves{\Psi'_v}>\numleaves{\Psi_w}$ and
  $\fbalance(\Psi'_v) \leq \fbalance(\Psi_v)$.

  It remains to show that no other nodes violate the invariant. This part
  of the proof is an induction over the recursion depth, analogous to the one in
  the proof of Lemma~\ref{lem:trh-reduceheight}. If there is no rotation, then there
  can be no other node that violates the invariant by the condition in the lemma
  statement.

  If there is a simple rotation in Line~\ref{alg:trh:rotate}, the only node
  except $v$ which can violate the invariant is the node $w'$ as denoted in
  Table~\ref{tab:rotations}. The induction hypothesis yields that $w'$ satisfies
  the invariant after the recursive call.

  It remains to discuss double rotations in Line~\ref{alg:trh:finalrotate}. We
  discuss the two rotations of the double rotations seperately, starting with
  the height preserving rotation, i.e., the first rotation of the double
  rotation. By Lemma~\ref{lem:hp-rotate}(e), we have that $w'$ is green. We
  remind that $u$ is at most 6 nodes above $v$ and $|\balance(v)| \leq 1$. The 
  argumentation for the second rotation of the double rotation is exactly as for
  simple rotations in Line~\ref{alg:trh:rotate}, as we also do a recursive call on $w'$.
  This concludes the proof.
\end{proof}

Now we show that Algorithm~\ref{alg:remove} maintains the deletion invariant:
\begin{lem}\label{lem:delete}
  The deletion invariant is maintained by Algorithm~\ref{alg:remove}.
\end{lem}
\begin{proof}
  We have to show that the deletion invariant is maintained for all nodes. It is
  maintained for all nodes $u$ of $\Psi$ that are not an ancestor of $v$, as the
  algorithm does not modify $\Psi_u$. It remains to show that the invariant is
  maintained for ancestors of $w$ and for nodes $w'$ that result from some
  rotation.

  We now show the following property for every ancestor $v$ of $w$ which will 
  yield that the invariant has been maintained for $v$:\\
  Either $v$ is green after the update, or
  \begin{itemize}
  \item the height of $\Psi_v$ stayed the same;
  \item the number of leaves of $\Psi_v$ decreased by exactly one; and
  \item $\fbalance(\Psi_v)$ improved by at least $21 \cdot \height(\Psi_v)$.
  \end{itemize}
  It is straightforward to verify that the property ensures that the
  deletion invariant is maintained. We would like to stress that we need that
  $\fbalance(\Psi_v)$ improves by $21 \cdot \height(\Psi_v)$, as we need $20
  \cdot \height(\Psi_v)$ to compensate for the deficit that increases by one,
  and another $\height(\Psi_v)$ to account for the number of leaves that
  decreases by one.

  Now, we prove the property.
  The height of $\Psi_v$ cannot increase. If the height of $\Psi_v$ decreases,
  then $v$ would be green afterwards. Thus if $v$ is still yellow\footnote{We note that $v$ cannot be red, as we only removed a single node from a formula satisfying the deletion invariant. The proof of Lemma~\ref{lem:delete-invariant} actually shows that there is a gap before the node turns red.}, the height of
  $\Psi_v$ has stayed the same. The number of leaves of $\Psi_v$ has decreased
  by one due to the deletion operation and cannot change afterwards. And from
  the fact that the height did not decrease despite $v$ being yellow, we can
  conclude that $\tryReduceHeight$ was never called for $v$. As the if-condition
  thus needs to be false, we can conclude that $\fbalance(\Psi)$ has been
  improved by at least $21 \cdot \height(\Psi)$. We note that
  $\height(\Psi_v) \leq \height(\Psi)$ and that the improvement of
  $\fbalance(\Psi_v)$ equals the improvement of $\fbalance(\Psi)$, as all
  balance changes are internal to $\Psi_v$ in the case that $\Psi_v$ does not
  change its height. We can conclude that $\fbalance(\Psi_v)$ improved by at
  least $21 \cdot \height(\Psi_v)$. 

  It remains to observe that Lemma~\ref{lem:trh-delete} ensures that the
  deletion invariant is satisfied by all nodes $w'$ that result from some
  rotation performed by \tryReduceHeight. This concludes the proof.
\end{proof}

It remains to show that also the insertion algorithm maintains the deletion
invariant. Of course, we already have shown that it even maintains a 
stricter invariant. However, if we arbitrarily mix insertions and deletions, we
can no longer guarantee that the strong invariant is satisfied before a call to
the insertion algorithm.

\begin{lem}\label{lem:insert-deletionInvariant}
  Algorithm~\ref{alg:insert} maintains the deletion invariant.
\end{lem}
\begin{proof}
  We first observe, that an insertion into $\Psi_v$, where $v$ is a yellow node,
  increases the number of nodes in $\Psi_v$ by one, but cannot change the height
  of $\Psi_v$: If the insertion is not on $\lp(v)$, the height cannot increase.
  Otherwise, if the insertion is on $\lp(v)$, the node $v$ will be
  yellow or red. Therefore, by
  lemmas~\ref{lem:noinvariant-unbalanced-path},~\ref{lem:noinvariant-rotation},
  and~\ref{lem:noinvariant-hr} we can conclude that---possibly after doing
  some height preserving operations---a height reducing
  rotation is possible on $\lp(v)$ and thus the insertion algorithm will perform a
  height reducing rotation somewhere on $\lp(v)$.

  As the deletion invariant did hold before the insertion, we
  know that before the insertion we had
  \[
    \deficit(v) \quad\leq\quad \frac{\height(\Psi_v) \cdot \numleaves{\Psi_v} -
    \fbalance(\Psi_v)}{20 \cdot \height(\Psi_v)}\;.
  \]
  After the insertion we have that $\numleaves{\Psi_v}$ is increased by 1 and
  $\fbalance(\Psi_v)$ is at most worsened by $\height(\Psi_v)$, as the insertion
  can only worsen $|\balance(w)|$ by at most one for each ancestor $w$ of the
  inserted node. We remind that all rotations performed by the insertion
  algorithm can only improve $\fbalance(\Psi_v)$. Altogether, the equation of
  the invariant still holds true, as $\deficit(v)$ shrinks by one while the
  right side of the equation in the invariant cannot shrink. The value
  $\fbalance(\Psi_v)$ can be worsened by at most $\height(\Psi_v)$, which is
  absorbed by the increment of $\numleaves{\Psi_v}$.
\end{proof}

\subsection*{Wrap-Up}
Altogether we have shown that given a tree $T$, we can compute a formula $\Psi$
that represents $T$ in linear time. This formula will be of logarithmic height.
Especially we have shown that this formula satisfies the deletion invariant and
that after any update to $T$, we can update $\Psi$ in logarithmic time to match
the new tree in such a way that the deletion invariant is maintained.
Theorem~\ref{thm:update} follows by the combination of
Theorem~\ref{thm:construct}, Lemma~\ref{lem:insert-runtime},
Corollary~\ref{cor:delete-runtime}, Lemma~\ref{lem:delete}, and
Lemma~\ref{lem:insert-deletionInvariant}.

\section{Stepwise Automata and Transition Algebras}\label{sec:automata}
In this section we introduce the tree automata that we use to represent MSO
queries. Furthermore, we also introduce their transition algebras which are
forest algebras that capture the whole behavior of a tree automaton on some
forest or context. Transition algebras are related to tree automata in the same
way as transition monoids are related to string automata.

\subsection*{Stepwise Tree Automata}
Stepwise tree automata were first described in~\cite{CarmeNT-rta04}
using a curry encoding of unranked trees. For convenience, we use the
definition from~\cite{MartensN-jcss07a} that directly works on
unranked trees.

A \emph{stepwise nondeterministic tree automaton} or \emph{\ta} is a tuple $N =
(Q, \Sigma, \delta, \init, q_I, q_F)$ where $Q$ is the finite set of states,
$\Sigma$ is a finite alphabet, $q_I$ and $q_F$ are the global initial and final
states, $\init \colon \Sigma \to 2^Q$ assigns a local set of
initial states to every symbol of $\Sigma$, and $\delta \subseteq Q \times Q
\times Q$ is a transition relation. We use $\transition{q_1}{q_2}{q_3}$ to
denote a transition that goes from $q_1$ to $q_3$ when reading state $q_2$.

Intuitively, a stepwise tree automaton computes a run bottom-up. After assigning
states $q_1,\dots,q_n$ to the $n$ children of some node $v$, it assigns a state
to $v$, by starting in some initial state (determined by the label of $v$) and
reading the string $q_1 \dots q_n$, i.e, it reads the states of all children.
The resulting state is assigned to $v$. Whether a run is accepting is determined
by the state assigned to the root.

This informal description of runs of a stepwise tree automaton actually
describes two different types of runs: a vertical run that assigns states to
nodes, and, for each node, a horizontal run along the children of a node. The
horizontal run corresponds exactly to the run of a string automaton over the
input alphabet $Q$. We place the state $q_2$ under
the arrow in the notation $\transition{q_1}{q_2}{q_3}$ because the
states $q_1$ and $q_3$ are part of the horizontal run of some node $v$ while the
state $q_2$ is the state of a child of $v$ in the vertical run. For our formal
description, we combine the vertical run and all horizontal runs into a
single run as follows.

Formally, a \emph{run} $\lambda$ of $N$ on a labeled tree $T$ is an assignment
of transitions to states $\lambda \colon \Dom(T) \rightarrow \delta$, where we
write $\lambda_\pre(v)$, $\lambda_\self(v)$, and $\lambda_\post(v)$ to denote
the individual components of the transition
$\lambda(v)=\transition{q_\pre}{q_\self}{q_\post}$. A run has to satisfy the
following conditions for every node $v$:
  \allowdisplaybreaks
\begin{align*}
  \lambda_\pre(v) \quad&=\quad q_I & \text{if $v$ is the root}\\
  \lambda_\pre(v) \quad&\in\quad \init(\lab(\parent(v))) & \text{if $v$ has no left sibling}\\
  \lambda_\pre(v) \quad&=\quad \lambda_\post(w) & \text{if $w$ is the left sibling of $v$}\\
  \lambda_\self(v) \quad&\in\quad \init(\lab(v)) & \text{if $v$ is a leaf}\\
  \lambda_\self(v) \quad&=\quad \lambda_\post(w) & \text{if $w$ is the last child of $v$}
\end{align*}

We note that our definition of a run is equivalent to the standard definition of
horizontal and vertical runs. The usual definition of a vertical run corresponds
to $\lambda_\self$, while the list of transitions assigned to the children of a
node $v$ corresponds to the list of transitions of the horizontal run as
defined in the literature. 

The rules above establish that $\lambda_\self(v)$ is computed, just as a
(nondeterministic) string automaton computes the final state when starting in
some initial state from $\init(\lab(v))$ and reading the string of states
$\lambda_\self(v_1) \cdots \lambda_\self(v_n)$ of the children $v_1,\dots,v_n$
of $v$.

A run is \emph{accepting} if $\lambda_\post(r)=q_F$, where $r$ is the root of $T$. A
tree $T$ is \emph{accepted} if there exists an accepting run on $T$. The set of
all accepted trees is denoted by $L(N)$. By $\states(\lambda)$ we denote
the image of $\lambda_\self$.

We note that our notation of acceptance is equivalent to the usual notation
where a tree is accepted if $\lambda_\self(\Root(T)) \in F$ for some set of
final states $F$. One simply has to add the transitions $\{\transition{q_I}{q}{q_F} \mid q
\in F\}$ to $\delta$. We prefer our acceptance model, as it will greatly simplify the
definition of transition algebras. Also this mode of acceptance is much
closer to the model of string automata and can be easily generalized to forest
languages.

\begin{exa}
  In Figure~\ref{fig:nfta-run}, we depict an NFTA that checks whether the number
  of $a$-nodes in the tree is equivalent to 0 modulo 3. The states $q_0$, $q_1$,
  and $q_2$ encode the number of $a$-nodes modulo 3. The initial state for
  $a$-nodes is $q_1$ to count the $a$-node itself even before starting the
  horizontal run. The initial states for all other symbols is $q_0$. Transitions between
  states encode modulo arithmetic and the transition $\transition{q_I}{q_0}{q_F}$
  encodes the acceptance condition. We also depict a tree $T$
  together with an accepting run $\lambda$. We explain the horizontal run of the
  root node. The run starts in $q_1$
  to count the $a$-label of the root. It then adds $2$ more $a$-nodes from the
  first subtree, 0 $a$-nodes for the $c$-node, the $a$-node, and finally again
  $2$ $a$-nodes from the last subtree. It reaches state $q_0$ as there are $6$
  $a$-nodes in total. On the right, we depict a decomposition $T=C
  \odott F$ of $T$ into a context $C$ and a forest $F$ together with a
  decomposition of $\lambda$ into (partial) runs for $C$ and $F$ as defined
  below.
\end{exa}

\newcommand{\myarrow}[1]{{{}\xrightarrow[{\raisebox{.3mm}[0mm][0mm]{\scriptsize $#1$}}]{}{}}}
\newcommand{\mytransition}[3]{{\scriptsize \mathllap{#1}
    \myarrow{#2} \mathrlap{#3}}}
\newcommand{\phantomtransition}{\phantom{$\mytransition{q_2}{q_2}{q_2}$}}

\begin{figure}
	\centering
  \begin{tikzpicture}[>=latex,level distance=15mm, sibling distance=9.2mm]
    \tikzstyle{every state}=[minimum size=7mm, inner sep=0pt]
    \scriptsize
		\node[state]	(q0) at (1,-2.23)	{$q_2$};
    \tikzset{initial text={$b,c$}}
		\node[state,initial]	(q1) at (0,-0.5)	{$q_0$};
    \tikzset{initial text={$\vphantom{b}a$}}
		\node[state,initial right]	(q2) at (2,-0.5)	{$q_1$};
    \tikzset{initial text={}}
		\node[state,initial]	(qI) at (0.1,-3.2)	{$q_I$};
		\node[state,accepting]	(qF) at (1.9,-3.2)	{$q_F$};

    \path[->] (q1) edge[out=30,in=150] node [above] {$q_1$} (q2)
              (q2) edge[out=180,in=0] node [below] {$q_2$} (q1)
              (q0) edge[out=140,in=280] node [left] {$q_1$} (q1)
              (q1) edge node [right] {$q_2$} (q0)
              (q0) edge node [left] {$q_2$} (q2)
              (q2) edge[out=-90,in=30] node [right] {$q_1$} (q0)
              (qI) edge node [above] {$q_0$} (qF)
              (q1) edge [loop above] node[above] {$q_0$} (q1)
              (q2) edge [loop above] node[above] {$q_0$} (q2)
              (q0) edge [loop right] node[right] {$q_0$} (q0);

    \node[align=center] at (5.2,0) {$\mytransition{q_I}{q_0}{q_F}$\\$\vphantom{b}a$}
    child { node[align=center] {{\scriptsize $q_1 \myarrow{q_2} q_0 \mathrlap{{} \myarrow{q_0} q_0 \myarrow{q_1} q_1 \myarrow{q_2} q_0 }$}\\$\vphantom{b}a$}
      child { node[align=center] {{\scriptsize $q_1 \myarrow{q_1} q_2 \mathrlap{{} \myarrow{q_0} q_2 }$}\\$\vphantom{b}a$} }
      child { node[align=center] {\phantomtransition\\$b$} }
}
    child { node[align=center] {\phantomtransition\\$\vphantom{b}c$} }
    child { node[align=center] {\phantomtransition\\$\vphantom{b}a$} }
    child { node[align=center] {\phantomtransition\\$b$} 
      child { node[align=center] {{\scriptsize $q_0 \myarrow{q_1} q_1 \mathrlap{{} \myarrow{q_1} q_2 }$}\\$\vphantom{b}a$} }
      child { node[align=center] {\phantomtransition\\$\vphantom{b}a$} } 
    };

    \node[align=center] at (9,0) {$\mytransition{q_I}{q_0}{q_F}$\\$\vphantom{b}a$}
    child { node[align=center] {{\scriptsize $q_1 \myarrow{q_\hole} q_1 \mathrlap{{} \myarrow{q_2} q_0}$}\\$\vphantom{b}\hole$}
    }
    child { node[align=center] {\phantomtransition\\$b$} 
      child { node[align=center] {{\scriptsize $q_0 \myarrow{q_1} q_1 \mathrlap{{} \myarrow{q_1} q_2 }$}\\$\vphantom{b}a$} }
      child { node[align=center] {\phantomtransition\\$\vphantom{b}a$} } 
    };

    \node[align=center] at (11.5,0) {{\scriptsize $q_1 \myarrow{q_2} q_0 \mathrlap{{} \myarrow{q_0} q_0 \myarrow{q_1} q_1}$}\\$\vphantom{b}a$}
      child { node[align=center] {{\scriptsize $q_1 \myarrow{q_1} q_2 \mathrlap{{} \myarrow{q_0} q_2 }$}\\$\vphantom{b}a$} }
      child { node[align=center] {\phantomtransition\\$b$} }
    ;
    \node[align=center] at (12.42,0) {\phantomtransition\\$\vphantom{b}c$};
    \node[align=center] at (13.34,0) {\phantomtransition\\$\vphantom{b}a$};
  \end{tikzpicture}

\caption{NFTA (left), tree with accepting run (middle), context and forest
    with runs (right)\label{fig:nfta-run}}
\end{figure}
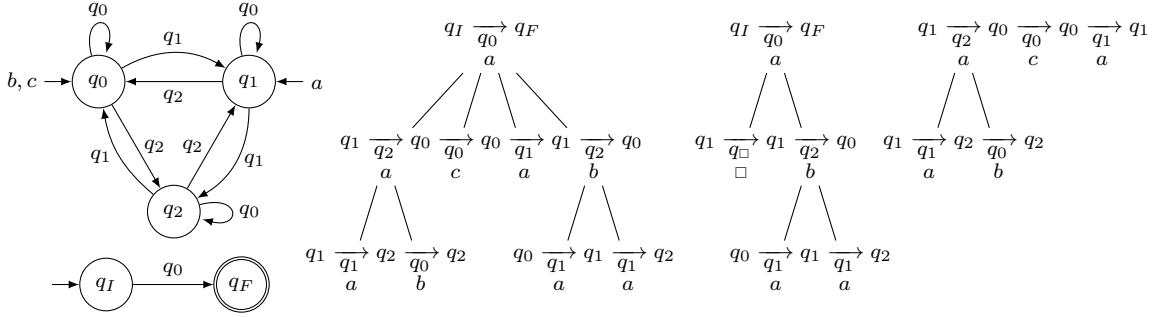

We want to be able to compose runs, just as we compose trees from forests and
contexts. Therefore, we define runs on forests and contexts as follows: A run of
$N$ on a forest $F$ is defined just as a run of $N$ on a tree, with the
following difference: $\lambda_\pre(v)$ is allowed to be any state, where $v$ is
the leftmost root, i.e. the unique root of $F$ that does not have a left
sibling. This modification is necessary, as there is no way of knowing how the
context where $F$ will be inserted in looks like.

To define runs on contexts, we add a new state $q_\hole$, define
$\init(\hole)=q_\hole$ and add transitions $(Q \setminus \{q_\hole\}) \times
\{q_\hole\} \times (Q \setminus \{q_\hole\})$ to $\delta$. Now runs on contexts
are defined just like runs on forests using the extended automaton. This has the
effect that a run on a context does not make any assumptions on what happens
inside the forest that will be inserted into the hole to form a tree or
forest.\footnote{Of course, we could restrict the additional transitions to only
  those that are possible by some forest, but it does not make any difference
  for asymptotic worst case complexity and therefore we stick with this simpler
  construction.} We note that accepting runs are only defined for trees, not for
forests or contexts.

\subsection*{Selecting Automata}
In order to evaluate non-Boolean queries, we utilize (node- and tuple\nobreakdash-)selecting
finite tree automata (see, e.g., \cite{FrickGK-lics03,neventhesis}) as formalism
for queries. It is well-known that these can express MSO queries with free node
variables over unranked trees~\cite[Theorem 7]{NiehrenPTT-dbpl05}.

For $k \in \nat$, a
\emph{$k$-ary non-deterministic finite selecting tree automaton ($k$-\sta)}
$M$ is a pair $(N,S)$, where $N$ is a \ta over $\Sigma$ with
states $Q$ and $S \subseteq Q^k$ is a set of \emph{selecting
  tuples}. The \emph{size} of $M$ is defined as $|Q| + |S|$.  When $M$ reads a tree $T$, it computes a set of tuples in $\nodes(T)^k$. More precisely,
we define 
\begin{multline*}
  M(T) =  \big\{\;(v_1,\ldots,v_k) \mid 
  \text{ there is an accepting run $\lambda$ of $N$ on $T$} 
  \text{ and a tuple} \\ (p_1,\ldots,p_k) \in S  \text{ such that } 
  \lambda_\self(v_\ell)=p_\ell\text{ for } \ell \in \{1,\dots,k\}\;\big\}\;.
\end{multline*}
Notice that, if $T \notin L(N)$ then $M(T) = \emptyset$.

\begin{figure}[t]
  \center
  \begin{tikzpicture}[>=latex,level distance=13mm, sibling distance=8.7mm]
    \tikzstyle{every state}=[minimum size=7mm, inner sep=0pt]
    \tikzstyle{selecting}=[state,fill=gray!30,minimum size=7mm, inner sep=0pt]
    \tikzset{initial text={$a,b$}}
\node[state,initial] (q1) at (-5,0) {$q_1$};
    \node[state] (q2) at (-3,0) {$q_2$};

    \tikzset{initial text=$a$}
    \node[state, initial,selecting] (q3) at (-5,-2) {$q_3$};

    \node[state, initial] (q4) at (0,0) {$q_4$};
    \node[state, selecting] (q5) at (2,0) {$q_5$};    

    \tikzset{initial text=$b$}

    \node[state, initial] (q6) at (0,-2) {$q_6$};
    \node[state] (q7) at (2,-2) {$q_7$};    

    \tikzset{initial text=}
    \node[state,initial] (q0) at (-5,-3.5) {$q_I$};
    \node[state,accepting] (qF) at (-3,-3.5) {$q_F$};
    
\path[->] (q1) edge [loop above] node [above] {$q_1$} (q1)
              (q2) edge [loop above] node [above] {$q_1$} (q2)
              (q3) edge [loop above] node [above] {$q_1$} (q3)
              (q4) edge [loop above] node [above] {$q_1$} (q4)
              (q5) edge [loop above] node [above] {$q_1$} (q5)
              (q6) edge [loop above] node [above] {$q_1$} (q6)
              (q7) edge [loop above] node [above] {$q_1$} (q7)
              (q1) edge node [above] {$q_2,q_5$} (q2)
              (q4) edge node [above] {$q_7$} (q5)
              (q6) edge node [above] {$q_3,q_7$} (q7)
              (q0) edge node [above] {$q_2,q_3$} (qF);

    \node[align=center] at (5,1) (b) {$\mytransition{q_I}{q_2}{q_F}$\\ $b$}
    child {
      node[align=center] (a1) {{\scriptsize $q_1 \myarrow{q_5} q_2 \mathrlap{\myarrow{q_1}q_2}$}\\ $\vphantom{b}a$ }
      child {
        node[align=center] (b1)  {$\mytransition{q_4}{q_7}{q_5}$\\ $b$}
        child {
          node[align=center] (b2) {$\mytransition{q_6}{q_7}{q_7}$\\ $b$}
          child {
            node[align=center] (a2) {{\scriptsize $q_6\myarrow{q_3}q_7\mathrlap{\myarrow{q_1}q_7}$} \\ ${a}$}
          }
          child {
            node[align=center] (a3) {\phantom{$\mytransition{q_4}{q_7}{q_5}$}\\${a}$}
          }
        }
      }
    }
    child {
      node[align=center] (b3) {\phantom{$\mytransition{q_4}{q_7}{q_5}$} \\[.5mm] $b$}
    };

  \end{tikzpicture}
  \caption{$2$-NFSTA $M$ with $S = \{(q_3,q_5)\}$ (left) and tree with accepting
    run $\lambda$ (right). The transition assigned to some node by $\lambda$ is
    denoted above its label. }
  \label{fig:example1}
\end{figure}
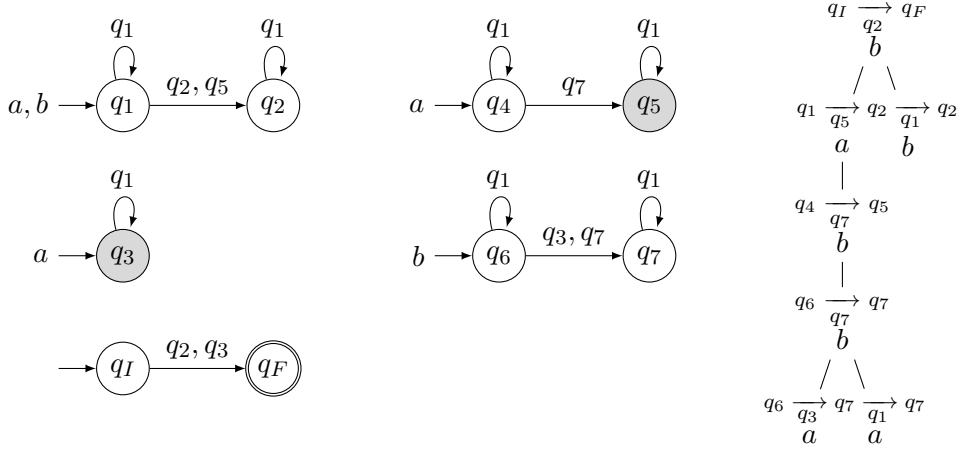
 
\begin{exa}
  Figure~\ref{fig:example1} illustrates a $2$-\sta $M$ over $\Sigma=\{a,b\}$
  that outputs each pair of $a$-labeled nodes that are connected by a path of
  $b$-labeled nodes. The states have the intended meaning as follows: $q_1$ and
  $q_2$ are states not belonging to the path, where $q_1$ is for nodes that do
  not have the path below them and $q_2$ is for nodes that have the path below
  them. Therefore, every state (except $q_I,q_F$) has a $q_1$-loop to ignore
  parts of the tree that are unrelated to the path. The states $q_3$ and $q_5$
  mark the lower end upper end of the path, respectively, while $q_7$ marks
  $b$-nodes on the path. The states $q_4$ and $q_6$ cannot appear as states of
  any node, as otherwise the run cannot be completed to an accepting run. They
  are guesses as initial states for the upper $a$-node of the path and $b$-nodes
  on the path. However, for an accepting run, there has to be a child with state
  $q_3$ or $q_7$ below. The only selecting tuple is $(q_3,q_5)$. We accept if
  the root has state $q_2$ or $q_3$, thus either it is some node with the path
  below it, or the top $a$-node of the path.

  Next to the automaton we depict a tree with an accepting runs that returns
  the $a$-node below the root and the left $a$-leaf. A symmetric run that assigns $q_3$ to the right $a$-leaf returns the
  pair of nodes consisting of the $a$-node below the root and the right $a$-leaf.
\end{exa}

\subsection*{Signatures of Runs}
The \emph{signature} of a run over a forest or context captures the states
before the first and after the last root. Furthermore, in case of a context, the
signature additionally captures the states before and after the hole. Let $v_1
\leq \dots \leq v_k$ be the list of roots. Formally, we define the signature of
a run $\lambda$ as follows:
\[
  \signature(\lambda) \quad=\quad \begin{cases}
    \big(\lambda_\pre(v_1),\lambda_\post(v_k)\big) & \text{if $\lambda$ is over a
      forest} \\
  \Big(\big(\lambda_\pre(v_1),\lambda_\post(v_k)\big),\big(\lambda_\pre(\hole),\lambda_\post(\hole)\big)\Big)
  & \text{if $\lambda$ is over a context}
  \end{cases}
\]
We denote the set of all possible signatures over the states $Q$ with
$\signatures=Q^2 \cup (Q^2)^2$.

\subsection*{Transition Algebra}
The \emph{transition algebra} of a stepwise tree automaton is the generalization
of the transition monoid of a finite string automaton. Indeed the horizontal
monoid is defined exactly like the transition monoid of a finite string
automaton over the alphabet $Q$. Each element of the transition algebra captures
the signatures of all possible runs over the underlying forest or context.

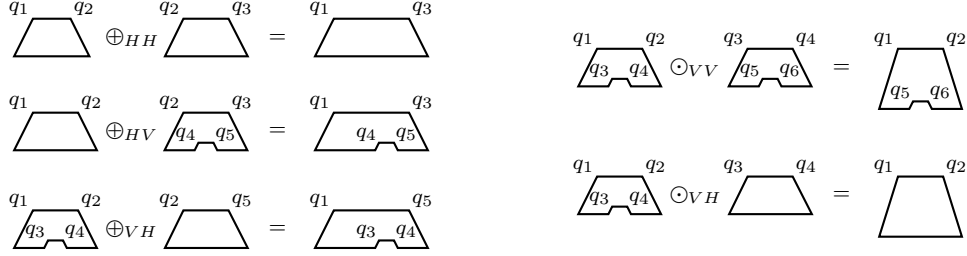
\begin{figure}
  \begin{tikzpicture}
    \tikzstyle{sl}=[inner sep=1pt]
    \tikzstyle{sm}=[inner sep=0.5pt]
    \scriptsize
\begin{scope}
\draw[thick]  (0,0) coordinate (x1l) -- ++(5mm,0mm) coordinate (x1r) -- ++(2.5mm,-5mm) -- ++(-10mm,0mm) -- cycle;

\draw[thick]  (2,0) coordinate (x2l) -- ++(6mm,0mm) coordinate (x2r) -- ++(2.55mm,-5mm) -- ++(-11mm,0mm) -- cycle;

\draw[thick]  (4,0) coordinate (x3l) -- ++(10mm,0mm) coordinate (x3r) -- ++(2.5mm,-5mm) -- ++(-15mm,0mm) -- cycle;
      
      \node at (1.1,-0.25) {$\concat_{\mathrlap{HH}}$};
      \node at (3.25,-0.25) {$=$};

      \node[sl,anchor=south east] at (x1l) {$q_1$};
      \node[sl,anchor=south west] at (x1r) {$q_2$};
      \node[sl,anchor=south east] at (x2l) {$q_2$};
      \node[sl,anchor=south west] at (x2r) {$q_3$};
      \node[sl,anchor=south east] at (x3l) {$q_1$};
      \node[sl,anchor=south west] at (x3r) {$q_3$};
    \end{scope}

\begin{scope}[shift={(0,-1.25)}]
\draw[thick]  (0,0) coordinate (x1l) -- ++(6mm,0mm) coordinate (x1r) -- ++(2.5mm,-5mm) -- ++(-11mm,0mm) -- cycle;

\draw[thick]  (2,0) coordinate (x2l) -- ++(6mm,0mm) coordinate (x2r) -- ++(2.5mm,-5mm) -- ++(-4mm,0mm) -- ++(-.5mm,1mm) coordinate (x2br) -- ++(-2mm,0mm) coordinate (x2bl) -- ++(-.5mm,-1mm) -- ++(-4mm,0mm) -- cycle;

\draw[thick]  (4,0) coordinate (x3l) -- ++(10mm,0mm) coordinate (x3r) -- ++(2.5mm,-5mm) -- ++(-4mm,0mm) -- ++(-.5mm,1mm) coordinate (x3br) -- ++(-2mm,0mm) coordinate (x3bl) -- ++(-.5mm,-1mm) -- ++(-8mm,0mm) -- cycle;

      \node at (1.1,-0.25) {$\concat_{\mathrlap{HV}}$};
      \node at (3.25,-0.25) {$=$};

      \node[sl,anchor=south east] at (x1l) {$q_1$};
      \node[sl,anchor=south west] at (x1r) {$q_2$};
      \node[sl,anchor=south east] at (x2l) {$q_2$};
      \node[sl,anchor=south west] at (x2r) {$q_3$};
      \node[sm,anchor=south east] at (x2bl) {$q_4$};
      \node[sm,anchor=south west] at (x2br) {$q_5$};

      \node[sl,anchor=south east] at (x3l) {$q_1$};
      \node[sl,anchor=south west] at (x3r) {$q_3$};
      \node[sm,anchor=south east] at (x3bl) {$q_4$};
      \node[sm,anchor=south west] at (x3br) {$q_5$};
    \end{scope}
    
\begin{scope}[shift={(0,-2.55)}]
\draw[thick]  (0,0) coordinate (x1l) -- ++(6mm,0mm) coordinate (x1r) -- ++(2.5mm,-5mm) -- ++(-4mm,0mm) -- ++(-.5mm,1mm) coordinate (x1br) -- ++(-2mm,0mm) coordinate (x1bl) -- ++(-.5mm,-1mm) -- ++(-4mm,0mm) -- cycle;

\draw[thick]  (2,0) coordinate (x2l) -- ++(6mm,0mm) coordinate (x2r) -- ++(2.55mm,-5mm) -- ++(-11mm,0mm) -- cycle;

\draw[thick]  (4,0) coordinate (x3l) -- ++(10mm,0mm) coordinate (x3r) -- ++(2.5mm,-5mm) -- ++(-4mm,0mm) -- ++(-.5mm,1mm) coordinate (x3br) -- ++(-2mm,0mm) coordinate (x3bl) -- ++(-.5mm,-1mm) -- ++(-8mm,0mm) -- cycle;

      \node at (1.1,-0.25) {$\concat_{\mathrlap{VH}}$};
      \node at (3.25,-0.25) {$=$};

      \node[sl,anchor=south east] at (x1l) {$q_1$};
      \node[sl,anchor=south west] at (x1r) {$q_2$};
      \node[sm,anchor=south east] at (x1bl) {$q_3$};
      \node[sm,anchor=south west] at (x1br) {$q_4$};

      \node[sl,anchor=south east] at (x2l) {$q_2$};
      \node[sl,anchor=south west] at (x2r) {$q_5$};

      \node[sl,anchor=south east] at (x3l) {$q_1$};
      \node[sl,anchor=south west] at (x3r) {$q_5$};
      \node[sm,anchor=south east] at (x3bl) {$q_3$};
      \node[sm,anchor=south west] at (x3br) {$q_4$};
    \end{scope}

\begin{scope}[shift={(7.5,-.4)}]
\draw[thick]  (0,0) coordinate (x1l) -- ++(6mm,0mm) coordinate (x1r) -- ++(2.5mm,-5mm) -- ++(-4mm,0mm) -- ++(-.5mm,1mm) coordinate (x1br) -- ++(-2mm,0mm) coordinate (x1bl) -- ++(-.5mm,-1mm) -- ++(-4mm,0mm) -- cycle;

\draw[thick]  (2,0) coordinate (x2l) -- ++(6mm,0mm) coordinate (x2r) -- ++(2.5mm,-5mm) -- ++(-4mm,0mm) -- ++(-.5mm,1mm) coordinate (x2br) -- ++(-2mm,0mm) coordinate (x2bl) -- ++(-.5mm,-1mm) -- ++(-4mm,0mm) -- cycle;

\draw[thick]  (4,0) coordinate (x3l) -- ++(6mm,0mm) coordinate (x3r) -- ++(2.5mm,-8mm) -- ++(-4mm,0mm) -- ++(-.5mm,1mm) coordinate (x3br) -- ++(-2mm,0mm) coordinate (x3bl) -- ++(-.5mm,-1mm) -- ++(-4mm,0mm)  -- cycle;

      \node at (1.1,-0.25) {$\conapp_{\mathrlap{VV}}$};
      \node at (3.25,-0.25) {$=$};

      \node[sl,anchor=south east] at (x1l) {$q_1$};
      \node[sl,anchor=south west] at (x1r) {$q_2$};
      \node[sm,anchor=south east] at (x1bl) {$q_3$};
      \node[sm,anchor=south west] at (x1br) {$q_4$};

      \node[sl,anchor=south east] at (x2l) {$q_3$};
      \node[sl,anchor=south west] at (x2r) {$q_4$};
      \node[sm,anchor=south east] at (x2bl) {$q_5$};
      \node[sm,anchor=south west] at (x2br) {$q_6$};

      \node[sl,anchor=south east] at (x3l) {$q_1$};
      \node[sl,anchor=south west] at (x3r) {$q_2$};
      \node[sm,anchor=south east] at (x3bl) {$q_5$};
      \node[sm,anchor=south west] at (x3br) {$q_6$};
    \end{scope}
    
\begin{scope}[shift={(7.5,-2.1)}]
\draw[thick]  (0,0) coordinate (x1l) -- ++(6mm,0mm) coordinate (x1r) -- ++(2.5mm,-5mm) -- ++(-4mm,0mm) -- ++(-.5mm,1mm) coordinate (x1br) -- ++(-2mm,0mm) coordinate (x1bl) -- ++(-.5mm,-1mm) -- ++(-4mm,0mm) -- cycle;

\draw[thick]  (2,0) coordinate (x2l) -- ++(6mm,0mm) coordinate (x2r) -- ++(2.55mm,-5mm) -- ++(-11mm,0mm) -- cycle;

\draw[thick]  (4,0) coordinate (x3l) -- ++(6mm,0mm) coordinate (x3r) -- ++(2.5mm,-8mm) -- ++(-11mm,0mm) -- cycle;

      \node at (1.1,-0.25) {$\conapp_{\mathrlap{VH}}$};
      \node at (3.25,-0.25) {$=$};

      \node[sl,anchor=south east] at (x1l) {$q_1$};
      \node[sl,anchor=south west] at (x1r) {$q_2$};
      \node[sm,anchor=south east] at (x1bl) {$q_3$};
      \node[sm,anchor=south west] at (x1br) {$q_4$};

      \node[sl,anchor=south east] at (x2l) {$q_3$};
      \node[sl,anchor=south west] at (x2r) {$q_4$};

      \node[sl,anchor=south east] at (x3l) {$q_1$};
      \node[sl,anchor=south west] at (x3r) {$q_2$};
    \end{scope}

    \scriptsize

  \end{tikzpicture}
  \caption{Graphical representation of the operations of the transition algebra.
  Trapezoids represent forests while trapezoids with a cutout trapezoid
  represent contexts.}\label{fig:transitionAlgebra}
\end{figure}

Formally, the \emph{transition algebra} of a given stepwise tree automaton
$N=(\Sigma,Q,\delta,\init,F)$ is defined as \[\tra_N=(H, V, \concat_{HH},
  \concat_{HV}, \concat_{VH}, \conapp_{VV}, \conapp_{VH}, \id_Q, \id_{Q^2})\;,\]
where $H=2^{Q^2}$ and $V=2^{(Q^2)^2}$ are the powersets of the possible
signatures of forests and contexts, respectively. The neutral elements $\id_Q$
and $\id_{Q^2}$ are the identity relations over $Q$ and $Q^2$, respectively.
That is  $\id_Q=\{(q,q) \mid q \in Q\}$
and $\id_{Q^2}=\{((q_1,q_2),(q_1,q_2)) \mid (q_1,q_2) \in Q^2\}$. All operations
are relational joins over those states that need to be identical in order for
the runs to be combined, followed by a projection onto the states relevant for
the signature of the combined run. The intuition of the operations are depicted
in Figure~\ref{fig:transitionAlgebra}, where we depict how signatures can be
combined by the five operations. While in the figure we sketch the operations
for a pair of signatures, in the algebra we have to join all compatible pairs of
signatures. Formally, the operations are given by the following equations, where
$q_1,\dots,q_6$ are states of the automaton:
\begin{align*}
  F_1 \concat_{HH} F_2 &\;\; = \;\; \big\{ (q_1,q_3) \mid (q_1,q_2) \in F_1, (q_2,q_3) \in F_2\big\}\\
  C_1 \conapp_{VV} C_2 &\;\; = \;\; \big\{ \big((q_1,q_2),(q_5,q_6)\big) \mid \big((q_1,q_2),(q_3,q_4)\big) \in C_1, \big((q_3,q_4),(q_5,q_6)\big) \in C_2\big\}\\
  C \conapp_{VH} F &\;\; = \;\; \big\{(q_1,q_2) \mid \big((q_1,q_2),(q_3,q_4)\big) \in C, (q_3,q_4) \in F \big\}  \\
  F \concat_{HV} C &\;\; = \;\; \big\{\big((q_1,q_3),(q_4,q_5)\big) \mid (q_1,q_2) \in F, ((q_2,q_3),(q_4,q_5)) \in C \big\}  \\
  C \concat_{VH} F &\;\; = \;\; \big\{\big((q_1,q_5),(q_3,q_4)\big) \mid \big((q_1,q_2),(q_3,q_4)\big) \in C, (q_2,q_5) \in F \big\} 
\end{align*}

\begin{lem}
  For every NFTA $N$, the corresponding transition algebra $\tra_N$ is a forest
  algebra.
\end{lem}
\begin{proof}[Proof sketch]
  We have to show that all axioms from Table~\ref{tab:axioms} hold. The axioms
  A1--A4 (neutral elements) hold, as we do joins with identity relations. All
  other axioms hold because of the associativity of the relational join.
\end{proof}

As all operations are defined by the means of the relational join, we have the
following upper bound on their complexity using the trivial join algorithm.
\begin{obs}\label{obs:transitionAlgebra}
  Given a transition algebra $\tra$ over a set of states $Q$, all operations in
  $\tra$ can be performed in at most $\cO(|Q|^6)$ time.
\end{obs}

The morphism from the free forest algebra $\FA_\Sigma$ over $\Sigma$ to the
transition algebra $\tra_N$ of $N$ is defined by
\[
  h(\context a) \;\; = \;\; \big\{ \signature(\lambda) \mid \lambda \text{ is a valid
    run of $N$ on } \context a \}
\]
for all symbols $a \in \Sigma$. While we define $h$ by specifying the mappings
for the atomic contexts, the intuition is that the homomorphism maps every
forest or context to the set of signatures for all possible runs. This intuition
is formalized in the following lemma.
\begin{lem}\label{lem:ta}
  For every forest or context $D$ it holds that 
  \[ h(D) \;\; = \;\; \big\{ \signature(\lambda) \mid \lambda \text{ is a valid
    run of $N$ on } D \}\;. \]
\end{lem}
\begin{proof}
  We show both directions by induction. We start with the if-direction. The
  statement holds for $D=\context a$ by definition. For $D=a$, let $\lambda$ be
  some run of $N$ over $D$ and $(q_1,q_2) = \signature(\lambda)$. By definition
  of signatures and runs, there has to be some state $q_3 \in \init(a)$, such
  that $\transition{q_1}{q_3}{q_2}=\lambda(v)$ where $v$ is the single $a$-node in $D$. As $\transition{q_1}{q_3}{q_2}$
  and $\transition{q_3}{q_\hole}{q_3}$ are both transitions in $\delta$, we
  can conclude that $\big((q_1,q_2),(q_3,q_3)\big) \in h(\context a)$ and thus
  $(q_1,q_2) \in h(a) = h(\context a \conapp_{VH} \varepsilon)$.

  We showcase the induction step for the operation $\conapp_{VH}$, i.e., for $D=
  C \conapp_{VH} F$. We have to show that for every run $\lambda$ on $C
  \odott F$ with signature $(q_1,q_2)$, it holds that $(q_1,q_2)\in h(C)
  \conapp_{VH} h(F)$. By definition of $\conapp_{VH}$, this boils down to
  showing that there are states $q_3$ and $q_4$ such that
  $\big((q_1,q_2),(q_3,q_4)\big)\in h(C)$ and $(q_3,q_4) \in h(F)$. By the
  induction assumption, this is the case if there are runs $\lambda_C$ on $C$
  and $\lambda_F$ on $F$ with signatures
  $\signature(\lambda_C)=\big((q_1,q_2),(q_3,q_4)\big)$ and
  $\signature(\lambda_F)=(q_3,q_4)$. We now show that there are such runs
  $\lambda_F$ and $\lambda_C$. We get $\lambda_F$ by restricting $\lambda$ to
  the nodes in $F$. We define $q_3 = \lambda_\pre(v_1)$ and
  $q_4=\lambda_\post(v_k)$, where $v_1,\dots,v_k$ are the roots of $F$. By
  definitions of signatures we have that $\signature(\lambda_F)=(q_3,q_4)$ as
  desired. Likewise, we get $\lambda_C$ by restricting $\lambda$ to the nodes of
  $C$ and using the transition $\transition{q_3}{q_\hole}{q_4}$ for the hole.
  The signature of $\lambda_C$ is indeed $\big((q_1,q_2),(q_3,q_4)\big)$ as
  desired. The induction step for the other four operations works analogously,
  where the existing run is split into two runs as indicated in
  Figure~\ref{fig:transitionAlgebra}. This concludes the proof of the
  if-direction.

  We continue with the only-if-direction. The case $D=\context a$ is again
  trivial, we continue with $D=a$. From $(q_1,q_2) \in h(a)$, we can conclude
  that there has to be a state $q_3$ such that $((q_1,q_2),(q_3,q_3)) \in
  h(\context a)$, as $h(a)=h(\context a) \conapp_{VH} \id_Q$. Therefore $q_3 \in
  \init(a)$. We can conclude that the run $\lambda$ that assigns
  $\transition{q_1}{q_3}{q_2}$ to the $a$-node is a valid run with
  $\signature(\lambda)=(q_1,q_2)$.

  Again, we showcase the induction step for $D=C \conapp_{VH} F$. We have
  to show that for every tuple $(q_1,q_2) \in h(C \conapp_{VH} F)$ there is a
  run $\lambda$ on $C \conapp_{VH} F$ such that $\signature(\lambda)=(q_1,q_2)$.
  By definition of $\conapp_{VH}$, there have to exist states $q_3$ and $q_4$ such
  that $\big((q_1,q_2),(q_3,q_4)\big) \in h(C)$ and $(q_3,q_4) \in h(F)$. By the
  induction assumption, there are runs $\lambda_C$ on $C$ with signature
  $\big((q_1,q_2),(q_3,q_4)\big)$ and $\lambda_F$ on $F$ with signature
  $(q_3,q_4)$. By the definition of signatures, the run $\lambda_C$ uses the
  transition $\transition{q_3}{q_\hole}{q_4}$ for the hole. The run $\lambda$ is
  then derived from $\lambda_C$ and $\lambda_F$ by taking the disjoint union
  $\lambda_C \cup \lambda_F$ of both runs and omitting the mapping for the hole.
  It is easy to verify that the result is a valid run for $C \conapp_{VH} F$.
  Again, the induction step for the other operations works analogously, where
  runs can be combined as indicated in Figure~\ref{fig:transitionAlgebra}. This
  concludes the proof.
\end{proof}

\subsection*{Evaluation and Enumeration Problems for NFTAs}
Let $M$ be a selecting automaton, $T$ the input tree for $M$, and
$M(T)$ be the answer of $M$ on $T$. We are interested in efficiently
maintaining $M(T)$ under updates of $T$. This means that we can have
an update $u$ to $T$, yielding another tree $T'$, and we wish to
efficiently compute $M(T')$. The latter cost should be more efficient
than computing $M(T')$ from scratch.

We allow a single preprocessing phase in which we can
compute an \emph{auxiliary data structure} Aux($T$) that we can use
for efficient query answering. When $T$ is updated to $T'$, we
therefore want to efficiently compute $M(T')$ and efficiently
update Aux($T$) to Aux($T'$).

If $M$ is simply an \ta (i.e., a $0$-ary \sta), then
this problem is known as \emph{incremental evaluation} and was studied
by, e.g., \cite{BalminPV-tods04}. Here, we perform
\emph{incremental enumeration}, meaning that we extend the setting of
Balmin et al.\ from $0$-ary queries to $k$-ary queries. We measure the
complexity of our algorithms in terms of the following parameters:
\emph{(i)} size of Aux($T$), \emph{(ii)} time needed to compute Aux($T$), \emph{(iii)} time
needed to update Aux($T$) to Aux($T'$), and \emph{(iv)} time
delay we can guarantee between answers of $M(T')$. The underlying
model of computation is a random access machine (RAM) with
uniform cost measure. 

In the remainder we use \inceval and \incenum to refer to
the incremental evaluation and enumeration problems, respectively.

\section{Incremental Evaluation}\label{sec:inceval}
In this section, we use the forest algebra framework of
Section~\ref{sec:maintaining} to prove the following theorem:
\begin{thm}\label{thm:inceval}
  \inceval for an NFTA $N=(\Sigma,Q,\delta,\init,F)$ and a tree $T$
  can be solved with a preprocessing phase of time
  $\cO(|Q|^6 \cdot |T|)$, auxiliary structure of size
  $\cO(|Q|^4 \cdot |T|)$, and with update time $\cO(|Q|^6\log(|T|))$
  after each new update.
\end{thm}
\begin{proof}
  We use the algorithms of Section~\ref{sec:maintaining} to compute and maintain
  a balanced representation $\Psi$ of the tree $T$. For each node $v$ of the
  parse tree of $\Psi$, we store the element $h(\Psi_v)$ of the transition
  algebra of $N$ that corresponds to the forest or context represented by
  $\Psi_v$. By Lemma~\ref{lem:ta}, the
  evaluation problem can be solved by looking whether $(q_I,q_F)$ is contained
  in the transition algebra element represented at the root of the parse tree.

  As a last step in the preprocessing, we can compute the algebra elements that
  we store at the nodes of $\Psi$ in a bottom-up way, computing one algebra
  operation at each inner node of $\Psi$. For the updates, we observe that after
  the update itself, we only need to update the algebra elements of the
  (logarithmically many) ancestors of the inserted/removed/relabeled node. In
  the case of deleting the last child of some node $v$, we also have to
  recompute the logarithmically many algebra elements of the ancestors of $v$ in
  the formula.
  
  Furthermore, for each performed rotation, we need to compute one algebra
  operation to compute the element represented at the node $w$ after the
  rotation (see Table~\ref{tab:rotations} for the definition of $w$).
  As the rotations are equivalence-preserving (Lemma~\ref{lem:rotations-sound}),
  the elements for all other nodes stay the same. Especially the element
  represented at $v$ does not change.

  We can therefore conclude all the run times from the theorem statement using
  Observation~\ref{obs:transitionAlgebra} and Theorem~\ref{thm:update}, as the
  number of rotations is clearly bounded by the runtime of the update algorithm.
\end{proof}

\section{Enumerating MSO Queries}\label{sec:enum}
This section extends the incremental evaluation result from the previous
section to enumeration of non-Boolean MSO queries. That is, we prove the
following theorem:
\begin{thm}\label{thm:enuminc}
  \enuminc for a $k$-NFSTA $M$ and a tree $T$ can be
  solved with auxiliary data of size
  $\cO(|Q^4|\cdot |S| \cdot 2^k \cdot |T|)$ which can be computed in
  time $\cO(|Q^6|\cdot |S| \cdot 2^k \cdot |T|)$, maintained within
  time $\cO(|Q^6|\cdot |S| \cdot 2^k \log(|T|))$ per update, and which
  guarantees delay $\cO(|Q^6| \cdot k \cdot |S| \cdot 2^k \cdot \log(|T|))$
  between answers.
\end{thm}

Towards the proof, we first give a high level algorithm for the enumeration in
Section~\ref{sec:highlevel}. In Section~\ref{sec:complete}, we finish the proof
by providing an implementation for the core method used in the algorithm and
proving its runtime.

\subsection{High Level Algorithm}\label{sec:highlevel}
The high level presentation of our enumeration algorithm is given as
Algorithm~\ref{alg:enum}. We assume a total order $\leq_T$ on the nodes of $T$
that can depend on our auxiliary data structure. This order is independent of
the sibling order of $T$. The algorithm then enumerates all answers in
lexicographic order according to $\leq_T$. To avoid some case distinctions, we
use a symbol $\bot$ such that $\bot \leq_T v$ for any node $v$.

To understand the algorithm, we need the notion of an incomplete answer: We
call a tuple $A \in \nodes(T)^\ell$ with $\ell \leq k$ an \emph{incomplete
  answer} if it is a prefix of some answer $B \in M(T)$. We assume that the
empty tuple $\emptyA$ is always an incomplete answer, i.e., even if
$M(T)=\emptyset$ to avoid some corner cases. We write $A \preceq B$ for two
(in-)complete answers $A$ and $B$, if $A$ is a prefix of $B$. By $|A|:=\ell$ we
denote the number of nodes of the incomplete answer $A$.

\algdef{SE}[DOWHILE]{Do}{doWhile}{\algorithmicdo}[1]{\algorithmicwhile\ #1}\begin{algorithm}[t]
 \caption{Enumeration of $M(T)$\label{alg:enum}}
 \begin{algorithmic}[1]
   \Require $k$-NFSTA $M = ((Q, \Sigma, \delta, F),S)$, tree $T$, incomplete answer $A$
   \Ensure Enumeration of all answers in $M(T)$ that are compatible with $A$

   \Function {\enum}{$M, T, A$}
     \If{$|A|=k$} $\out(A)$\label{line:out}
     \Else
       \State $A' \gets \complete(A,\bot)$ \label{line:complete1} \While{$A' \neq \bot$ } 
         \State $\enum(M,T, A')$ \label{line:recursiveEnum} \State $v \gets A'_{|A'|}$ \State $A' \gets \complete(A, v)$ \label{line:complete2} \EndWhile
     \EndIf
   \EndFunction
 \end{algorithmic}
\end{algorithm}

To enumerate all answers, \enum has to be called with the empty incomplete
answer~$\emptyA$. The sub-procedure $\complete$ extends a given incomplete
answer $A$ with another node according to the following definition.

\begin{defi}\upshape \label{def:complete}
  Let $A=(v_1,v_2,\dots,v_j)$ be an incomplete answer, then
  \[\complete(A,u) := (v_1,v_2,\dots,v_j,v)\;,\] where $v$
  is the smallest node such that $u <_T v$ and
  $(v_1,v_2,\dots,v_j,v)$ is an incomplete answer. If
  no such node exists, then we define $\complete(A,u):=\bot$.
\end{defi}

By definition of $\complete$, the lines~\ref{line:complete1}
to~\ref{line:complete2} iterate over all incomplete answers $A'$ that
result from $A$ by adding one additional node.
Before we show how to efficiently implement $\complete$, we prove
correctness of Algorithm~\ref{alg:enum}.

\begin{lem}\label{lem:correctness}
  $\enum(M, T, \emptyA)$ enumerates all answers in $M(T)$.
\end{lem}
\begin{proof} 
  We show that for every incomplete answer $A$, the function call
  $\enum(M, T, A)$ outputs exactly the answers $B$ such that $A$ is a
  prefix of $B$.  The lemma statement follows, as the empty answer
  $\emptyA$ is a prefix of every answer.

  We first observe, that every output of the algorithm is clearly a valid answer
  in $M(T)$. It remains to show that every answer is an output and that no
  answer is output twice.
 
  The proof is by induction over $|A|$. The base case is $|A|=k$. In this case,
  the only compatible answer is $A$, which is output in Line~\ref{line:out} of
  the algorithm. Let now $A=(v_1,\dots,v_\ell)$ be an incomplete answer and
  $B=(v_1,\dots,v_\ell,v_{\ell+1},\dots,v_k)$ be some answer compatible with
  $A$. Eventually some call to $\complete$ in Line~\ref{line:complete1}
  or~\ref{line:complete2} will return the incomplete answer
  $(v_1,\dots,v_\ell,v_{\ell+1})$. By the induction hypotheses, the recursive
  call in Line~\ref{line:recursiveEnum} will output $B$. Also no answer can be
  output twice, as the incomplete answers that are processed in the while-loop
  are strictly increasing according to the lexicographic order induced by
  $\leq_T$.
\end{proof}

\subsection{Implementation of Complete}\label{sec:complete}
We now present the implementation of $\complete$. Our auxiliary data structure
is a balanced forest algebra formula $\Psi$ that represents the tree $T$. We use
the \emph{extended transition algebra} that we define below. To allow
logarithmic delay, we need to know whether some node that can extend the
incomplete answer is contained in some subformula. Technically, we could extend
the signatures by the set of states that is visited in the run of the
automaton. However, this would be quite inefficient, as we are only interested
in states that occur together in the same selecting tuple.

Let $M=(N,S)$ be a $k$-NFSTA and for each selecting tuple
$s=(q^s_1,\dots,q^s_k)$ in $S$ let $Q_s=\{q^s_1,\dots,q^s_k\}$ be the set of
states that occur in $s$. We define the \emph{extended transition algebra}
$\tra_{M,s}^+$ for some selecting tuple $s \in S$ as follows:
\begin{align*}
  \tra_{M,s}^+&\quad=\quad(H^+,V^+, \concat^+_{HV}, \concat^+_{VH}, \conapp^+_{VH}) \\
  H^+&\quad=\quad(2^{Q^2 \times 2^{Q_s}}, \concat^+_{HH}, \id_Q \times \{\emptyset\}) \\
  V^+&\quad=\quad(2^{(Q^2)^2 \times 2^{Q_s}}, \conapp^+_{VV}, \id_{Q^2} \times \{\emptyset\})
\end{align*}

We recall, that an element of the transition algebra is a set of
possible signatures of runs. An element of the extended transition algebra is
a set of \emph{extended signatures}. An extended signature consists of a
regular signature---a pair of states in the horizontal monoid or a pair of
pairs of states in the vertical monoid---and a set $Q' \subseteq Q_s$ of those
states from $Q_s$ that are visited by the run. This is used in the enumeration
algorithm to evaluate whether some position of the selecting tuple can be bound
to a node inside the forest or context.

Formally, we call a tuple $(x,r)$ from $\signatures \times 2^{Q_s}$ an
extended signature and use the syntax $\extsig(\lambda)=(x,r)$.
The operations of the extended transition algebra are defined by:
\begin{multline*}
  d_1 \op^+ d_2 = \big\{ (x,r) \in \signatures \times 2^{Q_s} \mid \exists (y_1,r_1) \in d_1, (y_2,r_2) \in d_2. \\
  x \in \{y_1\} \op \{y_2\} \text { and } r=r_1 \cup r_2 \big\}\;,
\end{multline*}
where $\op^+$ is some operation of the algebra and $\op$ refers to the
corresponding operation in the transition algebra $\tra_N$ defined in
Section~\ref{sec:automata}.
We define an homomorphism $h^+$ from the free forest algebra over $\Sigma$ to
$\tra_{M,s}^+$ by
\[ h^+(\context a) \;\; = \;\;\big\{ \extsig(\lambda) \mid \lambda \text{ is a
    valid run on } \context a \big\}\;. \]
Again, the intuition is that the homomorphism maps all forests and contexts to
the set of extended signatures of their possible runs. We omit a proof, as it is
very similar to the proof of Lemma~\ref{lem:ta} and we do not need the result.

We would like to stress that in the case $k=0$, the extended transition algebra
$\tra^+_M$ is isomorphic to the transition algebra $\tra_N$, as
$2^\emptyset=\{\emptyset\}$. This reflects that a stepwise tree automaton is just the
special case of a node selecting stepwise tree automaton with a single empty
selecting tuple.

We note that the horizontal monoid $H^+$ works exactly, as illustrated
by Losemann and Martens~\cite{LosemannM-LICS14}, in the word case if there is only one selecting
tuple. Especially, our definition of $\concat_{HH}^+$ is equivalent to the
definition of $\bowtie$ in~\cite{LosemannM-LICS14}.

\begin{obs}\label{obs:traplus}
  Given a $k$-NFSTA $M=(N,S)$, operations in $\tra_M^+$ can be carried
  out in time $\cO(|Q|^6\cdot 2^k)$ using join operations.
\end{obs}

We define the total order $\leq_T$ on the nodes of $T$, that is used in
Algorithm~\ref{alg:enum}, as follows: $v \leq_T w$ if and only if $v$
occurs before $w$ in the parse tree of $\Psi$, reading the leaves from left to
right. We stress that the order $\leq_T$ depends on the formula $\Psi$.
Especially, the order can change in non-obvious ways during insertion and
deletion updates, as the structure of $\Psi$ may change, due to rotations. We
sketch how to achieve enumeration in pre- or post-order in
Section~\ref{sec:order}.

For the following presentation we would like to remind that $T_v$ denotes the forest
or context that results from $T$ by restricting $T$ to the nodes that occur in
$\Psi_v$, if $\Psi_v$ denotes a context, add a hole at the appropriate position.

We already know that elements of $\tra_M^+$ can be interpreted as sets
of extended signatures of possible runs. However, not all runs of $M$ on $T_v$
(and thus not all signatures in $h^+(T_v)$) are actually useful for
completing a given incomplete answer $A$. To be useful, an extended
signature $(x,r) \in h^+(T_v)$
has to satisfy two conditions: It has to be the signature of a run
$\lambda$ on $T_v$ that
\begin{enumerate}[(1)]
\item is compatible with $A$, i.e., for the nodes in $A$ it visits the states
  indicated by $s$ for some $s \in S$; and
\item can be extended to some accepting run $\lambda'$ over $T$ that visits all
  states in $Q_s$.
\end{enumerate}

Let now $A=(v_1,\dots,v_\ell)$ be an incomplete answer and
$Q_s=\{q^s_1,\dots,q^s_k\}$ for each $s=(q^s_1,\dots,q^s_k)$ in
$S$. We write $\lambda \models_s A$ for some run $\lambda$, if $\lambda_\self(v_i)=q_i^s$ for
$i \in \{1,\dots,\ell\}$.
Towards the above conditions, we define sets of relevant tuples of
each node $v$ of $\Psi$.
The sets $R^1_{A,s}(v)$ account for the first condition, and are
defined by
\[
  R^1_{A,s}(v) \;\; = \;\; \left\{\begin{array}{@{\hspace{2pt}}lr@{}} 
    h^+(T_v) \; \cap \; (\signatures \times \{\{q_i^s\}\}) &\text{if } v = v_i \\
    h^+(T_v) & \mathllap{\text{if $v \notin A$ is a leaf of $\Psi$}} \\
    R^1_{A,s}(\lc{v}) \op^+_v R^1_{A,s}(\rc{v}) & \quad\text{if $v$ is not a leaf of $\Psi$}
\end{array} \right.
\]
For leaf nodes of the formula that occur in $A$, we only keep those tuples from
$h^+(T_v)$ that are compatible with $A$, i.e., that assign the correct state to
$v$. We note that if $v$ occurs at two positions $i$ and $j$ in $A$, then
$q_i^s$ and $q_j^s$ are necessarily equivalent. For leaf nodes of the formula
that do not occur in $A$, we simply retain all possible signatures. Finally, for
inner nodes of the formula, we use $\tra_M^+$ to compute $R^1_{A,s}$, i.e., we
perform the same join operation that is used to compute $h(T_v)$ in the first
place. The only difference is that we apply the operation only to those
signatures in the two children of $v$ that are compatible with $A$.

The sets $R^2_{A,s}(v)$ additionally account for the second condition and are
computed top-down, starting with the set of signatures compatible with $A$ that
belong to accepting runs in the root:
\[
  R^2_{A,s}(v) \;\;=\;\;\begin{cases}
    R^1_{A,s}(v) \; \cap \; (\{(q_I,q_F)\} \times \{Q_s\}) & \text{if } v=\Root(\Psi) \\
    \{ x \in R^1_{A,s}(v) \mid \big(R^1_{A,s}(w) \op^+_u \{x\} \big) \cap
    R^2_{A,s}(u) \neq \emptyset \}  & \text{if $v=\rc{u}$} \\
    \{ x \in R^1_{A,s}(v) \mid \big(\{x\} \op^+_u R^1_{A,s}(w)\big) \cap
    R^2_{A,s}(u) \neq \emptyset \} & \text{if $v = \lc{u}$}
  \end{cases}
\]
Here, $w$ is always the sibling of $v$.
For the root, we simply drop all signatures that indicate that the run is either
not accepting or does not contain all states that are needed to produce an
answer. For inner nodes, we use again the operations of $\tra_M^+$ to compute
the desired signatures, but this time, we know the result of the operation (it
is stored at the parent $u$) and need to compute all those signatures from
$R^1_{A,s}(v)$ that can be combined with some signature of $v$'s sibling $w$ to
obtain some signature from $R^2_{A,s}(u)$. There are two cases, depending on
whether $v$ is left of $w$ or vice versa.

The computation of $R^1_{A,s}$ followed by $R^2_{A,s}$ is similar to
Yannakakis algorithm~\cite{Yannakakis81} for computing the result of acyclic relational joins.
This is not surprising at all, as essentially we are computing an acyclic join. The
only difference is that for the last component, i.e., for computing the subset
of $Q_s$ that occurs in the run, we compute the union.

The following two lemmas show that the definitions of $R^1_{A,s}$ and
$R^2_{A,s}$ work as expected.
\begin{lem}\label{lem:R1}
  It holds that $(x,r) \in R^1_{A,s}(v)$ if and only if there exists a
  run $\lambda$ on $T_v$ such that $\extsig(\lambda)=(x,r)$ and
  $\lambda \models_s A$.
\end{lem}
\begin{proof}
  The proof is by induction and very similar to the proof of
  Lemma~\ref{lem:ta}. We show both directions by induction starting with the if-direction.

  If there is a run $\lambda$ on $T_v$ that is compatible with $A$, then
  $(\signature(\lambda),\states(\lambda) \cap Q_s) \in R^1_{A,s}(v)$. For the
  base cases this follows directly from the definition of $R^1_{A,s}$. For the
  induction case, where $\Psi_v=\Psi_l \op^+ \Psi_r$ for some operation $\op^+$
  and some subformulas $\Psi_l$ and $\Psi_r$ this follows, as we can decompose
  $\lambda$ into two runs $\lambda_l$ and $\lambda_r$ (just as in the proof of
  Lemma~\ref{lem:ta}) and by the induction assumption we have
  $(\signature(\lambda_l),\states(\lambda_l)) \in R^1_{A,s}(\lc{v})$ and
  $(\signature(\lambda_r),\states(\lambda_r)) \in R^1_{A,s}(\rc{v})$.

  For the only-if-direction assume that $(x,r) \in R^1_{A,s}(v)$. Then there is
  a run $\lambda$ on $T_v$ that is compatible with $A$. For the base cases this
  follows again directly from the definition of $R^1_{A,s}$. For the induction
  case, we get from the definition of $R^1_{A,s}$ that there are extended
  signatures $(x_1,r_1) \in R^1_{A,s}(\lc{v})$ and $(x_2,r_2) \in
  R^1_{A,s}(\rc{v})$, such that $x \in \{x_1\} \op_v \{x_2\}$ and $r=r_1 \cup
  r_1$. By the induction assumption there are runs $\lambda_i$ such that
  $\extsig(\lambda_i)=(x_i,r_i)$ for $i \in \{1,2\}$. Following the proof of
  Lemma~\ref{lem:ta}, $\lambda_1$ and $\lambda_2$ can be combined into a run
  $\lambda$ over $T_v$ with $\signature(\lambda)=x$. As $r=r_1 \cup r_2$ it
  follows that $\extsig(\lambda)=(x,r)$.
\end{proof}

\begin{lem}\label{lem:R2}
  It holds that $(x,r) \in R^2_{A,s}(v)$ if and only if there exists a run
  $\lambda$ on $T$ such that $\lambda$ is accepting,
  $\extsig(\lambda_v)=(x,r)$ and $\lambda \models_s A$, where
  $\lambda_v$ is the restriction of $\lambda$ to $T_v$.
\end{lem}
\begin{proof}
  We again start with the if-direction. The proof is by a top-down induction. At the
  root, the claim holds by the definition of $R_{A,s}^2$ and Lemma~\ref{lem:R1}.
  We note that at the root $\lambda=\lambda_v$. Let now $v$ be a non-root node
  of $\Psi$. \mbox{W.l.o.g.}, we assume that $v$ is the left child of its parent
  $u$. The other case is symmetric. If there is an accepting run $\lambda$ on
  $T$ such that $\lambda \models_s A$, then by the induction assumption there is
  the signature $\extsig(\lambda_u) \in R^2_{A,s}(u)$, where $\lambda_u$ is the
  restriction of $\lambda$ to $T_u$. Let $w$ be the right sibling of $v$. By
  Lemma~\ref{lem:R1} it holds that $\extsig(\lambda_w) \in R^1_{A,s}(w)$, where
  $\lambda_w$ is the restriction of $\lambda$ to $T_w$. By the definition of the
  extended transition algebra and the fact that $\lambda_v$ and $\lambda_w$ are
  compatible and be combined into $\lambda_u$, we have that
  $\{\extsig(\lambda_v)\} \op^+_u \{\extsig(\lambda_w)\} =
  \{\extsig(\lambda_u)\}$. By the definition of $R^2_{A,s}$ we have that
  $\extsig(\lambda_v) \in R^2_{A,s}(v)$.

  We continue with the only-if-direction. Let $(x,r)$ be a signature from
  $R^2_{A,s}(v)$. By definition of $R^2_{A,s}$ we have that there are signatures
  $(x_u,r_u) \in R^2_{A,s}(u)$ and $(x_w,r_w) \in R^1_{A,s}(w)$. By the
  induction assumption there is an accepting run $\lambda$ such that $\lambda
  \models_s A$ and $\extsig(\lambda_u)=(x_u,r_u)$, where $\lambda_u$ is the
  restriction of $\lambda$ to $T_u$. By Lemma~\ref{lem:R1} and the definition
  of$ R^1_{A,s}$, we have that there is a run $\lambda_u'$ on $T_u$, such that
  $\lambda_u' \models_s A$, $\extsig(\lambda_u')=\extsig(\lambda_u)$ and the
  restrictions $\lambda_v'$ and $\lambda_w'$ of $\lambda_u'$ to $T_v$ and $T_w$,
  respectively are such that $\extsig(\lambda_v')=(x,r)$ and
  $\extsig(\lambda_w') \in R^1_{A,s}$. The run $\lambda'$, where we replace
  $\lambda_u$ with $\lambda_u'$ in $\lambda$ is an accepting run such that
  $\lambda' \models_s A$ and the restriction of $\lambda'$ to $T_v$ has
  signature $(x,r)$. We note that we can replace $\lambda_u$ by $\lambda_u'$, as
  both runs have the same signature. This concludes the proof.
\end{proof}

The definitions of $R^1_{A,s}$ and $R^2_{A,s}$ yield straightforward algorithms
to compute these sets. The computation of $R^1_{A,s}$ can be done bottom-up
(just as the computation of $\tra^+$), while the computation of $R^2_{A,s}$ can
be done top-down. We note that the runtime of the naive algorithm to compute $R^1_{A,s}(u)$
is linear in $||\Psi_u||$, as the computation is bottom-up. However, after
extending an incomplete answer $A$ with an additional node $v$ to an incomplete
answer $A'$, it is sufficient to compute $R^1_{A',s}(u)$ for those nodes $u$ in
$\Psi$ that are an ancestor of $v$, as---by the definition of the sets
$R^1_{A',s}$---it holds that $R^1_{A',s}(u)=R^1_{A,s}(u)$ for all nodes $u$ such
that $\Psi_u$ does not contain $v$. In fact $R^1_{A,s}(u)=h^+(\Psi_u)$ if no node
from $A$ occurs in $\Psi_u$.

We now have all ingredients for an implementation of the procedure
\complete that we present as Algorithm~\ref{alg:complete}. We first prove
correctness before we give an upper bound on the runtime. We use
$\states(R^2_{A,s}(v))$ to denote the set of states that occur in some tuple
of $R^2_{A,s}(v)$, i.e.,
\[\states(R^2_{A,s}(v))=\bigcup_{(x,r) \in R^2_{A,s}(v)} r \;.\]

\begin{algorithm}[t]
 \caption{Procedure \call{Complete} as used in Algorithm~\ref{alg:enum}\label{alg:complete}}
 \begin{algorithmic}[1]
   \Require incomplete answer $A=(v_1,v_2,\dots,v_j,\bot,\dots,\bot)$, node $u$
   \Ensure the answer $\complete(A,u)$ from Definition~\ref{def:complete}

   \Function {\complete}{$A, u$}
     \State \Return $\complete(A, \operatorname{Root}(\Psi),u)$
   \EndFunction

   \Function {\complete}{$A, v, u$}
\State compute $R^2_{A,s}(v)$ for $s \in S$\label{alg:complete:computeR2} \If{$\max(\nodes(T_v)) \leq u \;\text{ or } \; q_{j+1}^s \notin
       \states(R^2_{A,s}(v))$ for every $s \in S$} 
     \State \Return $\bot$ \label{alg:complete:returnBot}
     \EndIf
     \If{$\operatorname{isLeaf}(v)$}
     \State $A' \gets (v_1,v_2,\dots,v_j,v)$ \label{alg:complete:computeA}
\Else
       \State $A' \gets \complete(A,\lc v,u)$ \label{alg:complete:rec1}
       \If{$A' = \bot$} $A' \gets \complete(A, \rc v,u)$ \EndIf \label{alg:complete:rec2}
\EndIf
     \If{$A' \neq \bot$}  
     compute $R^1_{A',s}(v)$ for $s \in S$\label{alg:complete:computeR1}
     \EndIf
     \State \Return $A'$ \label{alg:complete:return}
\EndFunction
 \end{algorithmic}
\end{algorithm}

\begin{lem}\label{lem:comCorrect}
  The procedure \complete correctly computes the incomplete answer as required by
  Definition~\ref{def:complete}.
\end{lem}
\begin{proof}
  The main challenge of the procedure is to find a node $v_{j+1}$ that
  can be used to extend the incomplete answer $A$. As our order of the
  nodes of $T$ is induced by the order of the leaves of $\Psi$, we
  have to find the leftmost leaf of $\Psi$ that can be used to extend
  $A$. By the definition of $R^2_{A,s}$, this is the leftmost leaf $v$ with
  $v > u$ and $q_{j+1}^s \in \states(R^2_{A,s}(v))$ for some $s \in S$.

  The procedure returns in Line~\ref{alg:complete:returnBot}, only if it is sure
  that no such node $v_{j+1}$ can be found among the descendants of $v$, either
  because all nodes below $v$ are smaller or equal than $u$, or because $q_{j+1}
  \notin \states(R^2_{A,s}(v))$ and therefore also $q_{j+1} \notin
  \states(R^2_{A,s}(w))$ for any $w$ below $v$ by the definition of $R^2_{A,s}$.
  
  If $v$ is a leaf, the procedure either returns $\bot$ in
  Line~\ref{alg:complete:returnBot} or computes the correct incomplete answer
  $A'$. We note that if the algorithm does not return in
  Line~\ref{alg:complete:returnBot} and $v$ is a leaf, then $v$ is the desired
  node.

  If $v$ is not a leaf, the algorithm first descends into the left
  subtree and only if no appropriate node was found there descends
  into the right subtree. This way, the algorithm finds the leftmost leaf
  satisfying the conditions. 
\end{proof}

\begin{lem}\label{lem:comRuntime}
  The procedure \complete runs in time $\cO(\log(|T|) \cdot |Q|^6\cdot |S| \cdot 2^k)$.
\end{lem}
\begin{proof}
  The time spent in each invocation of \complete (excluding time spent in
  recursive calls) is dominated by the computation of $R^2_{A,s}(v)$ and
  $R^1_{A',s}(v)$ for each $s \in S$. Both operations can be carried out in time
  $\cO(|Q|^6\cdot 2^k)$ for each $s \in S$ using ordinary join algorithms. As observed above, we
  can compute $R^1_{A',s}(v)$ using the existing information for $R^1_{A,s}$ for
  the child of $v$ that was not used to extend $A$ to $A'$.

  It remains to show that the total number of recursive calls is bounded by
  $\cO(\log(|T|))$. Obviously it is enough to count non-tail calls, as for every
  non-tail call, there can be at most two tail calls. Calls that return in
  Line~\ref{alg:complete:returnBot} are clearly tail calls. Therefore, we only
  count calls that return in Line~\ref{alg:complete:return}. There can be at
  most $\height(\Psi)$ calls that return a value that is not $\bot$, as such a
  value prevents further recursive calls. And returning $\bot$ in
  Line~\ref{alg:complete:return} is only possible if $u \in \Psi_v$. If $u$ is
  smaller than all nodes in $\Psi_v$, then the definition of $R^2_{A,s}$ ensures
  that we can find a node to extend $A$ in $\Psi_v$. And if $u$ is larger than
  all nodes in $\Psi_v$, then the algorithm already returns in
  Line~\ref{alg:complete:returnBot}. Altogether we have shown that there can be
  at most $\cO(\log(|T|))$ many recursive calls.
\end{proof}

We now have all ingredients to show Theorem~\ref{thm:enuminc}.

\begin{proof}
  The delay follows from Lemma~\ref{lem:comRuntime} and the fact that we need at
  most $k$ calls to $\complete$ to compute the next answer. The preprocessing and
  update times follow from Theorem~\ref{thm:update} and
  Observation~\ref{obs:traplus}. We note that we need to compute one algebra
  operation for each inner node of the formula during preprocessing and we need
  at most logarithmically many algebra operations per update, exactly as in
  Theorem~\ref{thm:inceval}.
\end{proof}

\subsection{Enumerating in Pre-order and Post-order}\label{sec:order}
There is one unaesthetic detail in our algorithm, that can be fixed: Our
implementation enumerates the tree in the order in which nodes appear in the
formula, i.e., the enumeration order depends on the internal state of our data
structure and can change due to updates. With a small change in the forest
algebra and the enumeration algorithm, it is possible to enumerate $T$ in
pre-order or post-order. Algorithm~\ref{alg:complete} needs to be changed so
that it does three recursive calls at each inner node that represents a context
application. One to search for $v$ in the context among the nodes before the
hole, a second that searches the tree the context is applied to, and a third
searching the context again, but now on the nodes after the hole. Thus the first and third recursive call are on the left child and the second recursive call is on the right child.
To allow these
three calls, the vertical monoid of the forest algebra needs to be extended such
that it carries the information which states of $S$ are visited before and after
the hole, respectively. The vertical monoid thus has the elements $2^{(Q^2)^2 \times 2^{Q_s} \times 2^{Q_s}}$ and the operations that involve the vertical monoid have to be changed accordingly.
In the asymptotic runtime bounds this change leads to another factor of $2^k$ resulting from the larger monoid, i.e., the runtime of \complete is now $\cO(\log(|T|) \cdot |Q|^6\cdot |S| \cdot 2^k \cdot 2^k)$.

\section{Concluding Remarks and Further Directions}\label{sec:conclusion}
We presented a framework for representing trees by forest algebra formulas of
logarithmic depth. The algorithms for preprocessing and maintaining the formulas
unfortunately have some quirks in them, which are needed to prove our runtime bounds.
For example, we do not always apply all possible rotations after an update, as
we could not prove that this simpler algorithm still runs in logarithmic
time in the worst case. It would be nice to show the same bounds for simplified
versions of the algorithms.

We showed that by using our framework it is possible, after a linear preprocessing
phase, to enumerate MSO queries over trees with logarithmic delay and that we
can restart enumeration after logarithmic time after an update to the tree.
In~\cite{AmarilliBMN-pods19} it has already been shown that our framework is
usable for an entirely different enumeration approach that solves the same
problem. Using a more sophisticated enumeration algorithm on top of our
forest algebra framework, it has been shown that constant delay enumeration can
be achieved while the updates can still be carried out in logarithmic time. This
result is at least close to optimal. It is not possible that the delay and the
update time are both in $o(\log(n)/\log(\log(n)))$~\cite{AmarilliBMN-pods19}.

Possible follow up work is to look which other results for static trees can be
lifted to trees with updates using the framework, just
as~\cite{AmarilliBMN-pods19} did with the enumeration algorithm from
~\cite{AmarilliBJM-icalp17}, or which other results for dynamic strings can be
lifted to dynamic trees, just as we did with the enumeration algorithm
from~\cite{LosemannM-LICS14} in Section~\ref{sec:enum}.

A natural question is whether our approach can be generalized to a bigger class
of structures. As long as no updates are considered, many results for trees
carry over to structures of bounded treewidth. This usually works by computing a
tree decomposition for a graph of bounded treewidth and then solve the problem
on the tree decomposition. A very interesting but probably also very hard
question is, whether the algebraic approach of forest algebras can somehow be
extended to structures of bounded treewidth. More concretely: is there an
algebraic framework that can represent a graph by a formula $\Phi$ of
logarithmic height and---after a local update to the underlying graph---compute
an updated formula $\Phi'$ representing the updated graph in a reasonable time?

\section*{Acknowledgment}
We thank the anonymous reviewers for exceptionally detailed and profound reviews. The presentation of the article has been improved a lot.

\bibliographystyle{alphaurl}
\bibliography{references}

\end{document}